\documentclass[aps,preprint,showpacs]{revtex4}
\usepackage{epsfig,amssymb,amsmath,bm}

\begin{document}

\title{
 Spin effects
 and baryon resonance dynamics in $\bm\phi$-meson photoproduction at few GeV}
\author{A.I.~Titov$^{ab}$ and T.-S.H. Lee$^{c}$}
 \affiliation{
 $^a$Advanced Science Research Center, Japan Atomic Energy Research Institute,
 Tokai, Ibaraki,
 319-1195, Japan\\
 $^b$Bogoliubov Laboratory of Theoretical Physics, JINR,
  Dubna 141980, Russia\\
 $^c$Physics Division, Argonne National Laboratory, Argonne,
  Illinois 60439, USA}


\begin{abstract}
 The diffractive $\phi$-meson photoproduction amplitude is
 dominated by the Pomeron exchange process and contains
 the terms that govern the  spin-spin and spin-orbital interactions.
 We show that these terms  are responsible for the spin-flip transitions
 at forward photoproduction angles and appear
 in the angular distributions of $\phi\to K^+K^-$ -decay
 in reactions with unpolarized and polarized photon beams.
 At large momentum transfers, the main contribution to the
 $\phi$-meson photoproduction is found to be due to
 the excitation of nucleon resonances.
 Combined  analysis of $\omega$ and $\phi$ photoproduction
 indicates strong OZI-rule violation in $\phi NN^*$ - couplings.
 We also show that the spin observables are
 sensitive to the dynamics of $\phi$-meson photoproduction at large angles
 and  could help distinguish  different
 theoretical models of nucleon resonances.
 Predictions for spin effects in $\phi$-meson
 photoproduction are presented for future experimental tests.

\end{abstract}


 \maketitle

\section{Introduction}

 The photoproduction of light vector meson ($\rho,\omega,\phi$)
 is an interesting reaction for many reasons. At high energies
 ($W\equiv \sqrt{s}\gtrsim 10$ GeV) it brings
 information on the dynamics of the Pomeron
 exchange~\cite{DL84-92,LN87,Gol93,LM95,PL96,HL98,CKK97,Soloviev}.
 At low energies ($W\sim 2$ GeV), its observables
 are  sensitive to the resonance
 channel and can be used to obtain some unique information about the
 structure of baryon resonances, properties of
 $VNN^*$-interactions~\cite{ZLB98,Zhao:phiomega,Zhao:omega,PM00,OTL01,Zhao:phi,LWF01,TL02},
 and possible manifestation of the so-called
 "missing resonances"~\cite{IK77-80,Caps92,CR94}.
 Therefore,  the study of vector-meson photoproductions
 is  an important component of the
 experimental programs at the electron and photon facilities
 like Thomas Jefferson National Accelerator Facility, the LEPS at SPring-8.
 ELSA-SAPHIR at Bonn and  GRAAL at Grenoble.
 The $\phi$-meson photoproduction at relatively low energies $E_\gamma\simeq
 2-3$ GeV plays particularly important role. It is expected
 that in the diffractive region,
 the dominant contribution comes from the Pomeron exchange, since
 trajectories associated with conventional meson exchanges
 are suppressed by the OZI-rule. The exception is the finite contribution
 of the pseudoscalar $\pi$,$\eta$-meson-exchange channel,
 but its properties are quite well understood~\cite{TLTS99,Laget2000}.
 Therefore, the low-energy $\phi$-meson photoproduction may be
 used for studying the presence of additional trajectories.
 Candidates are trajectories associated with
 a scalar meson~\cite{TLTS99,Will98} and $f_2'$-meson~\cite{Laget2000}
 containing a large amount
 of strangeness,  glueball~\cite{NakanoToki}
 or other exotic channels~\cite{Kochelev}.
 But the relative contributions of these additional processes
 can not be well defined
 within  Regge phenomenology and must be determined from comparisons
 with experimental data. Spin observables
 are of  crucial importance for such studies. As a matter of fact,
 the Pomeron exchange amplitude which is inspired by  multi-gluon
 exchange~\cite{DL84-92} contains specific spin-dependence
 terms  which are negligible in $\phi$-meson photoproduction
 at high energies but become important  at a few GeV.
 These interactions lead to spin-flip processes and give non-trivial
 behaviour of
 the spin-density matrix elements even at forward angles.
 Therefore, the angular distribution of
 $\phi\to K^+K^-$ decay can be used as a tool to study the diffractive
 mechanism, and is complementary to the measurement of unpolarized
 cross section.

 Another subject is related to the strange degrees of freedom
 in a nucleon. Analysis of magnetic~\cite{GenzHolner,Jaffe}
 and electroweak~\cite{Gerasimov} moments of baryons
 show that the $\phi$-meson couples more strongly to the
 nucleon than expected on the basis of the OZI-rule~\cite{OZI}.
 The presence of the strange-quark content in the nucleon
 was indicated by measurements of the $\pi$-nucleon
 term~\cite{Sigmaterm},
 $\phi$-meson production in proton annihilation
 at rest~\cite{ASTERIX91,CBC95,OBELIX96}
 and deep-inelastic  electroweak lepton-nucleon scattering
 (see Ref.~\cite{Ellis} for references and a compilation of the data).

 The $\phi$-meson photoproduction seems to be an effective and
 promising candidate process for studying  the hidden strangeness in a nucleon.
 The  backward-angle photoproduction  is dominated by $s$ and $u$
 channels of the nucleon and resonant amplitudes and directly
 related to the
 strength of $\phi NN$ and $\phi NN^*$-interactions.
 The finite - strange content
 leads to an increase of this strength compared to the
 expectation based on the standard OZI-rule
 violation (OZI-rule-evading interaction~\cite{Will98}).
 This effect must be seen in both unpolarized and spin
 observables at large momentum transfers.

  At forward angles, the nucleon and resonant
  contributions become negligible  for $\phi$-meson photoproduction
  and OZI-rule violation could appear
  as  direct $s\bar s$-knockout~\cite{Henley} from a nucleon.
  The most promising here is a measurement of spin observables
  which represent the interference of the weak $s\bar s$-knockout
  and strong vector-meson dominance photoproduction
  amplitudes~\cite{Tabakin,TOY97,TOYM98}.
  It is clear that for this purpose the diffractive amplitude must be
  established unambiguously.

 The purpose of this paper is to investigate the problems mentioned above.
 The main differences with the previous studies of
 the conventional non-strange amplitude
 of $\gamma p\to \phi p$-reaction~\cite{TLTS99,Will98} are in giving a
 detailed analysis of the spin properties of
 the amplitude in diffractive region.
 We will present a comprehensive analysis of all spin-density
 matrix elements which are responsible for
 the angular distributions of $K^+K^-$
 in the reaction $\gamma p\to\phi p$ with unpolarized and polarized
 photons at a few GeV. For the most important matrix elements we give
 an estimation in an explicit analytical form,
 which is useful for the qualitative analysis.

 The backward angle photoproduction
 is described  by the nucleon resonance excitations.
 For the latter, we use an effective Lagrangian approach developed for
 $\omega$-meson photoproduction~\cite{TL02}, where all
 known nucleon resonances listed in
 Particle Data Group~\cite{PDG} are included.
 This resonant model is  different from the  approach of
 Ref.~\cite{Zhao:phi}, which results in giving significantly
 different predictions of some spin observables.

 This paper is organized as follows. In Section II we define the
 kinematics and observables.  Formula for
 calculation various spin observables are also introduced
 here. The basic amplitudes for the  conventional
 processes, such as Pomeron exchange, Reggeon exchanges, and
 resonance excitations, are given  in Section III.
 In Section~IV we discuss results and make predictions for the future
 experiments.  The summary is given in Section V.
 In Appendix A we discuss an extreme case when the exotic trajectories
 become dominant in the near-threshold  energy region.

\section{Kinematics and Observables}

The scattering amplitude $T$ of the $\gamma p \rightarrow V p$
reaction (where $V$ can be $\phi$ or $\omega$) is related to the
$S$-matrix by
\begin{equation}
S^{}_{fi} = \delta^{}_{fi}
- i(2\pi)^4 \delta^4(k + p - q - p') T^{}_{fi},
\label{S:conv}
\end{equation}
where $k$, $q$, $p$ and $p'$ denote the four-momenta of the
incoming photon, outgoing vector meson, initial nucleon, and final
nucleon, respectively. The standard Mandelstam variables are
defined by  $t = (p - p')^2 = (q-k)^2$, $s \equiv W^2 = (p+k)^2$,
and the vector meson  production angle $\theta$ by $\cos\theta
\equiv {\bf k} \cdot {\bf q} / |{\bf k}| |{\bf q}|$.
 We use convention  of Bjorken and Drell to define the $\gamma$ matrices;
the Dirac spinors are normalized as $\bar u(p)\gamma_\alpha
u(p)=2p_\alpha$.

The scattering amplitude is written as
\begin{equation}
T^{}_{fi} = \frac{I_{fi}} {(2\pi)^6\,
\sqrt{ 2E^{}_\omega({\bf q})\,2|{\bf k}|
2E_{N}({\bf p}) 2E_{N}({\bf p}')} },
\label{T:conv}
\end{equation}
 where $E^{}_i({\bf p}) = \sqrt{M^2_i + {\bf p}^2}$ with
 $M^{}_i$ denoting the mass of the particle $i$.
 In the c.m.s. the quantization axis ($\bf z$) is
 chosen along the beam momentum, and the $\bf y$-axis is
 perpendicular to the production plane:
 $\bf y=p\times p'/|p\times p'|$.
 The differential cross section is related to the invariant
 amplitude by
\begin{eqnarray}
\frac{d\sigma_{fi}}{dt}=
\frac{1}{64\pi(W^2-M_N^2)^2}\sum_{m_i\,m_f\lambda_\gamma\lambda_V}
|I_{fi}|^2, \label{cs}
\end{eqnarray}
 where $m_i,m_f$ are the proton spin projections in the initial
 and final state, respectively, and
 $\lambda_\gamma\,\lambda_V$ are the helicities of the
 incoming photon and  outgoing  vector meson, respectively.
 In this paper
 we will also investigate some of the single and double spin
 observables~\cite{Tabakin}.

 The considered  beam
 asymmetry ($\Sigma_x$) for the linearly polarized photons reads
  \begin{eqnarray}
  \Sigma_x&=&
   \frac{d\sigma_{{\bf y}} - d\sigma_{{\bf x}}}
  {d\sigma_{{\bf y}} + d\sigma_{{\bf x}}}=
   \frac{{\rm
  Tr}\left[I_{fi}\sigma_\gamma^x\,I_{fi}^\dagger\right]} {{\rm
  Tr}\left[I_{fi}\,I_{fi}^\dagger\right]},
  \label{sigma_x}
\end{eqnarray}
 where the subscript ${\bf y}\,({\bf x})$ corresponds to a photon linearly
 polarized along the ${\bf y}$ (${\bf x}$) axis.
 In the case of a circularly-polarized photon beam, the double
 beam-target (recoil) asymmetry is very sensitive
 to the production mechanism~\cite{TOYM98}. Therefore, in the present
 work we analyzes the beam-target asymmetry
\begin{equation}
 C^{BT}_{zz} = \frac{
 d\sigma(\rightrightarrows)-d\sigma(\leftrightarrows)}
  {d\sigma(\rightrightarrows) + d\sigma(\leftrightarrows)},
 \label{C-BT}
\end{equation}
where the arrows represent the spin projections of the incoming
photon and the target protons: $(\rightrightarrows)$ and
($\leftrightarrows$) thus  correspond to the initial states with
the total spin equal to $\frac32$ and $\frac12$, respectively.

 The double polarization observables related to the beam polarization and
 polarization of the outgoing vector mesons are described in terms
 of spin-density matrices $\rho_{ij}$, which determine the vector-meson decay
 distributions in its rest frame~\cite{SS68} and are defined by
 \begin{eqnarray}
 \rho^{0}_{\lambda\lambda'}&=&
 \frac{1}{N}\sum_{\alpha,\lambda_\gamma}
 I_{\alpha;\lambda,\lambda_\gamma}\,I^\dagger_{\alpha;\lambda',\lambda_\gamma},
 \nonumber\\
 \rho^{1}_{\lambda\lambda'}&=&
 \frac{1}{N}\sum_{\alpha,\lambda_\gamma}
 I_{\alpha;\lambda,-\lambda_\gamma}\,I^\dagger_{\alpha;\lambda',\lambda_\gamma},
 \nonumber\\
 \rho^{2}_{\lambda\lambda'}&=&
 \frac{i}{N}\sum_{\alpha,\lambda_\gamma}\lambda_\gamma
 I_{\alpha;\lambda,-\lambda_\gamma}\,I^\dagger_{\alpha;\lambda',\lambda_\gamma},
 \nonumber\\
 \rho^{3}_{\lambda\lambda'}&=&
 \frac{1}{N}\sum_{\alpha,\lambda_\gamma}\lambda_\gamma
 I_{\alpha;\lambda,\lambda_\gamma}\,I^\dagger_{\alpha;\lambda',\lambda_\gamma},
  \label{rhomatrix}
\end{eqnarray}
 where the symbol $\alpha$ includes the polarizations of the
 incoming and the outgoing baryons and the normalization factor reads
 \begin{eqnarray}
 {N}=\sum_{\alpha,\lambda,\lambda_\gamma}
 I_{\alpha;\lambda,\lambda_\gamma}\,I^\dagger_{\alpha;\lambda,\lambda_\gamma}.
 \label{Norm}
 \end{eqnarray}

 The $\phi\to K^+K^-$ - decay distribution as a function of
 the polar ($\Theta $) and azimuthal ($\Phi $) angles  is expressed
 through the spin-density-matrix elements and depends on the beam
 polarization. The  polarization vectors of the linear
 ($\bm \varepsilon $) and circular ($\bm{\varepsilon}^\lambda,\,\lambda=\pm1 $)
 photon polarizations read
 \begin{eqnarray}
 \bm{\varepsilon} &=& (\cos\Psi,\,\sin\Psi,\,0),\nonumber\\
 \bm {\varepsilon}^\lambda &=&-\frac{\lambda}{\sqrt{2}}
                        (1, \,i\lambda{}, \, 0).
 \label{epsilon}
 \end{eqnarray}
 For easy reference we list here the explicit form
 of the decay angular distribution $W(\cos\Theta,\Phi,\Psi)$
 for various photon polarizations in the rest frame of the outgoing
 $\phi$-meson.
 For unpolarized photons it reads
 \begin{equation}
 {W}_{\rm unpol.}(\cos\Theta, \Phi) =
 { W}^0(\cos\Theta,
 \Phi),
  \label{UDD0}
\end{equation}
with
\begin{eqnarray}
{W}^0 (\cos\Theta, \Phi) &=& \frac{3}{4\pi} \left\{ \frac12 ( 1 -
\rho^0_{00} ) + \frac12 ( 3 \rho^0_{00} - 1 ) \cos^2\Theta \right.
\nonumber \\ && \qquad \left. \mbox{} - \sqrt2 \, \mbox{Re}
\rho^0_{10} \sin2\Theta \cos\Phi - \mbox{}\rho^0_{1-1}
\sin^2\Theta \cos2\Phi \right\}. \label{UDD}
\end{eqnarray}
For the circularly-polarized photons of helicity $\lambda_\gamma =
\pm 1$ the angular distribution has the following form
 \begin{eqnarray}
 {W}^\pm (\cos\Theta, \Phi) &=& {W}^0(\cos\Theta, \Phi)\nonumber\\
 &\pm& \frac{3}{4\pi} P_\gamma \left\{ \sqrt2 \, \mbox{Im} \rho^3_{10}
  \sin2\Theta \sin\Phi + \mbox{Im} \rho^3_{1-1} \sin^2\Theta
  \sin2\Phi \right\},
  \label{CDD}
 \end{eqnarray}
 where $P_\gamma$ is the strength of polarization
 ($0\leq P_\gamma \leq 1$).

 In the case of the linearly-polarized photons
 the decay distribution is defined as
\begin{eqnarray}
 {W}^L(\cos\Theta, \Phi, \Psi) &=& {W}^0(\cos\Theta, \phi)\nonumber\\
 &-& P_\gamma \left({W}^1 (\cos\Theta, \Phi)\,\cos 2\Psi
                  + {W}^2 (\cos\Theta, \Phi)\,\sin 2\Psi\right),
\label{LDD}
\end{eqnarray}
where $\Psi$ denotes the angle between the photon-polarization
vector and $\phi$-meson production plane (cf. Eq.~\ref{epsilon}).
The partial distributions $W^{1,2}$ read
\begin{eqnarray}
 {W}^1 (\cos\Theta, \Phi) &=& \frac{3}{4\pi} \left\{ \rho^1_{11}
 \sin^2\Theta + \rho^1_{00} \cos^2\Theta \right. \nonumber \\ &&
 \qquad \left. \mbox{} - \sqrt2 \, \mbox{Re} \rho^1_{10}
 \sin2\Theta \cos\Phi - \mbox{}\rho^1_{1-1} \sin^2\Theta
 \cos2\Phi \right\}\nonumber\\
 {\cal W}^2 (\cos\Theta, \Phi) &=& \frac{3}{4\pi} \left\{ \sqrt2 \,
 \mbox{Im} \rho^2_{10} \sin2\Theta \sin\Phi + \mbox{Im}
 \rho^2_{1-1} \sin^2\Theta \sin2\Phi \right\}
 \label{W12}
\end{eqnarray}
 Spin-density matrix elements depend on the choice of the
 quantization axis
 ($\bf z'$) which defines the reference frame of the
 vector meson-decay distribution.
 There are several choices of the quantization axis $\bf z'$
 in the vector-meson rest frame:
 the helicity system with  $\bf z'$
 opposite to the velocity  of the recoiling nucleon,
 the Gottfried-Jackson system (GJ) with  $\bf z'$  parallel to the momentum
 of the photon, and the Adair  system with  $\bf z'$ parallel to the photon
 momentum in the c.m.s. Although the general formalism for the
 analysis of the vector-meson decay does not depend on the
 system, all our calculations are done in the GJ system, where  some of
 the amplitudes have a simple helicity-conserving form regardless
 of the momentum transfers.

  Using the two-dimensional decay distribution
  of Eq.~(\ref{UDD}),
  one can get the one-dimensional distributions after integrating
  over the remaining  variables
\begin{eqnarray}
 {W}^0 (\cos\Theta) &=&\frac32\left(
 \frac{1}{2}(1-\rho^0_{00})\sin^2\Theta
  + \rho^0_{00}\cos^2\Theta\right) ,\nonumber\\
 {W}^0 (\Phi)&=& \frac{1}{2\pi}\left(1-2{\rm Re}\rho^0_{1-1}\cos2\Phi
 \right).
 \label{W0}
 \end{eqnarray}
  In the case of the linearly polarized beam, the distributions
  depend additionally on the direction of the polarization vector
 \begin{eqnarray}
 {W}^L (\cos\Theta,\Psi) &=& W^0(\cos\Theta)
 -\frac32(\rho^1_{11}\sin^2\Theta +\rho^1_{00}\cos^2\Theta)
  P_\gamma\cos(2\Psi),
 \nonumber\\
 {W}^L(\Phi,\Psi) &=& W^0(\Phi)+\frac{1}{\pi}
 P_\gamma\left(\overline{\rho}^{1}_{1-1}\cos[2(\Phi-\Psi)]
 +\Delta_{1-1}\cos[2(\Phi+\Psi)]\right),
 \label{WL-0}
 \end{eqnarray}
 where
 \begin{eqnarray}
 \overline{\rho}^{1}_{1-1}=
 \frac{1}{2}(\rho^{1}_{1-1}-{\rm Im}\rho^{2}_{1-1}),\qquad
 \Delta_{1-1}=
  \frac12(\rho^{1}_{1-1}+{\rm Im}\rho^{2}_{1-1}).
  \label{rho11-1}
  \end{eqnarray}
The  averaging  over the angle between polarization  and
production planes, at fixed $\Phi-\Psi$ results in the following
one-dimensional distributions
 \begin{eqnarray}
 {W}^L (\cos\Theta) &=& W^0(\cos\Theta),
 \nonumber\\
 {W}^L(\Phi-\Psi) &=& \frac{1}{2\pi}\left(1 +
 2P_\gamma\overline{\rho}^{1}_{1-1}\cos[2(\Phi-\Psi)]\right).
 \label{WL}
 \end{eqnarray}
The integration over $\Theta$ and $\Phi$ gives dependence of the
total decay distribution as a function on $\Psi$
 \begin{eqnarray}
 {W}^L(\Psi) = 1 -
  P_\gamma (2\rho^{1}_{11}+\rho^1_{00})\cos2\Psi.
 \label{WLPSI}
 \end{eqnarray}
For the circularly-polarized beam the distributions read
 \begin{eqnarray}
 {W}^{\pm} (\cos\Theta) &=& W^0(\cos\Theta),\nonumber\\
 {W}^{\pm}(\Phi)&=& W^0(\Phi) \pm\frac{1}{\pi}P_\gamma\,
 {\rm Im}\rho^3_{1-1}\sin2\Phi.
 \label{WC}
  \end{eqnarray}
 We will also discuss the vector-meson decay asymmetry which is
 related to the matrix elements $\rho^{0,1}_{11}, \rho^{0,1}_{1-1}$
 \begin{eqnarray}
 \Sigma_V =
 \frac{\rho^{1}_{11} + \rho^{1}_{1-1}}
 {\rho^{0}_{11} + \rho^{0}_{1-1}},
 \label{sigma_v}
 \end{eqnarray}
 and has a meaning of the asymmetry between the two angular distributions
 when the decay angles are fixed and equal
 $\Theta=\frac\pi2$ and $\Phi=\frac\pi2$, and $\Psi=\frac\pi2,0$
\begin{eqnarray}
 \Sigma_V =\frac{1}{P_\gamma}
 \frac{W^L(\cos\Theta,\Phi,\Psi=\frac\pi2)
       - W^L(\cos\Theta,\Phi,\Psi=0)}
{W^L(\cos\Theta,\Phi,\Psi=\frac\pi2)
       + W^L(\cos\Theta,\Phi,\Psi=0)}.
 \label{sigma_v2}
 \end{eqnarray}

\section{The Amplitude}

\subsection{diffractive photoproduction}

 The invariant amplitude  in the region
 of small momentum transfers $I^{\rm Diff}$
 can be considered in frame of Regge
 phenomenology as a sum of the Pomeron
 and other Regge - trajectories. For $E_\gamma\simeq 2\div3$ GeV
 ($s\simeq 5\div 7$ GeV$^2$)
 this region is limited by the forward angle photoproduction with
 $|t|  \lesssim 0.5\div0.7$ GeV$^2$,
  where $|t|/s \lesssim 0.1\ll1$~\cite{Collins}.
 As we will see later, the employment of conventional residuals
 in corresponding amplitudes expressed
 through the isoscalar nucleon form factors leads to fast
 decreasing of $I^{\rm Diff}$, so that it becomes rather small
 in the region beyond its validity. On the other hand  the meson
 photoproduction at low energy and large momentum transfers
 with $|t|\sim |t|_{\rm max}$
 ($\theta\sim \pi$) can be described successfully in terms of
 the  nucleon and resonance exchange amplitudes $I^{\rm B}$.
 In diffractive region of the $\phi$-meson photoproduction
 $I^{\rm B}$ is suppressed by the OZI-rule and is negligible compared
 to $I^{\rm Diff}$. The total amplitude may be written as
 \begin{equation}
 I_{fi} =I^{\rm Diff}_{fi} \oplus I^{\rm B}_{fi}
 \label{T:twoterms}
 \end{equation}
 where $\oplus$ means that the above two components
 are defined  and operate in different regions of $t$ and simultaneous account
 of the Regge-amplitude and the resonant part leads, strictly speaking
 to double counting~\cite{Collins}. But since in
 considered case the interference of
 $I^{\rm Diff}$ and  $I^{\rm B}$
 at forward and backward angle photoproduction is negligible
 we can  substitute $\otimes\to + $. This leads to so called
 "interference model" which was widely used in the resonance region
 (see for example~\cite{DK69}). But taking into account the
 problem of double counting
 at the region of
 $\theta\sim\frac\pi2$ when interference of the resonant and  Regge-parts
 may be sizeable, predictions for a such model
 must be considered as a very qualitative estimations, especially for spin
 variables.

 In the diffractive region of the $\phi$-meson photoproduction
 the two processes
 are reasonably well established; the Pomeron exchange which is dominant
 and relatively weak  pseudoscalar $\pi,\eta$-mesons-exchange.
    \begin{figure}[ht]
    \includegraphics[width=0.7\columnwidth]
    {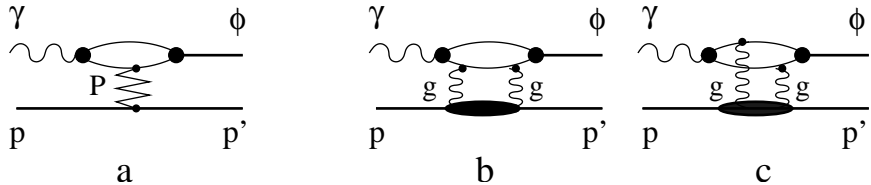}
    \caption{Diagrammatic  representation of
    (a) Pomeron exchange and (b,c) two gluon-exchange
    contributions in the $\gamma p\to \phi p$ reaction.}
    \label{fig:1}
    \end{figure}
 For the Pomeron-exchange process, depicted in Fig.~{\ref{fig:1}}(a),
 we use the Donnachie-Landshoff model~\cite{DL84-92}, based on the
 Pomeron-isoscalar-photon  analogy. It gives
 the amplitude in the following form
 \begin{eqnarray}
  I_{fi}^P &=& - M^{}_P (s,t) \,\Gamma^P_{fi},
 \nonumber\\
  \Gamma^P_{fi}&=&
  \varepsilon^*_{\mu}(\lambda_V)\,\bar{u}^{}_f\,h^{\mu\nu}_P\,
   u^{}_i
  \varepsilon_{\nu}(\lambda_\gamma),
 \label{IfiP}
 \end{eqnarray}
 where $\varepsilon^{}_{\mu} (V)$ and $\varepsilon^{}_{\nu}
 (\gamma)$ are the polarization vectors of the vector meson
 ($\omega,\phi$) and photon, respectively, and $u_i$=$u^{}_{m_i}(p)$
 ($u_f$=$u^{}_{m_f}(p')$)  is the
 Dirac spinor of the nucleon with momentum $p$ ($p'$) and spin projection
 $m_i$ ($m_f$).

  The scalar function $M_P(s,t)$ is described by the following Regge
 parameterization
 \begin{equation}
 M^{}_P (s,t)= C^{}_P \, F^{}_1(t) \, F^{}_V (t)\,\frac{1}{s}
 \left(\frac{s}{s_P} \right)^{\alpha_P^{} (t)} \exp\left[ -
 \frac{i\pi}{2}\alpha_P^{}(t) \right],
 \label{Pom:MP}
 \end{equation}
 where $F_1^{}(t)$ is the isoscalar electromagnetic form factor of
 the nucleon and $F_V^{}(t)$ is the form factor for the
 vector-meson--photon--Pomeron coupling. We also follow Ref.
 \cite{DL84-92} to write
 \begin{eqnarray}
  F_1^{} (t) = \frac{ 4 M_N^2-2.8t }{ (4M_N^2-t) (1-t/t_0)^2  },\qquad
  F_V^{} (t) = \frac{2\mu_0^2}{(1 - t/M_V^2)(2 \mu_0^2 + M_V^2 - t)},
 \end{eqnarray}
 where $t_0 = 0.7$ GeV$^2$. The Pomeron trajectory is known to be
 $\alpha_P^{} (t) = 1.08 + 0.25\,t$. (See also Ref. \cite{CKK97})
 The strength factor $C_P^{}$ equal
\begin{eqnarray}
  C_P^{} = \frac{6g^2
 \sqrt{4\pi\alpha_{\rm em}}}{\gamma_V^{}},
 \label{Cp}
 \end{eqnarray}
 where $\gamma_V$ is the vector meson
 decay constant  ($2\gamma_\omega=17.05$ and $2\gamma_\phi=13.13$), and
 $\alpha^{}_{\rm em} = e^2/4\pi$. The parameter
 $g^2$ is the product of two dimensionless  coupling constants
 $g^2\equiv g_{Pss}\cdot g_{Pqq}=
 (\sqrt{s_P}\beta_s)\cdot(\sqrt{s_P}\beta_u)$, where
 $g_{Pss}$ and $g_{Pqq}$
 have a meaning of the
 Pomeron coupling with the strange quarks in a $\phi$-meson and
 the light in a proton, respectively.
 For the $\omega$-meson  photoproduction $g^2\equiv q_{Pqq}^2$.

 The  vertex  function $h^{\mu\nu}_P$ is defined by a trace  calculation
 of the quark loop in the diagram of Fig.~\ref{fig:1}(a) with
 the non-relativistic vector-meson wave function and the vector coupling for the
 Pomeron-quark-quark vertex ($V_{Pqq}\sim\gamma_\alpha$). The net result reads
 \begin{eqnarray}
  {h}^{\mu\nu}_P= k \!\!\!/\,(g^{\mu\nu}-\frac{q^\mu q^\nu}{q^2})
  -\gamma^\nu (k^\mu - \frac{q^\mu k\cdot q}{q^2})
  - q^\nu(\gamma^\mu-\frac{q \!\!\!/\ q^\mu}{q^2}),
  \label{h-munu}
 \end{eqnarray}
where we keep the explicit gauge-invariant terms and skip the term
proportional to $k^\nu$, which, after  being  multiplied by the
photon polarization vector does not contribute. One can see that
the last term violates  gauge invariance. That is a serious
problem and its solution requires a more detailed description of
the gluon-exchange mechanism and vertex
functions~\cite{PL96,HL98,CR97}, which is beyond the scope of this
work. Here, we perform gauge-invariance restoration by gauge
transformation, since the above described simple model has been so
successful in description of many processes. The easiest way is
transformation of $q^\nu$ in the last term of Eq.~(\ref{h-munu})
 \begin{eqnarray}
   q^\nu\to  \bar q^\nu = q^\nu -\bar p^\nu\frac{k\cdot q}{\bar p\cdot k}.
  \label{q-nu}
 \end{eqnarray}
 where the vector $\bar p$ must be fixed via an  additional
 assumption. Using this transformation the vertex function
${\Gamma}^P_{fi}$ has the following form
 \begin{eqnarray}
  \Gamma^P_{fi}=
  \bar{u}^{}_f k \!\!\!/\, u^{}_i
 (\varepsilon^*_{\lambda_V}\cdot \varepsilon_{\lambda_\gamma})
  -\bar{u}^{}_f\varepsilon_{\lambda_\gamma}\!\!\!\!\!\!\!\!/\,\,\,\,u^{}_i
  (\varepsilon^*_{\lambda_V}\cdot k)
 -\bar{u}^{}_f\varepsilon^*_{\lambda_V}\!\!\!\!\!\!\!\!\!/\,\,\,\,\,u^{}_i
   \left(\varepsilon_{\lambda_\gamma}\cdot q
   -\frac{(\varepsilon_{\lambda_\gamma}\cdot \bar p)(k\cdot q)}{\bar p
 \cdot k}\right).
  \label{G-fi}
 \end{eqnarray}
 To find the vector $\bar p$ we take into account that it must lie in the
 production plane and be constructed from the three linear independent vectors
 $p,p'$ and $q$, be different from $q$ and, it must has a proper
 "high energy limit".
 This limit may be find using the using the Pomeron -
 two-gluon-exchange analogy~\cite{Dael}, since
 it is now generally believed that the Pomeron exchange is generated by the
 gluon exchange~\cite{LowNuss,CDL89} and
 the non-perturbative two-gluon-exchange process~\cite{LN87}
 justifies the vector type of coupling in $Pqq$-vertex.

 The two-gluon exchange amplitude for the vector meson photoproduction
 is depicted in Fig.~\ref{fig:1}(b, c). The amplitude, where the two
 gluons
 interact with the same quark (Fig.~\ref{fig:1}(b))
 generates the vertex $h^{\mu\nu}_{2g}$ with the same
 structure as the Pomeron-exchange mechanism. The amplitude where the two
 gluons interact with different quarks (Fig.\ref{fig:1}(c))
 contains  an additional term proportional to
 $\bar{u}^{}(p')\varepsilon_{\lambda_\gamma}\!\!\!\!\!\!\!\!/\,\,\,\,u^{}(p_1)
\cdot
\bar{u}^{}(p_1)\varepsilon^*_{\lambda_V}\!\!\!\!\!\!\!\!\!/\,\,\,\,\,u^{}(p)$,
 where $p_1$ is the momentum of the quark in the intermediate state. This term
 restores the gauge invariance. At
 high energies and small momentum transfers where $p'\simeq p_1\simeq
 p$ and $\bar u(p')\gamma_\alpha u(p)\simeq 2p_\alpha$,
 the two-gluon-exchange model results~\cite{Ryskin,CR97}
\begin{eqnarray}
 \Gamma^{2g}_{fi}\sim
  (k\cdot p)
 (\varepsilon^*_{\lambda_V}\cdot \varepsilon_{\lambda_\gamma})
  -(\varepsilon_{\lambda_\gamma}\cdot p)
  (\varepsilon^*_{\lambda_V}\cdot k)
 - (\varepsilon^*_{\lambda_V}\cdot p)
   \left(\varepsilon_{\lambda_\gamma}\cdot q
   -\frac{(\varepsilon_{\lambda_\gamma}\cdot p)(k\cdot q)}{p\cdot
   k}\right).
  \label{G2g-fi}
 \end{eqnarray}
 One can see the identity of Eqs.~(\ref{G-fi}) and(\ref{G2g-fi}) and  if
 $\bar p=p$ (cf. Ref.~\cite{Laget2000}). But at relatively
 low energies  where $p'\neq p$, a more reasonable choice is
 $\bar p =\frac{1}{2}(p+p')$, symmetrical with respect of $p$ and $p'$,
 because of an approximate estimates
 $\bar u(p')\gamma_\alpha u(p)\simeq (p+p')_\alpha$.
 This value of $\bar p$, together with
 Eq.~(\ref{G-fi}) will be used in our calculations.
 Note, that at large photon energies, the last two terms in~(\ref{G-fi})
 is negligible, but at a few  GeV  and finite $|t|$ they generate
 spin-flip transitions and become important.

 By fitting all available  total cross
 section data for $\omega$, $\rho$, and $\phi$ photoproduction at
 high energies, the remaining parameters of the model are
 determined: $\mu_0^2 = 1.1$ GeV$^2$,  $s_P = 4$ GeV$^2$, and
 $\beta_s = 1.61$ and $\beta_u=2.05$ GeV$^{-1}$, which give
 $ g_{Pqq}= 4.1$ and $g_{Pss}=3.22$.

The pseudoscalar-meson-exchange amplitude may be expressed
 either in terms of the one-boson-exchange
 model~\cite{Zhao:phiomega,TLTS99,Will98,TOYM98}
 or using the Regge model~\cite{Laget2000,Kochelev}.
 The pseudoscalar-meson exchange amplitude in
 the one-boson-exchange (OBE) approximation is evaluated from the following
 effective Lagrangians
\begin{eqnarray}
{\cal L}_{V \gamma \varphi}^{} &=& \frac{e g_{V\gamma
\varphi}}{M_V} \epsilon^{\mu\nu\alpha\beta}
\partial_\mu V_\nu \partial_\alpha A_\beta\, \varphi,\qquad
\nonumber\\
{\cal L}_{\varphi NN}^{} &=& -i g_{\pi NN} \bar N \gamma_5\tau_3 N
\pi^0 -i g_{\eta NN} \bar N \gamma_5 N \eta,
\end{eqnarray}
where $\varphi = (\pi^0,\eta)$ and $A_\beta$ is the photon field.
The resulting invariant amplitude is
\begin{eqnarray}
I^{ps}_{fi} = - \sum_{\varphi=\pi,\eta} \frac{iF_{\varphi NN}(t)
F_{V\gamma\varphi}(t)}{t-M_\varphi^2} \frac{e
g^{}_{V\gamma\varphi} g^{}_{\varphi NN}}{M_V} \,
\bar{u}_{m^{}_f}(p') \gamma_5 u_{m^{}_i}(p) \,
\varepsilon^{\mu\nu\alpha\beta} q_{\mu}^{} k^{}_\alpha
\varepsilon_\nu^* (V) \varepsilon^{}_\beta (\gamma). \label{T:ps}
\end{eqnarray}
In the above, we have followed Ref. \cite{FS96} to include the
following form factors to dress the $\varphi NN$ and
$V\gamma\varphi$ vertices,
\begin{equation}
 F_{\varphi NN}^{} (t) = \frac{\Lambda_\varphi^2 - M^2_\varphi}
 {\Lambda_\varphi^2 -t},  \qquad F_{V\gamma\varphi}^{} (t) =
 \frac{\Lambda_{V\gamma\varphi}^2-M_\varphi^2}
 {\Lambda_{V\gamma\varphi}^2-t} .
 \label{PS:FF}
\end{equation}
 We use $g_{\pi NN} = 13.26$ and  $g_{\pi NN} = 3.527$ for the $\pi
 NN$ and $\eta NN$ coupling constants, respectively~(cf. Ref.~\cite{TLTS99}
 for discussion and references).
 The coupling constants $g_{V\gamma\varphi}$ can
 be estimated through the decay widths of $V \to \gamma\pi$
 and $V \to \gamma \eta$ \cite{PDG} which lead to
 $g_{\omega\gamma\pi} = 1.823$, $g_{\phi\gamma\pi} =-0.141$,
 $g_{\omega\gamma\eta} = 0.416$ and $g_{\phi\gamma\eta} =-0.707$.
 The cutoff parameters $\Lambda_\varphi$ and
 $\Lambda_{\omega\gamma\varphi}$ in Eq.~(\ref{PS:FF}) are chosen
 to reproduce the $\omega$-meson photoproduction at low energies~\cite{TL02}:
 $\Lambda_\pi=0.6$ GeV$^2$, $\Lambda_\eta=0.9$ GeV$^2$,
 $\Lambda_{V\gamma\pi}=0.6$ GeV$^2$ and $\Lambda_{V\gamma\eta}=1.0$
 GeV$^2$.

 In Refs.~\cite{Laget2000,Kochelev,GudalLagetVander} it is suggested to
 describe pseudoscalar-meson (PS) exchange with a
 $\pi$-meson-exchange trajectory by making use of Reggeization
 of the Feynman propagator in the following way
 \begin{equation}
 \frac{1}{t-m_{\pi}^2}
 \Longrightarrow
 \left(\frac{s}{s_1} \right)^{\alpha_{\pi}(t)}
 \frac{(1+{\rm e}^{-i\pi\alpha_{{\pi}(t)}})\pi\alpha_{\pi}'}
 {2\sin(\pi\alpha_{\pi})\,\Gamma(\alpha_{\pi}(t)+1)},
 \label{Regge-ps}
 \end{equation}
where  the trajectory is given by
\begin{eqnarray}
\alpha_{\pi}(t)=\alpha_{\pi}'(t-m^2_{\pi}),
\end{eqnarray}
with $\alpha'_{\pi}=0.7$ GeV$^{-2}$.
 Using properties of the $\Gamma$-function: $\Gamma(1+z)=z\Gamma(z)$
 and $\Gamma(z)\Gamma(1-z)\sin\pi z=\pi$,
 and assuming that $|\alpha_\pi(t_{\rm max})|\ll1$ one can see that the
 Reggezaition of Eq.~(\ref{Regge-ps})
 does not modify the pole $t$-dependence
 of the OBE amplitude at forward-angle photoproduction.
 The Regge phenomenology does not define $t$-dependence
 and phases of residuals. In practice, they  are determined by comparison of
 the Regge and OBE-amplitudes at low energies where
 the lowest $t$-channel resonances
 are expected to be dominant~\cite{CollinsSquires,GudalLagetVander}.
 In the present paper we analyze photoproduction in narrow  energy region
 for $W=2\sim3$ GeV and in order to avoid
 consideration of additional parameters
 we will use directly the OBE model for
 the pseudoscalar-meson-exchange processes,
 with the parameters taken from independent studies. In this case
 our result coincides with the Regge model at small $|t|$ and
 slightly overestimates it at backward-angle photoproduction where
 the contribution of pseudoscalar-meson-exchange becomes negligible
 relative to the other channels.

 Together with these conventional
 processes, several other diffractive channels are
 discussed in literature. One of them is OZI-rule allowed $f_2'$
 meson exchange or contribution of the $f_2'$-meson Regge
 trajectory~\cite{Laget2000}.
 Significant strangeness content of $f_2'(1525)$  supports
 existence of this channel.
 But, on the other hand, the absence of a finite $f_2'\to \gamma\phi$ decay
 width~\cite{PDG} is a real problem for applying
 of a $f_2'$-trajectory in $\phi$-meson photoproduction
 and the only argument for its  use is
 to obtain agreement with the low-energy  data.

 Another possible low-energy channel is related to the pure gluon
 dynamics like a glueball exchange or the contribution of a glueball
 ($J^\pi=0^+, M^2_{\rm gl}{\rm \simeq 3}$ GeV$^2$) inspired
 trajectory~\cite{NakanoToki}. The idea is based on assumption
 that the Pomeron trajectory is also inspired by the glueball
 exchange,  being the leading glueball Regge trajectory. Other
 trajectories may appear as its daughter trajectories. This
 picture has been justified recently by calculations within the
 string model~\cite{Soloviev,Talalov}. The slope of the
 glueball trajectory is found to be much smaller than slope of
 conventional mesonic trajectories and close to the Pomeron
 trajectory.

 Since both $f_2'$ and glueball trajectories are not forbidden,
 we also include them into the total amplitude.
 To fix the vertex ($h^{\mu\nu}$) we use the simplest covariant and gauge invariant
 effective Lagrangians for $VV\xi$ interacting were $V$ and $\xi$ are the
 vector  and boson ($0^+,\, 2^+$) fields, respectively:
 \begin{subequations}
 \label{LB+}
\begin{eqnarray}
{\cal L}_{0^+}&=&
\frac14\,g_{\alpha\beta}\left(\Lambda^{\alpha\beta}  + \Lambda^{\beta\alpha}\right)
                \xi\label{L0+}\\
 {\cal L}_{2^+}&=&
 \frac14
 \left(\Lambda^{\alpha\beta}  + \Lambda^{\beta\alpha}\right)\xi_{\alpha\beta}
+ \left(\Lambda^{\alpha\beta} - \Lambda^{\beta\alpha}\right)\xi_{\alpha\beta}, \label{L2+}
\end{eqnarray}
\end{subequations}
with
\begin{eqnarray}
 \Lambda^{\alpha\beta}=
      \partial_\mu V_1^\alpha\partial^\mu V_2^\beta
  +   \partial^\alpha V_1^\mu\partial^\beta {V_2}_\mu
   -  \partial^\alpha V_1^\mu\partial_\mu V_2^\beta
   -  \partial_\mu V_1^\alpha\partial^\beta V_2^\mu.
  \label{LL2+}
\end{eqnarray}
 Taking the corresponding fermion-boson interactions as $\bar\psi\psi\xi$
 and $\bar\psi\gamma^\alpha\gamma^\beta\psi\xi_{\alpha\beta}$,
 we get the vertices ($h^{\mu\nu}$) as following
 \begin{subequations}
 \label{h-B+}
  \begin{eqnarray}
  && h^{\mu\nu}_{0^+}= g^{\mu\nu}k\cdot q - k^\mu q^\nu,
     \label{h-gl}\\
   && h^{\mu\nu}_{2^+}=h^{\mu\nu}_{0^+}
    - 2i\sigma_{\alpha\beta}\left[
  {g}^{\alpha\mu}{g}^{\beta\nu} \,k\cdot q+
  {q}^\alpha {k^\beta}g^{\mu\nu}+
  {g}^{\alpha\nu} {q}^\beta k^\mu
  + {g}^{\beta\mu} {k}^\alpha {q}^\nu
  \right]\label{h-f2}
  \end{eqnarray}
  \end{subequations}
  Note, that the similar expressions can be also found using the loop integration
  within the same prescription as for the Pomeron exchange in
  the Donnachie-Landshoff model with the Reggeon-quark-quark vertices taken
  as 1 and $\gamma_\alpha\gamma_\beta$ for $0^+$ and $2^+$-exchanges,  respectively.
  The difference is in additional gauge breaking term which appears in (\ref{h-f2}):
  $ {g}^{\alpha\mu} {q}^\beta {q}^\nu$. It brings some ambiguity for this method,
  but for us it is important that the structure of all other terms,
  including their mutual signs is identical to (\ref{h-f2}).

The scalar $(M(s,t))$- functions read
  \begin{eqnarray}
   M^{}_{R} (s,t)= C^{}_R \, F^{}_1(t) \, F^{}_V (t)\,
  \frac{1}{N_R}   \left(\frac{s}{s_R} \right)^{\alpha_{R}(t)}
  \frac{\eta_R(1+{\rm e}^{-i\pi\alpha_{R(t)}})\pi\alpha_{R}'}
  {2\sin(\pi\alpha_{R})\,\Gamma(\alpha_{R}(t))},
  \label{Pom:R}
  \end{eqnarray}
 where $R=f_2',\, gl$  and   $C^{}_{R}$  differ  from $C^{}_{P}$
 in (\ref{Pom:MP}) by substitution
 $g^2_P\to g^2_R$.  The normalization factors $N_R$ in
 (\ref{IfiP}) are defined by $h_R$: $ N_{2^+}\simeq 2sM_V,\,
 N_{0^+}\simeq M_NM_V^2$.
  Following Ref.\cite{Laget2000} we choose $f_{2}'$ trajectory as
  $\alpha_{f_2'}=0.55 + \,\alpha_{f_2'}'t$, with
  $\alpha_{f_2'}=0.7$ GeV$^{-2}$ and the mass scale $s_{f_2'}=1$ GeV$^2$.
  For the glueball trajectory we will use parameters of Ref~\cite{NakanoToki}
  with $\alpha_{\rm gl}(0)=-0.75$ and $\alpha_{\rm gl}'=0.25$
  GeV$-2$
  and $s_{gl}= {\alpha_{\rm gl}'}^{-1}$.
  Phase $\eta_R=\pm1$ and strength $g_R$ are not defined in the Regge model
  and  have to be found to bring the model calculation close to the data
  for unpolarized total cross section.

  Finally, we note that sometimes in literature
  the Regge amplitude at low energy
  is chosen with an additional threshold factor to get a better shape
  of the energy dependence in the near-threshold region,
  which modifies  the standard parameterization as follows
 \begin{eqnarray}
 \left( \frac{s}{s_R}\right)^{\alpha(t)}\to \left(
 \frac{s-a_R}{s_R}\right)^{\alpha(t)}.
 \label{s_thresh}
 \end{eqnarray}
 It is obvious, that relative contribution of each trajectory in
 the threshold region strongly depends  on the threshold
 parameter $a_R$, which is not defined by the Regge model.
 Thus, the finite value of $a_P$ in Pomeron exchange leads
 to a decrease of the contribution of the Pomeron exchange-amplitude,
 which must be compensated by the increase of strength
 in additional trajectories~\cite{TLTS99,Will98,NakanoToki}.
 The shape of the energy dependence of the total cross section
 is sensitive to the choice of $a_P,\, a_R$
 at  energies close to the threshold.
 In the present paper we  choose the  value $a_R=0$ for all
 trajectories. This choice  corresponds to the upper limit
 for contribution of Pomeron exchange and the lower limit
 for the additional Regge trajectories in near-threshold region.
 Its validity must be checked in study of the polarization
 observables in diffractive region, because the vertex functions
 for the Pomeron and  other trajectories in Eqs.~(\ref{G-fi}),
 (\ref{h-gl}) and (\ref{h-f2}) lead to quite different predictions.
 In Appendix A  we illustrate this point for two other extreme
 cases when the diffractive amplitude is dominated by the $f_2'$
 and $0^+$ (glueball)-trajectories, respectively. They can be realized
 at large $a_P$: $a_P\sim s_P$.

\subsection{baryon and baryon resonances exchange}

 The amplitude $I^{\rm B}$ in Eq.~(\ref{T:twoterms}) consist of baryon
 $I^{N}$ and baryon-resonance-exchange $I^{N^*}$ terms
 \begin{eqnarray}
 I^{\rm B} =I^{N}+I^{N^*}.
 \label{TB:2terms}
 \end{eqnarray}
 To evaluate these channels we use the
 effective Lagrangian approach, developed for $\omega$-meson
 photoproduction and discussed
 in our recent paper~\cite{TL02}. Here,
 we restrict the consideration to a brief description,
 given below.
 We consider all isospin $I=1/2$ nucleon resonances
 listed by PDG~\cite{PDG} with empirically known
 helicity amplitudes of $\gamma N \rightarrow N^*$ transitions.
 We thus have
 contributions from 12
 resonances: $P_{11}(1440)$,
 $D_{13}(1520)$, $S_{11}(1535)$,
 $S_{11}(1650)$, $D_{15}(1675)$, $F_{15}(1680)$, $D_{13}(1700)$, $P_{11}(1710)$,
 $P_{13}(1720)$, $F_{17}(1990)$, $D_{13}(2080)$, and $G_{17}(2190)$.

 In Ref.~\cite{TL02} we found that the contribution of the nucleon
 term is much smaller than a resonant part. Therefore,
 in present paper we only consider the
 resonance-exchange contribution.

 For the $N^{*}$ with spin $(J)$ and parity
 $(P)$ $J^P={\frac{1}{2}^\pm, \frac{3}{2}^\pm,\frac{5}{2}^\pm,\frac{7}{2}}^\pm$,
 we use the following effective Lagrangians:
 \begin{eqnarray}
 {\cal L}_{\gamma NN^*}^{\frac{1}{2}^{\pm}}
 &=&
 \,\, \frac{eg_{\gamma NN^*}}{2M_{N^*}} \bar\psi_{N^*} \,\Gamma^{(\pm)}
 \sigma_{\mu\nu} F^{\mu\nu} \psi_{N}\qquad +\qquad{\rm h.c.},
  \label{R1/2}\\
 {\cal L}_{\gamma NN^*}^{\frac{3}{2}^\pm}
 &=&
 i\frac{eg^{}_{\gamma NN^*}}{M_{N^*}}
 \bar\psi_{N^*}^\mu \,\gamma_\lambda\Gamma^{(\mp)}
 F^{\lambda\mu}
 {\psi_{N}}\qquad + \qquad{\rm h.c.},
 \label{R3/2}\\
 {\cal L}_{\gamma NN^*}^{\frac{5}{2}^\pm}
 &=&
 \frac{eg^{}_{\gamma NN^*}}{M_{N^*}^{2}}
 \bar\psi_{N^*}^{\mu\alpha} \,\gamma_\lambda\Gamma^{(\pm)}
 (\partial_\alpha F^{\lambda\mu})
 {\psi_{N}}\qquad + \qquad{\rm h.c.},
 \label{R5/2}\\
 {\cal L}_{\gamma NN^*}^{\frac{7}{2}^\pm}
 &=&-i
 \frac{eg^{}_{\gamma NN^*}}{M_{N^*}^{3}}
 \bar\psi_{N^*}^{\mu\alpha\beta} \,\gamma_\lambda\Gamma^{(\mp)}
 (\partial_\beta\partial_\alpha F^{\lambda\mu})
 {\psi_{N}}\qquad + \qquad{\rm h.c.},
 \label{R7/2}
 \end{eqnarray}
  where $\psi_{N}$ is the nucleon fields,
 $\psi_{N^*},\psi_\alpha$, $\psi_{\alpha\beta}$,
 and $\psi_{\alpha\beta\gamma}$
 are the Rarita-Schwinger spin
 $\frac12,\frac32$,  $\frac52$ and $\frac72$
 field, respectively, $M_{N^*}$ is the resonance mass;
 $A_\mu$ is the photon field,
 and $F^{\mu\nu}=\partial^\nu A^\mu - \partial^\mu A^\nu$.
 The coupling
 $\Gamma^+ =1(\Gamma^-=\gamma_5$) defines the $N^*$ excitations
 with different parity.

 We define the $\omega NN^*$ interactions by using the vector-dominance
 model. This assumes that the effective
 ${\cal L}_{\omega NN^*}$ Lagrangian has the same form as
 the corresponding ${\cal L}_{\gamma NN^*}$ with substitutions
 $A_\mu\to \omega_\mu$ and  $eg_s{_{\gamma N N^*}}\to  f_\omega$,
 where  $f_\omega=2g_s\gamma_\omega$.
 The isoscalar-coupling constant $g_s$ is related to the
 strengths of the $N^*$ excitations
 on the proton ($g_p = g_{\gamma p N^*}$)
 and on the neutron ($g_n=g_{\gamma n N^*}$) and equals
 $g_s=(g_p + g_n)/{2}$.

 The $\gamma NN^*$ and $\omega NN^*$ vertices  are
 regularized by the form factor
 $F_{N^*}(r^2) = {\Lambda^4}/(\Lambda^4 + (r^2-M^2_{N^*})^2)$, where
 $r$ is  the four-momentum  of the intermediate
 baryon state, the cut-off parameter $\Lambda=0.85$ GeV is chosen  to
 be the same for all resonances.
 More detailed discussion of the
 effective Lagrangians, propagators,
 fixing the $\gamma NN^*$ couplings
 and comparison with other approaches is given in Ref.~\cite{TL02}.
  \begin{figure}[ht]
  \centering
  \includegraphics[width=51.75mm]{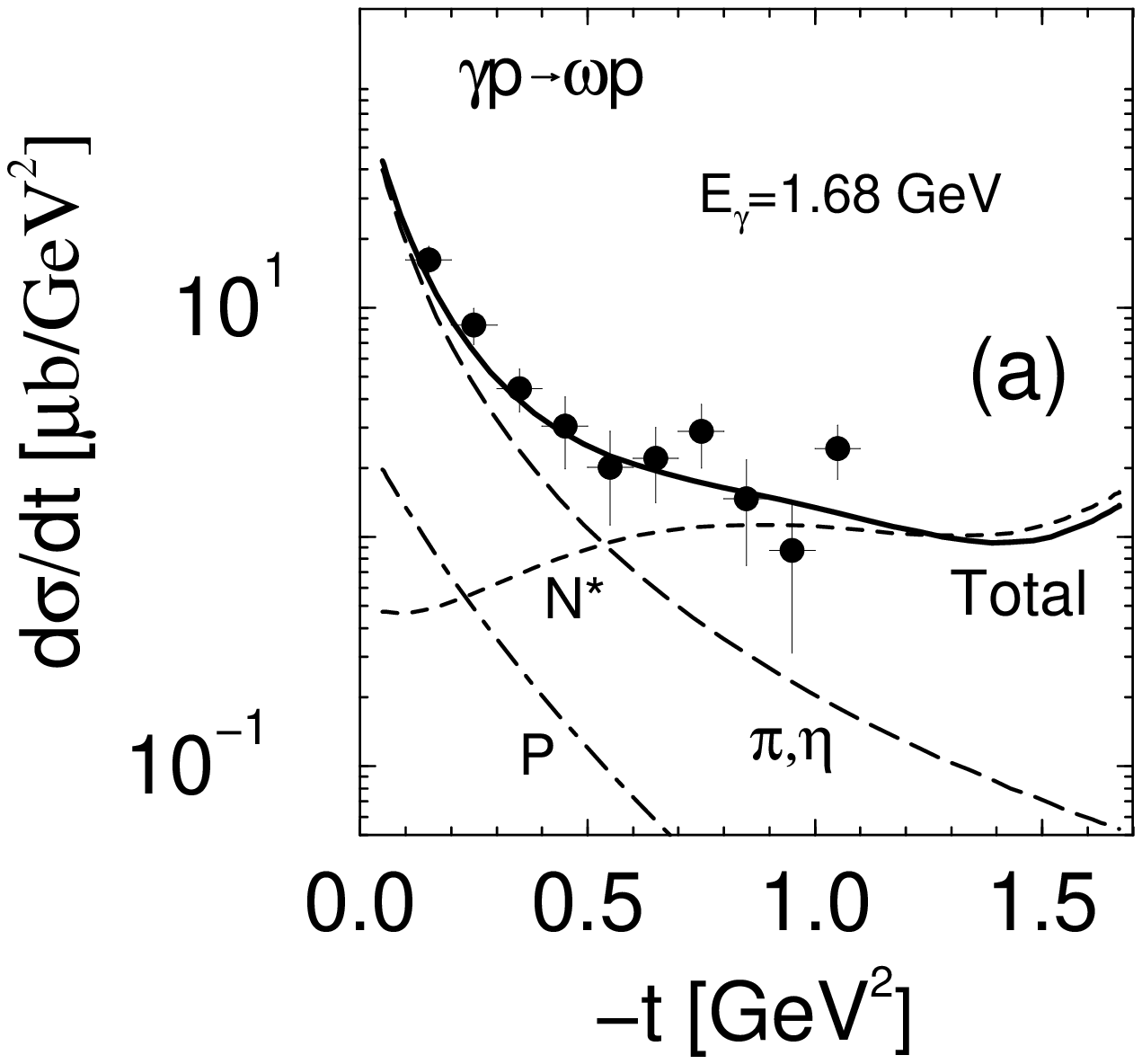}\qquad\qquad
  \includegraphics[width=55mm]{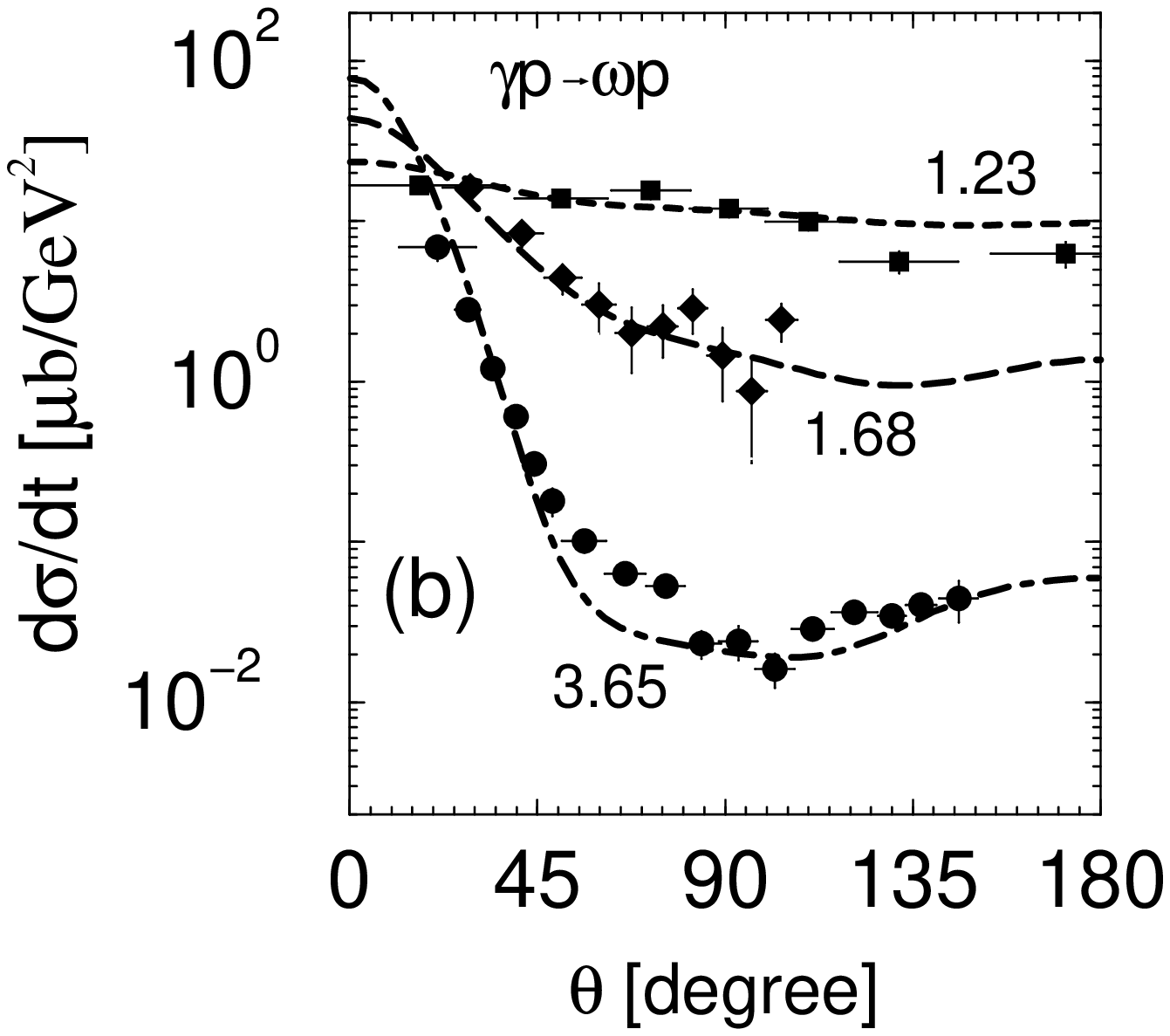}
  \vspace{0.5cm}
  \caption{
  (a): The differential cross section of  $\gamma p\to \omega p$
   reaction as a function of $-t$ at $E_\gamma =1.68$ GeV.
  The lines denote  the pseudoscalar-meson exchange (long dashed),
   Pomeron exchange (dot-dashed), resonant
   channel (dashed), and the full amplitude (solid), respectively.
  (b): The differential cross sections a function of
  the $\omega$-meson production angle at $E_\gamma=1.23$, $1.68$ GeV
  and 3.65 GeV.
   Data are taken from Refs.
   \protect\cite{Klein96-98,JLabomega}. }
 \label{fig:2}
\end{figure}
 Validity of the model is illustrated by Fig.~\ref{fig:2}, where
 we show result of our calculation for the differential cross section
 of $\gamma p\to \omega p$ reaction together
 with experimental data at $E_\gamma=$ 1.23, 1.68 GeV~\cite{Klein96-98}
 and 3.65 GeV~\cite{JLabomega}.
 In Fig.~\ref{fig:2}(a) we show
 the differential cross section as a function of $t$
 at $E_\gamma=1.68$ GeV (solid curve) together with the partial
 contribution of each of the main channels: pseudoscalar-meson
 exchange, Pomeron exchange,  and  resonance excitation,
 depicted by long-dashed, dot-dashed, and dashed curves, respectively.
 In Fig.~\ref{fig:2}(b) we show the differential cross
 section as a function of the $\omega$ production angle in c.m.s.
 at $E_\gamma=1.23,\,1.68$ and $3.65$ GeV by the dashed,
 dot-dashed and solid lines, respectively.
 The contribution of the resonance excitations
 is important at backward angles and it results in the agreement
 of data with our calculation at
 large momentum transfers. We also found that at $E_\gamma$=3.65 GeV
 the Pomeron-exchange contribution only is not sufficient
 to get agreement at forward angle photoproduction with $|t|<1$
 GeV$^2$.
 Therefore, following Ref.~\cite{Laget2000},  we include also
 a $f_2$-meson trajectory. It is calculated similarly to the
 $f_2'$-meson trajectory (Eqs.(\ref{Pom:R})) with
 $g^2_{f_2}\simeq 3g^2_P$ and $\eta_{f_2}=+1$.
 One can see, that the
 model  satisfactory reproduces both the  energy and the angular distribution
 of the $\omega$ photoproduction in the considered energy region.
 However, it leaves some windows for additional processes,
 like two-step photoproduction mechanisms~\cite{OhLee}
 and direct-quark exchange, which is expected to be important at
 large energy and $\theta\sim \pi/2$~\cite{JLabomega} and
 must be subject for special detailed analysis.
 There, it is most important that
 the model reproduces the cross
 section at backward angle photoproduction, which allows us
 to fix the resonant part of the $\phi$-meson photoproduction.

 For the $\phi$-meson photoproduction we assume
 the same $N^*$-excitation mechanism with the substitution
 $f_{\omega N N^*}\to f_{\phi N N^*}$.
 The key problem is how to fix these coupling constants.
 For this aim  we use the "minimal" parameterizations of the
 $\phi NN^*$ coupling constants
 \begin{eqnarray}
 {f_{\phi NN^*}}= -\tan\Delta\theta_V\, x_{\rm OZI}\,{f_{\omega NN^*}},
 \label{OZI3}
 \end{eqnarray}
 where $\Delta\theta_V$ is  the deviation of the $\phi-\omega$ mixing
 angle from the ideal mixing ($\Delta\theta_V\simeq
 3.7^0$~\cite{PDG}) and $x_{\rm OZI}$ is  the OZI-rule
 evading parameter~\cite{Will98}, which will be found from comparison
 of calculation and data at large momentum transfers.
 Enhancement of $x_{\rm OZI}$ from 1 determines  the
 scale of the OZI-rule violation in interaction of the $\phi$-meson with baryons.

\section{Results and Discussions}

\subsection{unpolarized cross sections}

  We first consider the total cross section. Its dominant
  contribution comes from the diffractive channels at $|t| < 1$ GeV$^2$.
  The  resonance excitations which are important at large $|t|$
  do not affect the total cross section. The total cross section data allow
  a wide range of OZI-rule evading parameter of the $\phi NN^*$
  couplings defined in Eq.(44). Therefore,
  for definiteness sake, we  take $x_{\rm
  OZI}=4$, which is close to its low bound  reported
  in~\cite{ASTERIX91,CBC95,OBELIX96} and as we will see later,
  this value is in
  agreement with the available data on $\phi$-meson photoproduction at
  large momentum transfers.

 For the analysis of the diffractive mechanism we have to keep in mind that
 the Pomeron (P) and pseudoscalar (PS) meson-exchange are
 well established.
 Therefore, any "exotic" process is
 included  as a supplementary channel. We will
 analysis three possibility: the model~I includes the Pomeron (P) and
 pseudoscalar  meson exchange (PS) (i.e. without exotic
 channel);
 model~II includes P and PS-meson exchange  and $f_2'$
 Regge trajectory;  model~III includes P+PS-meson exchange
 and the glueball inspired trajectory.

  All formula for the models~I-III have been described
 above. The strength parameter $g_R$ and phase factor $\eta_R$
 of $f_2'$-meson and glueball trajectories
 are fixed by fitting the
 available data: $g_{f_2'}=1.87,\eta_{f_2'}=+1$;
 $g_{gl}=7.66,\eta_{gl}=-1$.

 \begin{figure}[ht] \centering
 \includegraphics[width=51mm]{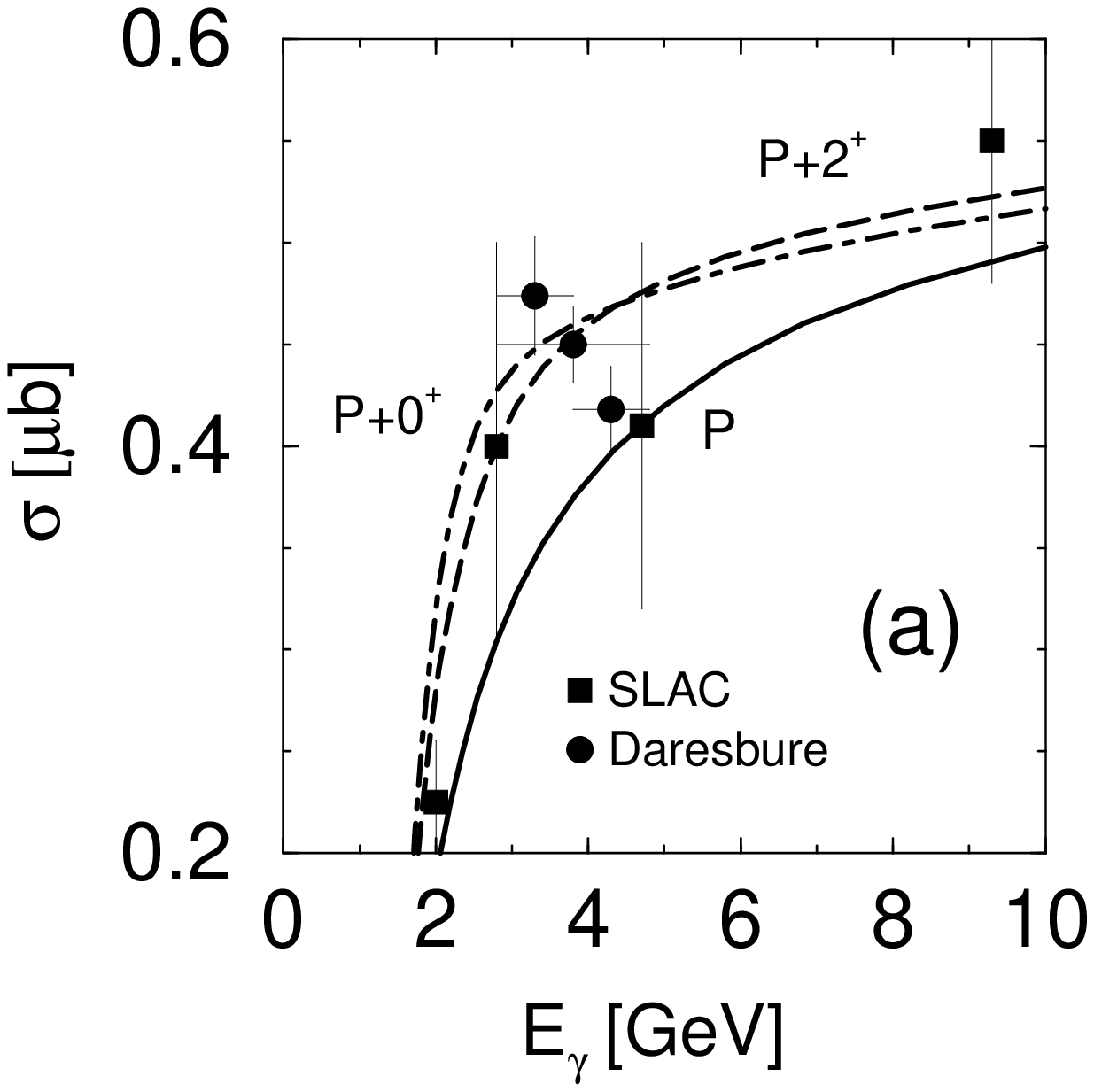}\qquad\qquad
 \includegraphics[width=51mm]{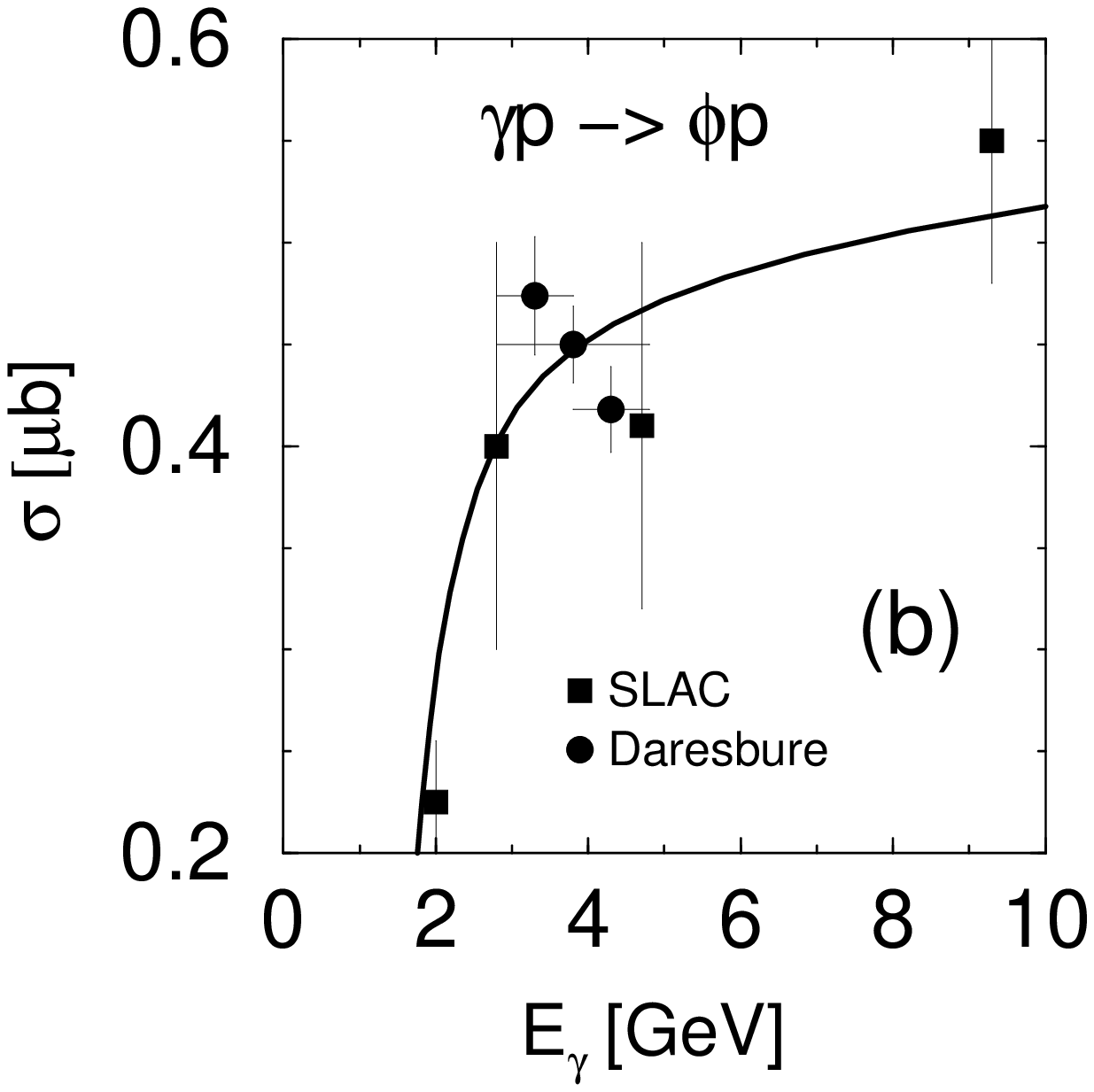}
\vspace{0.5cm}
 \caption{
 (a): The total cross section of  $\gamma p\to \phi p$
  reaction as a function of the photon energy $E_\gamma$ for the models
  I, II and III indicated by dashed,
  long-dashed and dot-dashed curves, respectively.
  Data are taken from Ref.
  \protect\cite{Ballam73,Barber82}.
 (b): The total cross section for the hybrid model.}
 \label{fig:3}
\end{figure}
 In Fig.~\ref{fig:3}(a) we show the total cross section of
 $\gamma p\to \phi p$ reaction as a function of the photon energy for
 three models together with the available experimental
 data~\cite{Ballam73,Barber82}.
  Comparison with the data slightly favors for the models II and III,
 but unfortunately, the accuracy of the
 data is not sufficient to make a definite conclusion about
 the preference for one of the models.
 The difference between the models disappears
 at high energy with $W\gtrsim 10$ GeV ($E_\gamma \gtrsim 50$ GeV),
 were only the Pomeron trajectory
 is important.  At low energy, high precision data are required
 to select the favored diffractive mechanism. Taking into account some
 ambiguity for the reaction mechanism, all our calculations will be
 done using the "hybrid" model, where the amplitude is taken
 as a  sum of the Pomeron and PS-meson-exchange amplitudes and
 small contribution of the
  $f_2'$ and glueball trajectories  taken with the equal weights
  with:
 ($g_{f_2'}=1.32$ and $g_{gl}=5.42$). The total
  cross sections calculated from this hybrid model is shown in
  Fig.~\ref{fig:3}(b). In the further discussions, for simplicity,
  we will denote the sum of Pomeron-, $f_2'$- and
  glueball-exchange amplitudes as the "diffractive" amplitude.
  The opposite case when the $f_2'$ or $0^+$-exchange trajectories
  are dominant is discussed in Appendix A.

\begin{figure}[ht] \centering
 \includegraphics[width=55mm]{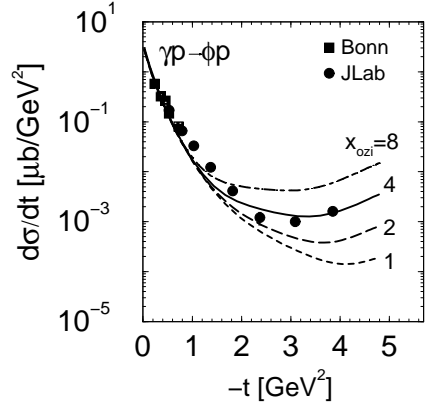}\vspace{0.5cm}
 \caption{ The differential cross section of  $\gamma p\to \phi p$ reaction
  as a  function of $t$ at $E_\gamma=3.6$ GeV
  for different values of the OZI-rule evading parameter $x_{\rm OZI}$.
  Data are taken from Refs.~\protect\cite{JLab00}.}
 \label{fig:4}
\end{figure}

 In $\gamma p\to \omega p$ reaction at low energies
 the backward-angle photoproduction is dominated by the
 resonant channel (cf. Fig.~\ref{fig:2}) and
 we expect a similar picture for $\gamma p\to \phi p$ reaction.
 The global structure of $N^*$-exchange amplitude is fixed
 by the $\omega$-meson photoproduction and the remaining parameter for
 the $\phi$-meson photoproduction is the OZI-rule evading parameter
 $x_{\rm OZI}$ in Eq.~(\ref{OZI3}). Fig.~\ref{fig:4} shows the
 differential cross section at $E_\gamma=3.6$~GeV as a function of
 $t$ for different values of $x_{\rm OZI}$,  together with experimental
 data~\cite{JLab00}.
 The calculation brings agreement with  data at $x_{\rm OZI}\simeq 4$.
 This value is consistent with results reported
 in Refs.~\cite{ASTERIX91,CBC95,OBELIX96} and the estimation of
 Refs.~\cite{Jaffe,Gerasimov}. It would be interesting to check
 this prediction at lower energies,  where the relative contribution of the
 resonant channels is expected to be stronger.
 \begin{figure}[ht] \centering
 \includegraphics[width=53mm]{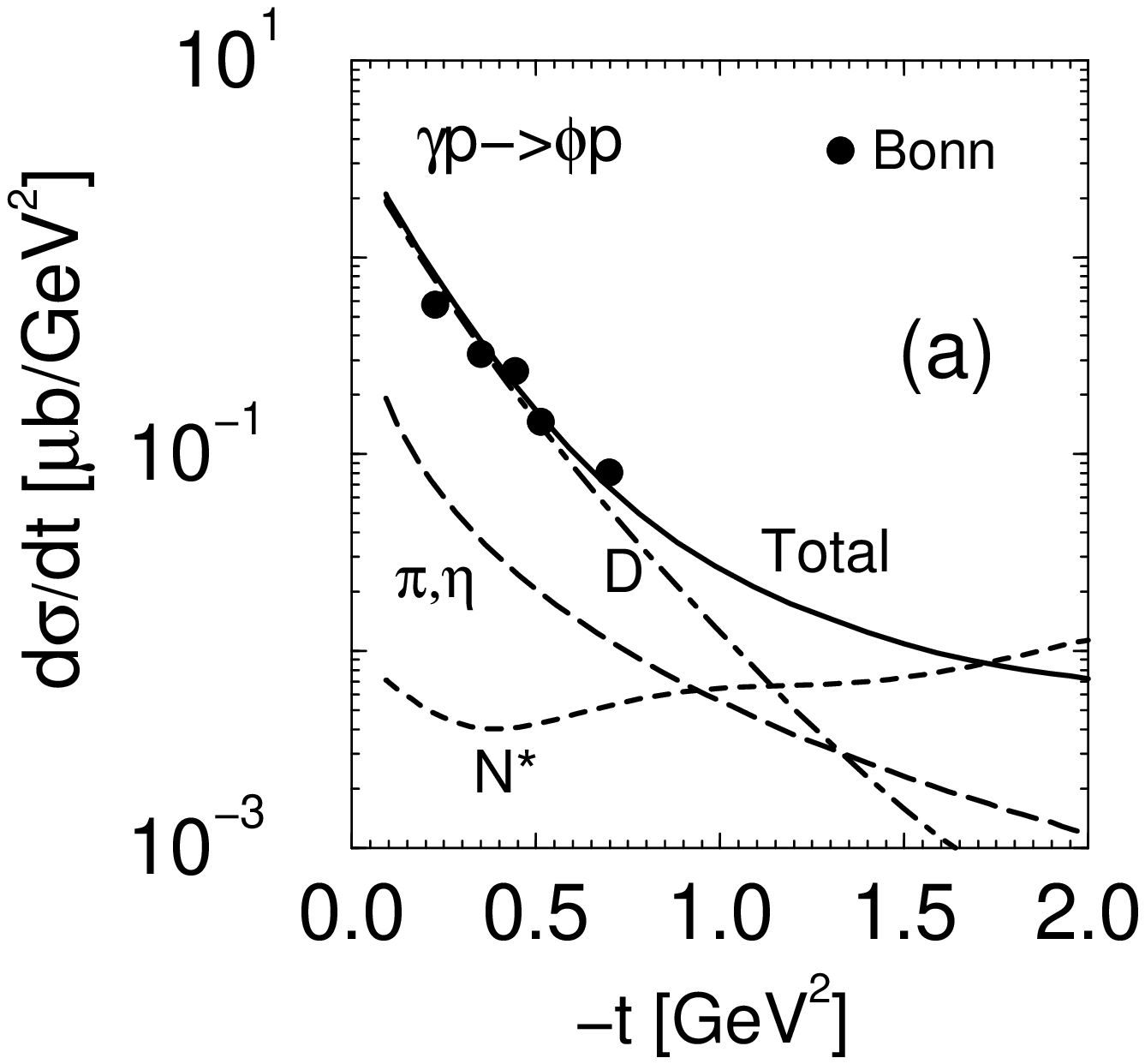}\qquad\qquad
 \includegraphics[width=53mm]{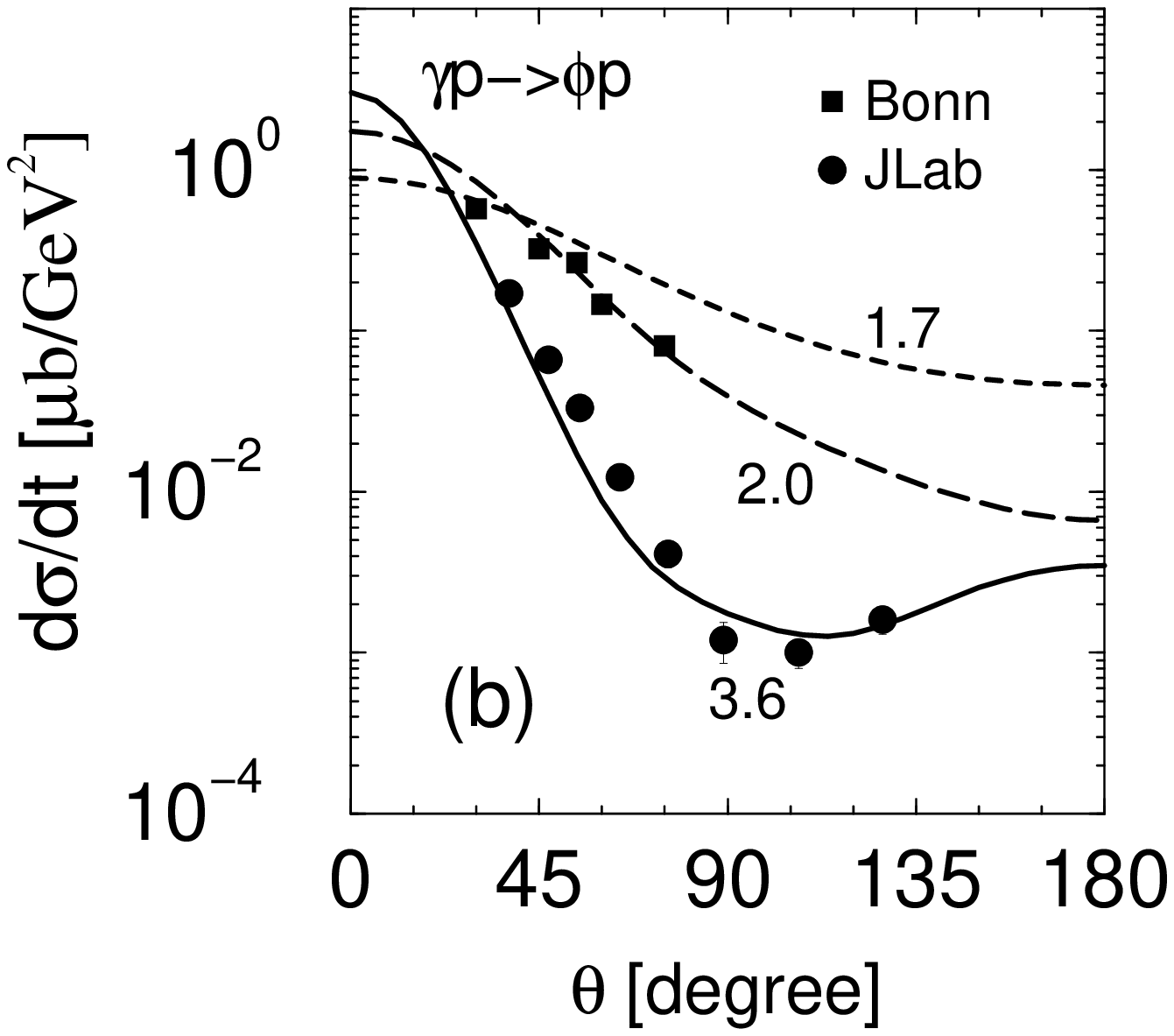}\vspace{0.5cm}
 \caption{(a): The differential cross section of  $\gamma p\to \phi p$
  reaction as a function of $-t$ at $E_\gamma =2.2$ GeV.
 The results are the pseudoscalar-meson exchange (long dashed),
 diffractive channels
 (dot-dashed), and resonance
 excitation (dashed), and the full amplitude (solid).
 (b): The differential cross section a function of
 the $\phi$-meson production angle at $E_\gamma=1.7, 2.0$ and $3.6$ GeV.
 Data are taken from Ref.
  \protect\cite{Bonn,JLab00}.}
 \label{fig:5}
\end{figure}

 In Fig.~\ref{fig:5}(a) we show the differential cross section as
 function of $t$ at $E_\gamma=2.0$~GeV  together with the partial
 contribution of main channels with $x_{\rm OZI}=4$.
 The diffractive part is described by the hybrid model.
 One can see that the forward-angle photoproduction is completely defined
 by the diffractive amplitude; the contribution from the diffractive channels
 exceeds the pseudoscalar meson exchange by an order of magnitude.
 Backward-angle photoproduction is dominated by the
 $N^*$-exchange channel, while in the central region
 ($0.7 \lesssim|t|\lesssim 1.4$ GeV$^2$), the coherent interference
 of all processes becomes important. Unfortunately, our model is not
 very well defined in this region. Fig.~\ref{fig:5}(b) shows
 the differential cross
 section as a function of the $\phi$-meson production angle in the c.m.s.
 at $E_\gamma=1.7,\,2.0$ and $3.6$ GeV (dashed,
 long dashed and dot-dashed lines, respectively).
 The calculations are in agreement with available data.
 Since the data at $E_\gamma=3.6$ GeV are used
 to fix $x_{\rm OZI}$, the other curves represent our
 prediction which would be interesting to check.

\subsection{spin observables}

 Spin observables can be used as a powerful tool to test
 the photoproduction mechanismis in detail. We first consider
 the spin-density matrix elements $\rho^{0-3}_{\lambda\lambda'}$,
 which are planned to be measured in near future at JLab~\cite{JLab}
 and at LEPS/SPring-8~\cite{LEPS}. All our calculation has been
 done in the Gottfried-Jackson system. For simplicity, we show
 our prediction  at all momentum transfers, however the
 applicability of the model at $E_\gamma\sim 2-3$ GeV  is limited by
 the  forward and backward
 photoproduction  with $|t_{\rm min}|\leq |t|\leq |t_l|$ and
 $|t_l|\leq |t|\leq |t|_{\rm max}$, respectively, were $|t_l|\simeq
 0.5-0.7$ GeV$^2$, depending on the energy.

 First of all, we remind that the non-zero spin-density matrix  elements for
 the  pure helicity conserving amplitude:
\begin{eqnarray}
 I_{fi}\sim \delta_{\lambda_i\lambda_f}\delta_{m_i\,m_f}
\label{delta-fi}
\end{eqnarray}
have the following values
\begin{eqnarray}
 \rho^0_{11}&=&\rho^0_{-1-1}=\frac12,\qquad
 \rho^1_{1-1}=\rho^1_{-11}=\pm\frac12,
\nonumber\\
 {\rm Im}\rho^2_{-11}&=-&{\rm Im}\rho^2_{1-1}=\pm\frac{1}{2},\qquad
 \rho^3_{11}=-\rho^3_{-1-1}=\pm\frac12,
 \label{rho0123}
\end{eqnarray}
 where the upper and lower signs in $\rho^{1,2,3}$ correspond to the
 amplitudes with natural ($I^{\rm N}$)
 and unnatural ($I^{\rm U}$) parity exchange,
 respectively.
 The typical example of the natural and unnatural parity exchange
 amplitude in our case are the scalar
 and the pseudoscalar-meson exchange amplitudes, respectively.
 For the forward-angle
 photoproduction they can be expressed  as
\begin{eqnarray*}
  I^{{\rm N}\atop{\rm U}}_{m_f\,m_i;\lambda_\phi\lambda_\gamma}(t)=
  \left({{1}\atop{2m_i\lambda_\gamma}}\right)
  \delta_{m_i m_f}\delta_{\lambda_\gamma\lambda_\phi}
    I^{{\rm N}\atop{\rm U}}_0 (t),
   \label{I_NU}
 \end{eqnarray*}
  where ${I^{{\rm N}\atop{\rm U}}_0(t)}$ is the spin-independent part of the
 corresponding amplitudes.

 The Pomeron-exchange amplitude in GJ-system has the following
 structure
 \begin{eqnarray}
  I^P_{fi}\sim
  -\delta_{\lambda_\phi\,\lambda_\gamma}\,\bar{u}^{}_f k \!\!\!/\, u^{}_i
  + \delta_{\lambda_\phi\,0}\,k_\gamma\,
   \bar{u}^{}_f\varepsilon_{\lambda_\gamma}\!\!\!\!\!\!\!\!/\,\,\,\,u^{}_i
  +  \sqrt{2}\lambda_\gamma\,p_x\,
   \frac{k\cdot q}{2p\cdot k-  k\cdot q}
  \bar{u}^{}_f\varepsilon^*_{\lambda_\phi}\!\!\!\!\!\!\!\!/\,\,\,\,u^{}_i,
  \label{P-GJ}
 \end{eqnarray}
 where $k_\gamma$ and $p_x$ are the photon momentum and the $x$-component
 of the proton ($p_x=p_x'$ in GJ-system) momentum, respectively.
 One can see, that only the first term  satisfies
 (\ref{delta-fi}). The second term describes the interaction of the
 photon and  nucleon spins and the interaction of the $\phi$-meson
 spin and the orbital momentum in the initial state. The third term is
 responsible for the interaction of the $\phi$-meson and nucleon
 spins and for the interaction of the photon spin with the orbital momentum
 in the final state.
 At $E_\gamma=2-3$ GeV, contribution
 of these two terms is finite and must be taken into account.
 Thus, the second and third terms in Eq.~(\ref{P-GJ}) are
 responsible for the spin-flip
 transitions $\lambda_\gamma\to\lambda_\phi=0$ and
 generate finite value of $\rho^0_{00}$. The contribution of the
 second term is dominant and it can be estimated as
 \begin{eqnarray}
\rho^0_{00}&\simeq& \frac{k^2_\gamma(|t|+ 2p_x^2)}{{\bar s}^2},\nonumber\\
{\bar s}^2&=& (s-M_N^2)^2\left( 1- \frac{M_\phi^2+|t|}{s-M_N^2}
\right).
 \end{eqnarray}
  This equation shows that $\rho^0_{00}$ increases monotonically with $|t|$
  and at $E_\gamma=$2.2 GeV and at $\theta=\pi$ reaches  large  value
  of $\rho^0_{00}\simeq0.6$.

 The interaction of the photon spin with the orbital momentum is
 responsible for
 so-called double spin-flip  transition $\lambda_\gamma\to
 \lambda_\phi=-\lambda_\gamma$ and generates $\rho^0_{1-1}$, which
 is defined by the interference of the first and third terms in
 Eq.~(\ref{P-GJ})
\begin{eqnarray}
\rho^0_{1-1}&\simeq& \frac{p_x^2(M_\phi^2 +|t|)}{{\bar s}^2}.
\end{eqnarray}
 This matrix element reaches its maximum value
 $\rho^{0}_{1-1}\simeq 0.2$ at $|t|\simeq 1$ GeV$^2$ and $E_\gamma=2.2$ GeV.
 Note, that this matrix element  depends on the choice of the gauge
 parameter $\bar p=ap+bp'$ in Eq.~(\ref{q-nu}) with constrains $a,b>0$ and
 $a+b=1$, which provide the proper high energy limit. Our choice is
 $a=b=\frac12$.  $\rho^0_{1-1}$ is proportional
 to $\varepsilon_\lambda(\gamma)\cdot\bar p/(s-M_N^2 - b(M_\phi^2
 +|t|))$. Since in GJ-system $p_x=p_x'$, then the product $\varepsilon_\lambda(\gamma)\cdot\bar
 p= \sqrt{2}\lambda p_x$ does not depend on this choice. The rest dependence
 on choice of $\bar p$ is rather weak. Thus, at $E_\gamma=2.2$ GeV
 the choice of $a=1, b=0$ $(a=0, b=1)$ results in decrease (increase)
 of  $\rho^0_{1-1}$ by a factor of 20\%.

 The resonant channel  and $f_2'$-trajectory
(first and third terms in the brackets in(\ref{h-f2}))
 also generate the finite spin-flip matrix
 elements,  but in the region with $|t|< 1$ GeV$^2$ their
 contribution to $\rho^0_{00},\,\,\rho^{0}_{1-1}$ is about of an
 order of
 magnitude smaller than the contribution from the  Pomeron exchange
 amplitude of Eq.~(\ref{P-GJ}). At large momentum transfers with
 $|t|\sim |t|_{\rm max}$  the resonant channel becomes essential.

 \begin{figure}[ht] \centering
 \includegraphics[width=50mm]{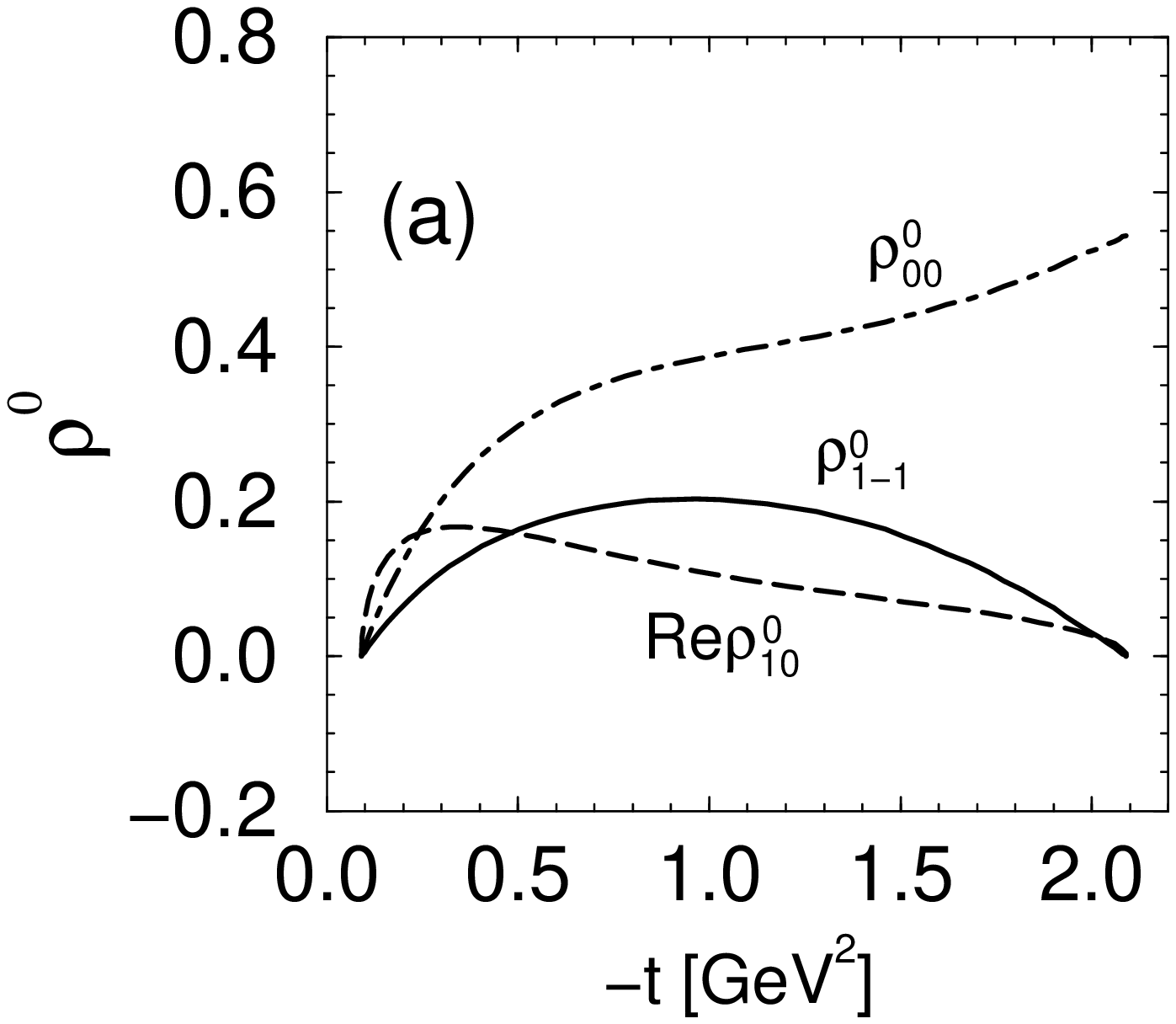}\qquad\qquad
 \includegraphics[width=50mm]{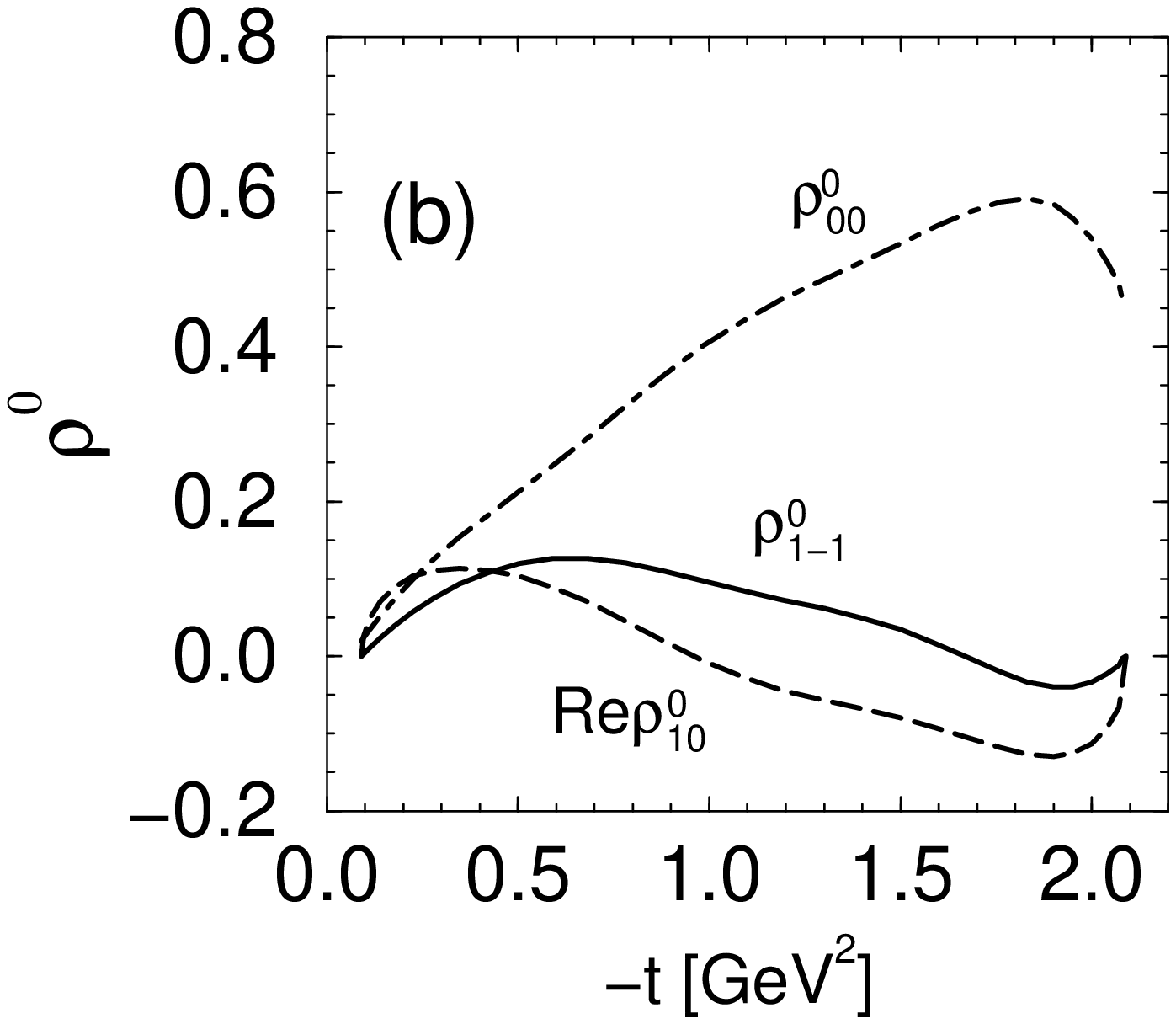}
 \caption{Spin-density matrix elements
 $\rho^0_{00}$, $\rho^0_{10}$ and $\rho^0_{1-1}$
 for reaction $\gamma p\to\phi p$ as a function of $-t$
 at $E_\gamma=2.2$ GeV shown as dot-dashed, solid and dashed
 lines, respectively. (a): Result for the Pomeron exchange amplitude;
 (b): Result for the full model.}
 \label{fig:6}
\end{figure}

 In Fig.~\ref{fig:6} we show the $t$-dependence of the three  $\rho^0$
 matrix elements at $E_\gamma=2.2$ GeV.  $\rho^0$ defines
 the angular distribution
 of $K^+K^-$-mesons in reactions with unpolarized
 photons (Eqs.~(\ref{UDD}), (\ref{W0})). Fig.~\ref{fig:6}(a) is for
 the pure Pomeron exchange amplitude (Eq.~(\ref{IfiP})) and Fig.~\ref{fig:6}(b)
 for the full model which includes diffractive part, PS-meson-exchange
 and resonance excitations.
  One can see that for the full model
 the non-zero values of these  spin-density matrix
 elements at forward angles photoproduction ($|t|<1$ GeV$^2$)
 are  mostly determined  by the Pomeron-exchange contribution. At large
 momentum transfers  the resonant excitations play a key role.

 The one-dimensional angular distributions $W^0(\cos\Theta)$
 and $W^0(\Phi)$ at three values of $|t|=0.2,\,0.5$ and $1.8$ GeV$^2$
\begin{figure}[ht] \centering
 \includegraphics[width=50mm]{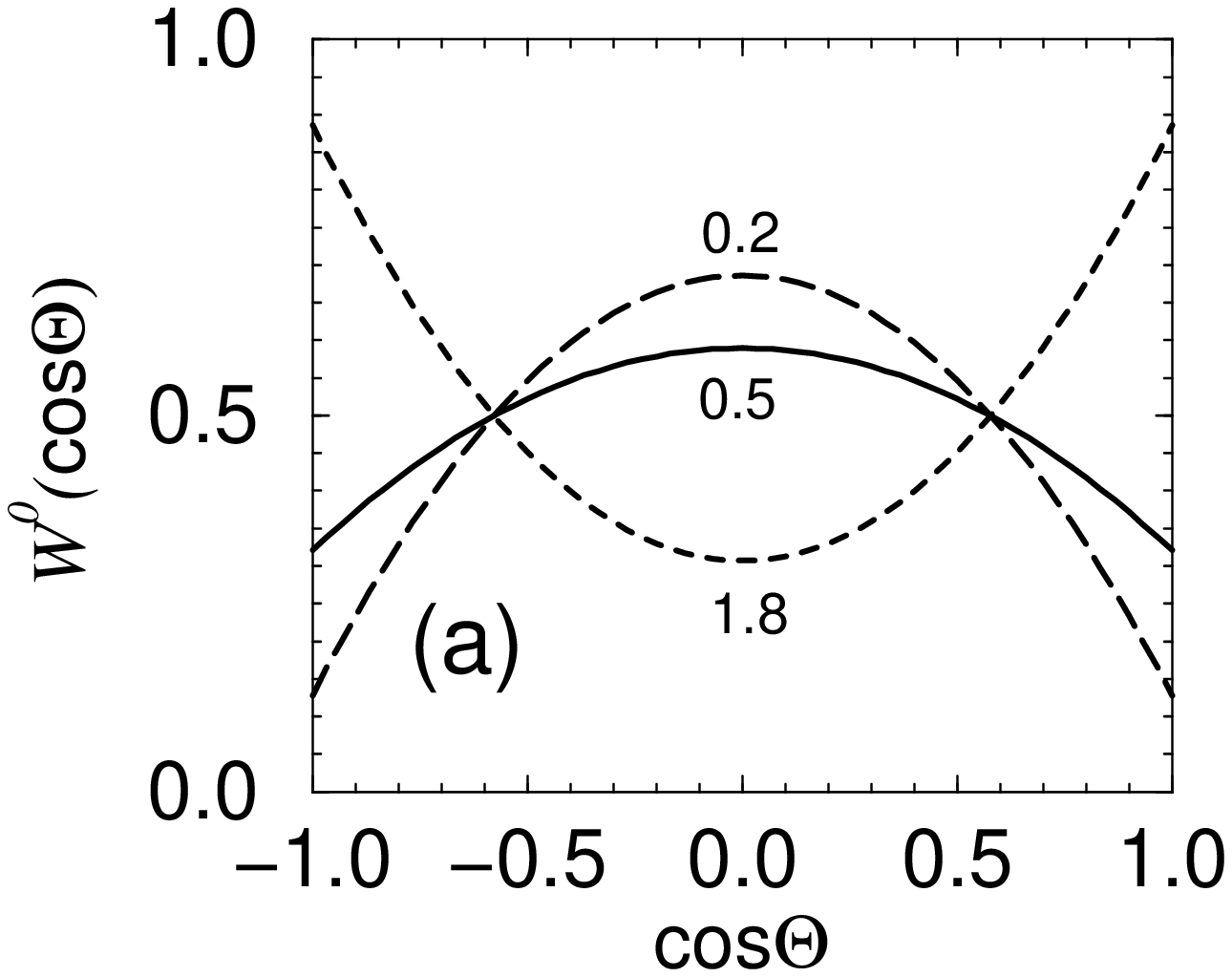}\qquad
 \includegraphics[width=50mm]{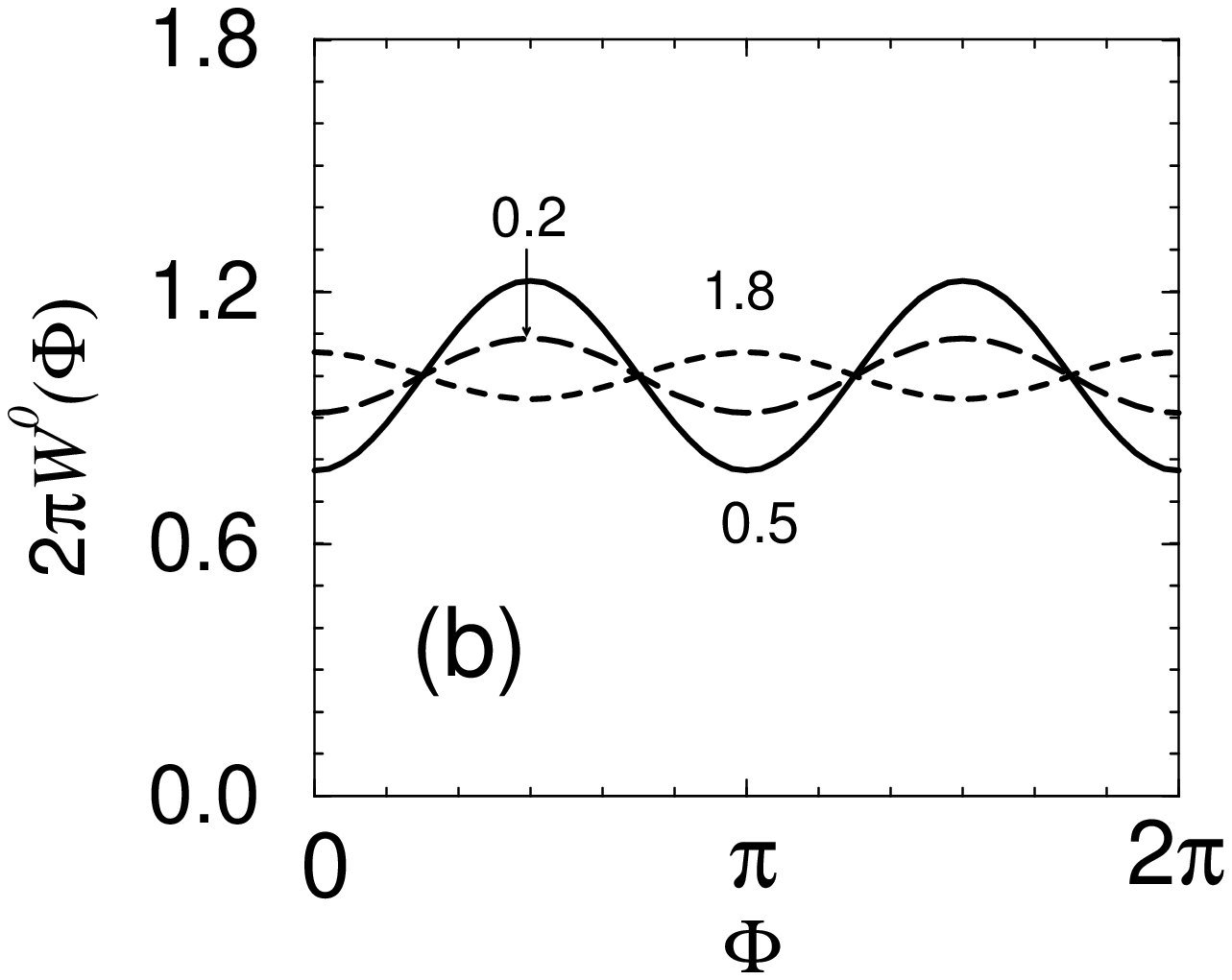}\qquad
 \caption{The angular distribution of the $\phi$-meson
 decay in reaction $\gamma p\to\phi p$ with unpolarized photon beam
 at $E_\gamma=2.2$ GeV
 and  $|t|=$ 0.2, 0.5 and 1.8 GeV$^2$
 (a): The dependence on $\cos\Theta$ (integrated over
 the azimuthal angle $\Phi$);
 (b): The dependence on $\Phi$ (integrated over
 $\cos\Theta$).}
 \label{fig:7}
\end{figure}
 are shown in Figs.~\ref{fig:7}(a) and (b), respectively.
 For small momentum transfers
 ($|t|\lesssim 0.2$ GeV$^2$),  the $\phi$-mesons
 are produced transversely with its spin  aligned
 along the quantization axis ${\bf z}'$ ($\rho^0_{00}\lll 1$).
 This results in the angular distribution
 $W^0(\cos\Theta)\simeq \sin^2\Theta$.
 When $|t|$ increases, the spin-flip processes generate
 longitudinally-polarized $\phi$-mesons ($\rho^0_{00}$ becomes finite).
 Initially, this leads
 to depolarization of the $\phi$-mesons with  $W^0(\cos\Theta)\simeq 0.5$
 at $|t|\sim 0.5-0.6$ GeV$^2$,
 and then, at large momentum transfers,
 to a predominance of the longitudinal polarization
 with $W^0(\cos\Theta)\simeq a + b\cos^2\Theta$, where  $a,b>0$.

 The  $\Phi$-dependence of $W^0$ is determined completely by
 the double-spin-flip processes. The amplitude of
 the  azimuthal-angle modulation is proportional to
 $2\rho^0_{1-1}$. It is exactly zero at $\theta=0$ ($|t|=|t|_{\rm min}$),
 increases with increasing $|t|$ and reaches its maximum value
  at $|t|\simeq 0.6$ GeV$^2$.
 It  goes down  at large momentum transfers, as it is shown
 in Fig.~\ref{fig:7}(b). It is important to note, that
 the spin-conserving scalar and pseudoscalar-exchange processes
 do not contribute to  $\rho^0_{1-1}$. The contribution of the
 tensor part of $f_2'$ (square brackets in (\ref{h-f2})) and the resonant
 channel  at $|t|\lesssim 0.8$
 GeV$^2$ are rather small.
 Therefore, the distribution $W^0(\Phi)$  may
 be used as a tool to study dynamics of the spin-orbital
 interaction generated by the gluon-exchange processes in
 diffractive amplitude. At large momentum transfers  $W^0(\cos\Theta)$,
 and $W^0(\Phi)$ are sensitive to the resonant channel.

\begin{figure}[ht] \centering
 \includegraphics[width=50mm]{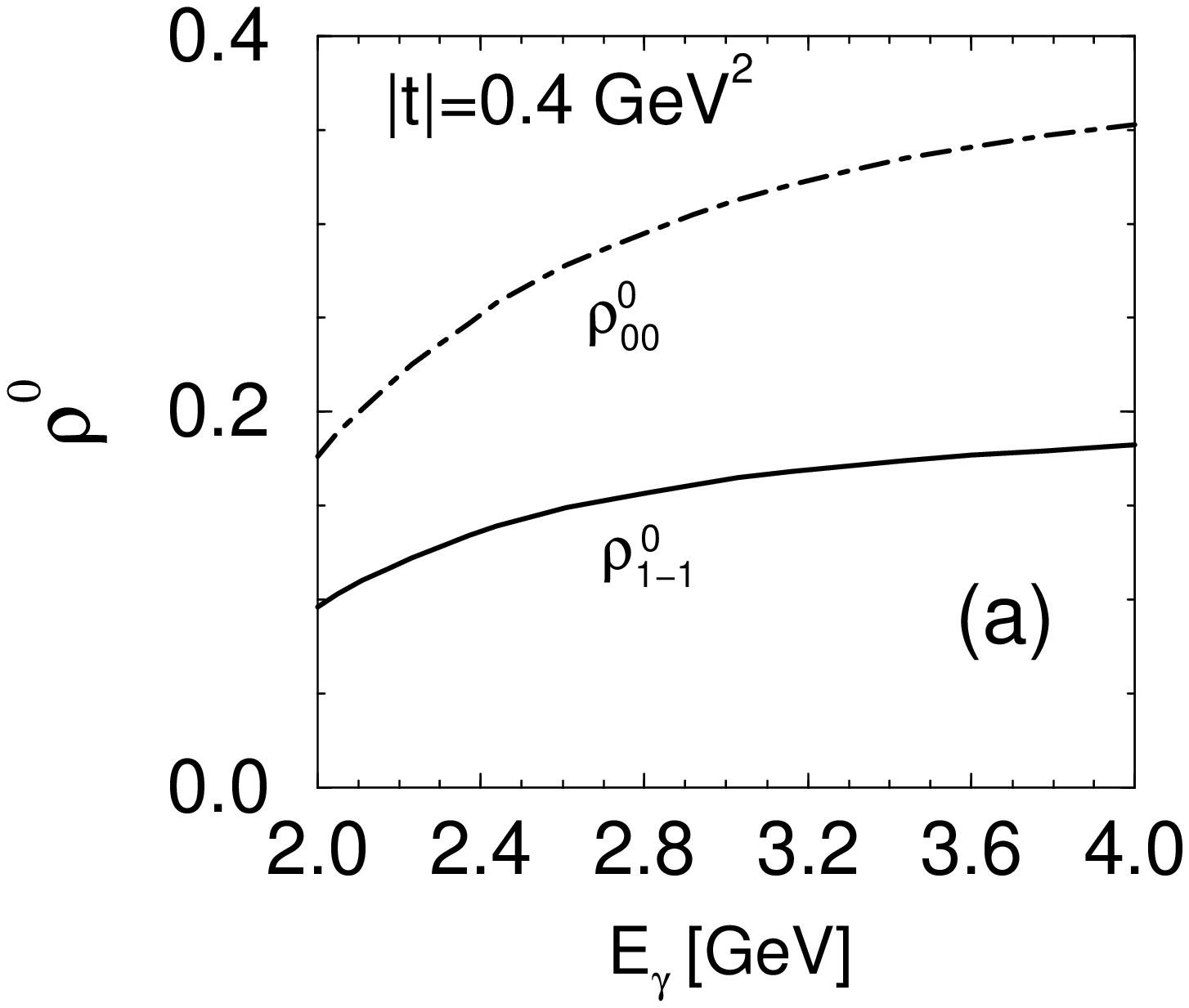}\qquad\qquad
  \includegraphics[width=50mm]{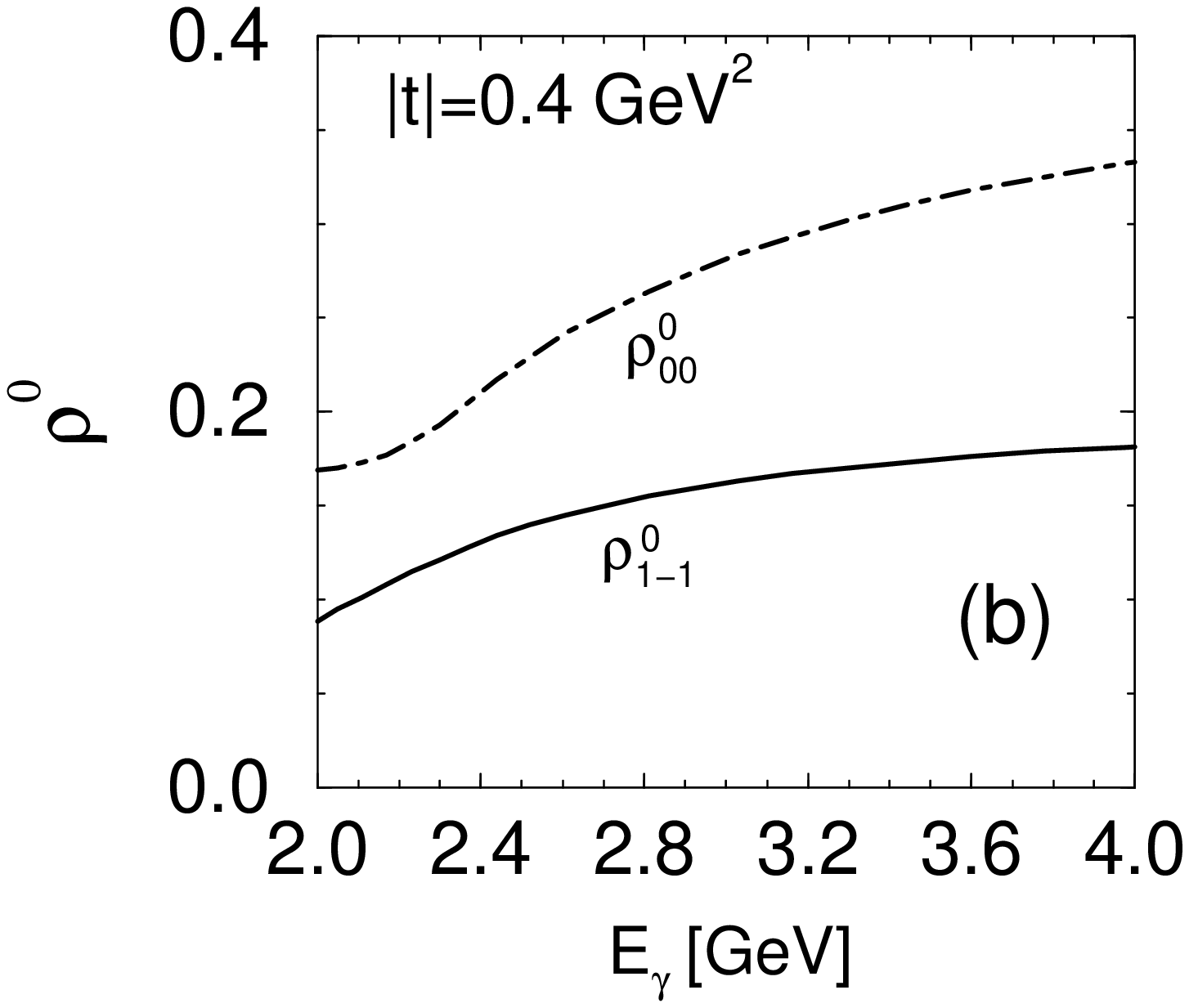}
 \caption{Spin-density matrix elements
 $\rho^0_{00}$ and  $\rho^0_{1-1}$
 for reaction $\gamma p\to\phi p$ as a function of $E_\gamma$
 at $|t|=0.4$ GeV$^2$.
 (a): Result for the Pomeron exchange amplitude;
 (b): Result for the full model. }
 \label{fig:13}
\end{figure}

 Fig.~\ref{fig:13} displays the energy dependence of  the matrix
 elements $\rho^0_{00}$  and $\rho^0_{1-1}$ which define the
 one-dimensional distributions $W^0(\cos\Theta)$ and $W^0(\Phi)$,
 respectively, at fixed $|t|$: $|t|=0.4$ GeV$^2$. The left (a)
 and right (b) panels correspond to calculation for the Pomeron-exchange
 and for the full amplitudes, respectively. One can see that the
 energy dependence of the matrix elements in both cases is very
 similar to each other.
 The increase of $\rho^0_{00}$ with energy reflects definite increase
 of amount of the longitudinally polarized $\phi$-mesons
 with energy. The increase of $\rho^0_{1-1}$ with energy results
 in increasing the amplitude of the $\Phi$-modulation in
 $W(\Phi)$.

\begin{figure}[ht] \centering
 \includegraphics[width=50mm]{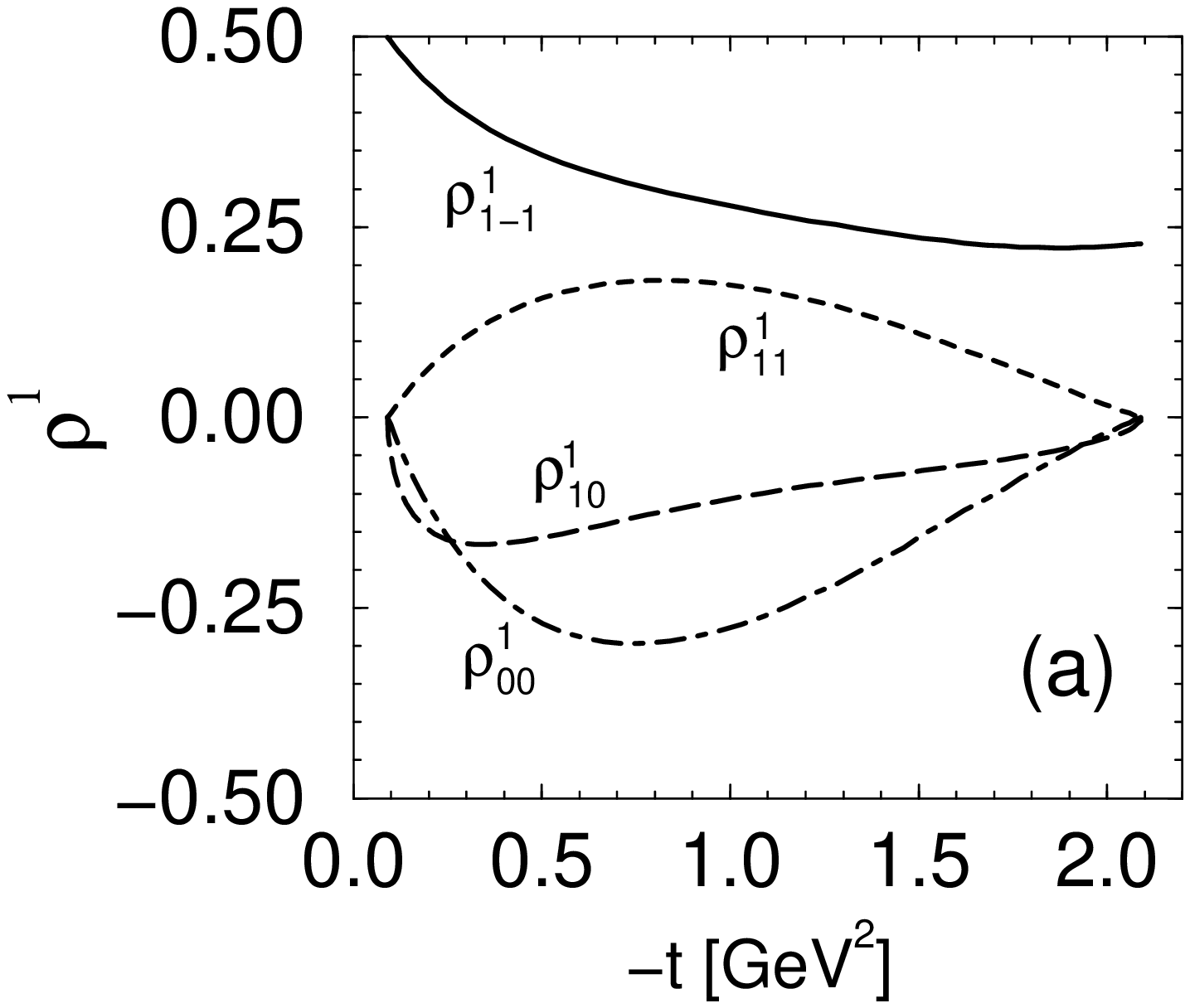}\qquad\qquad
 \includegraphics[width=50mm]{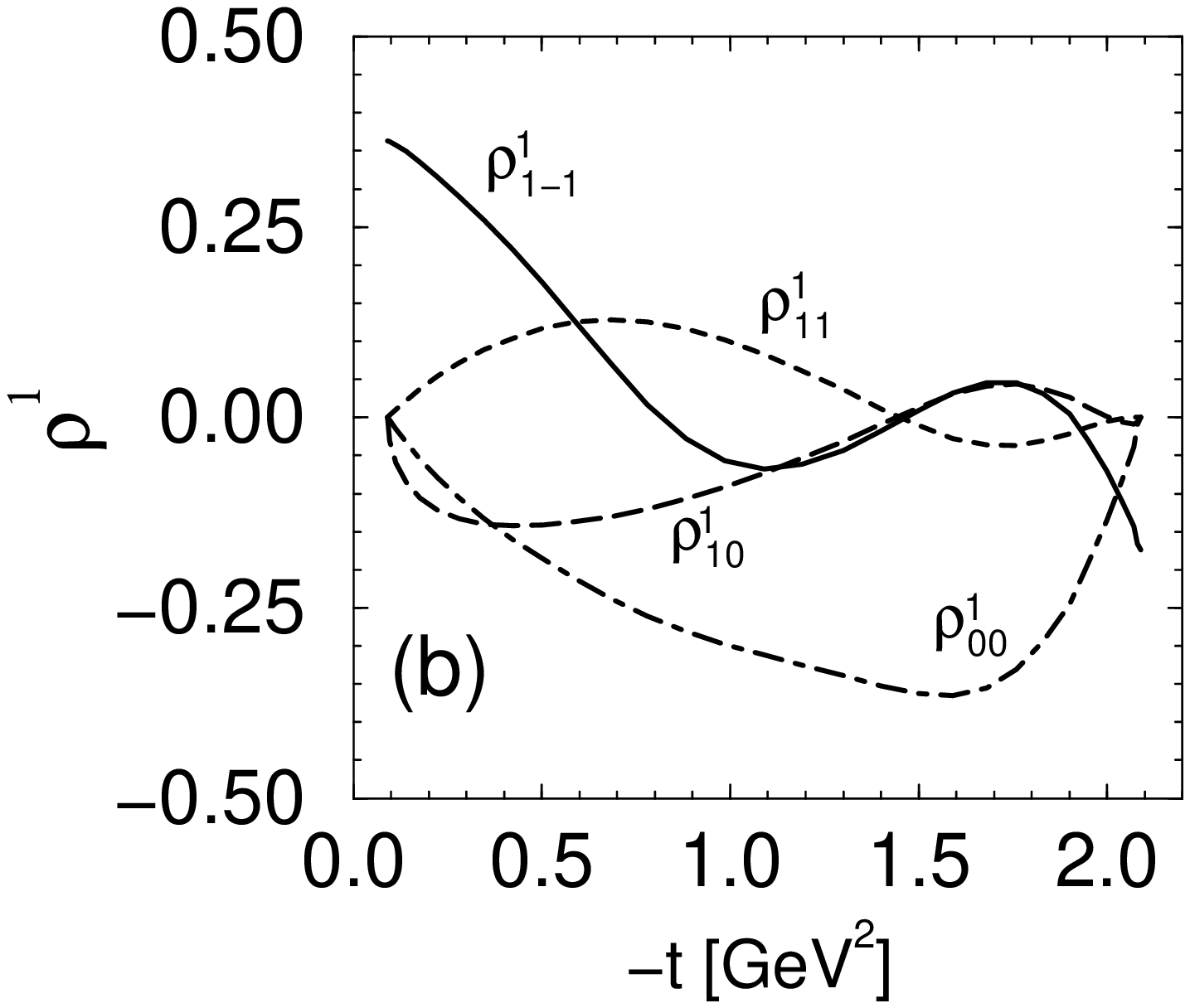}
 \caption{Spin-density matrix elements
 $\rho^1_{00}$,  $\rho^1_{11}$, $\rho^1_{10}$ and $\rho^1_{1-1}$
 for reaction $\gamma p\to\phi p$ as a function of $-t$
 at $E_\gamma=2.2$ GeV. (a): Result for the Pomeron-exchange amplitude;
 (b): Result for the full model.}
 \label{fig:8}
\end{figure}
\begin{figure}[ht] \centering
 \includegraphics[width=50mm]{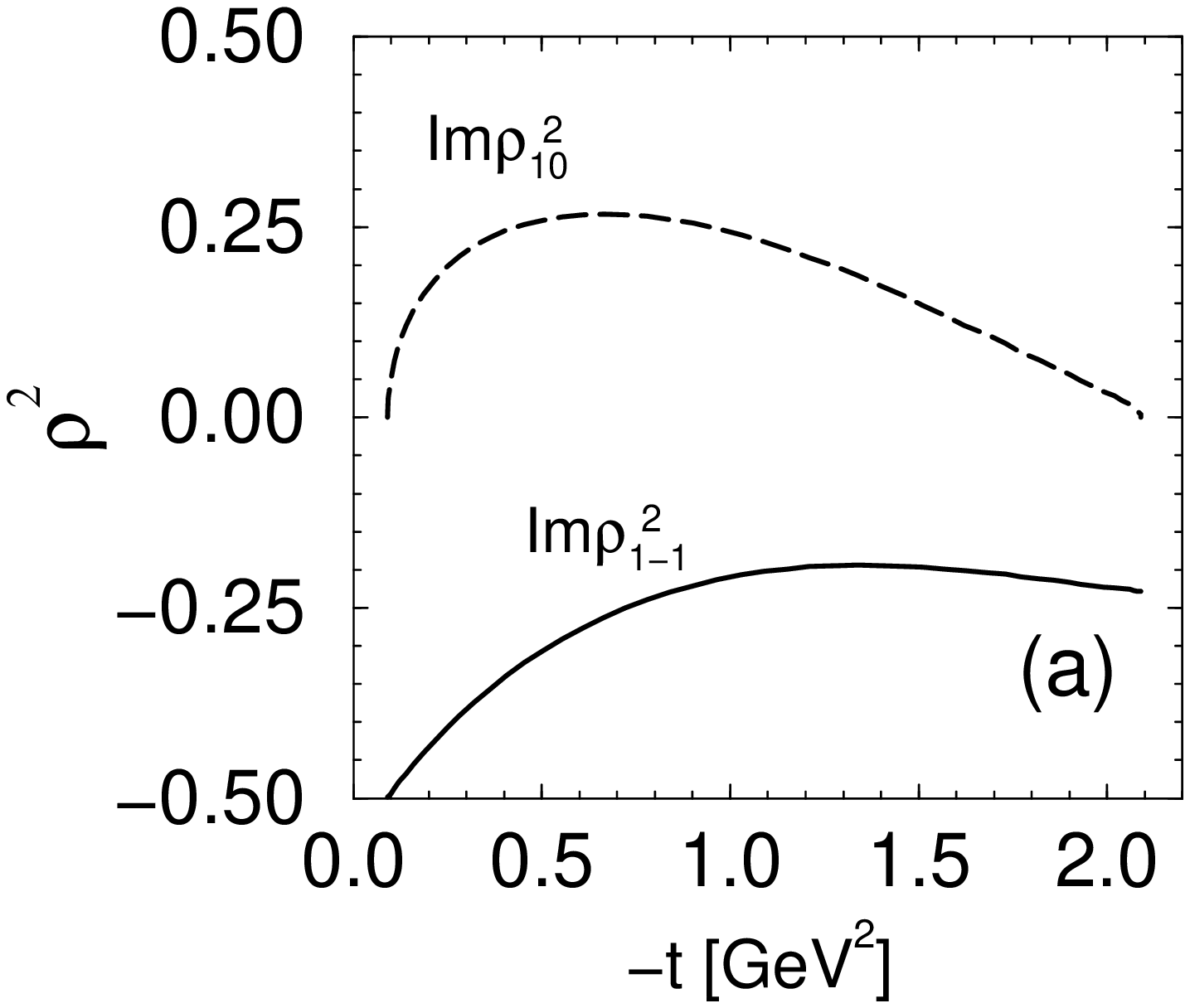}\qquad\qquad
\includegraphics[width=50mm]{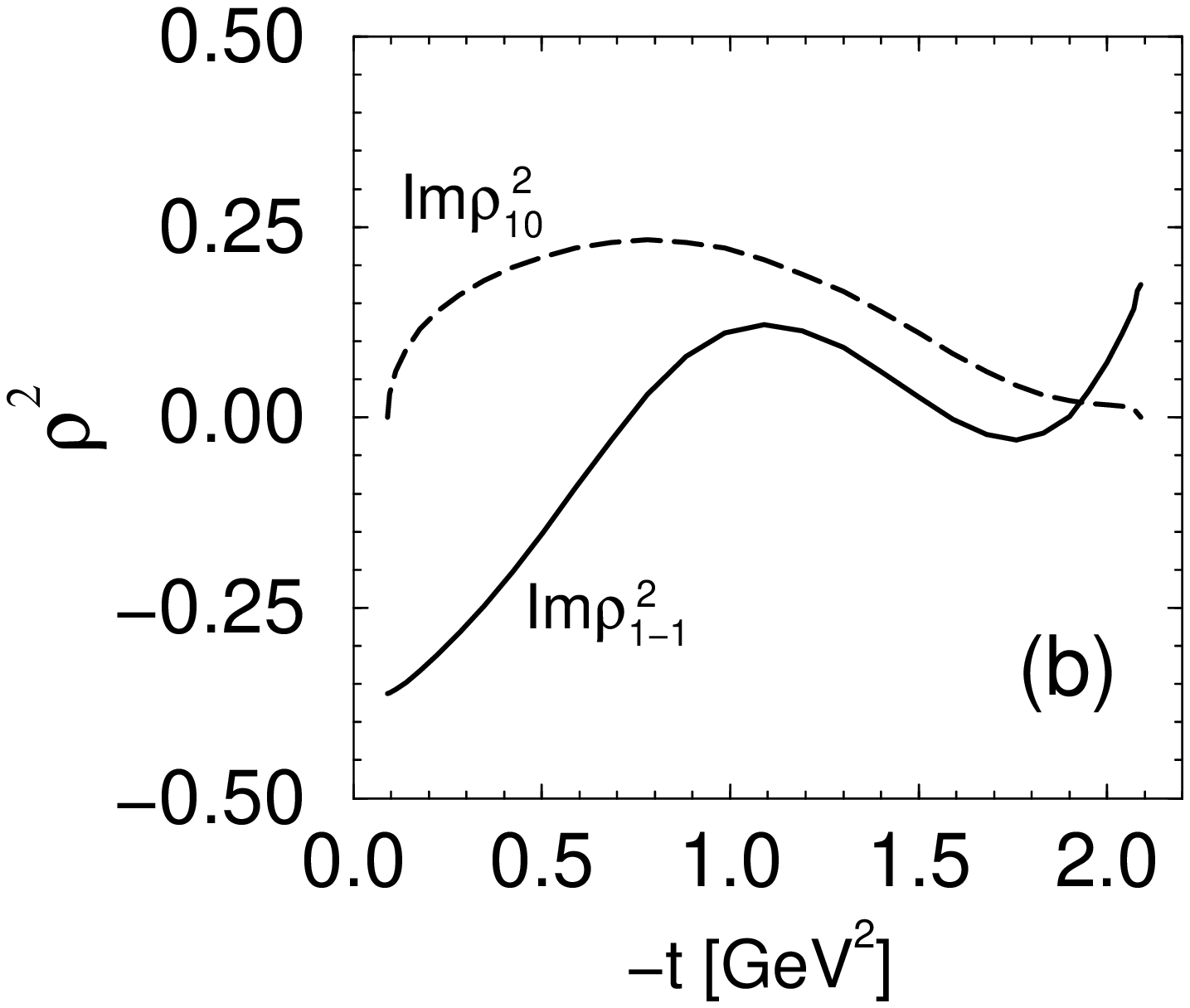}
 \caption{Spin-density matrix elements
 Im$\rho^2_{10}$ and  Im$\rho^2_{1-1}$
 for reaction $\gamma p\to\phi p$ as a function of $-t$
 at $E_\gamma=2.2$ GeV.
 (a): Result for the Pomeron exchange amplitude;
 (b): Result for the full model.}
 \label{fig:9}
\end{figure}

 Figs.~\ref{fig:8} and \ref{fig:9}
 display the $t$-dependence of the  matrix elements of
 $\rho^1$ and $\rho^2$ matrices, respectively.
 These matrix elements represent the angular distribution
 of $K^+K^-$-mesons in reactions with the linearly  polarized
 photons (cf. Eqs.~(\ref{LDD}), (\ref{WL})). Notation is the same as
 in Fig.~\ref{fig:6}.
 Consider first the matrix element $\rho^1_{1-1}$. In the case of
 the helicity conserving model of Eq.~(\ref{I_NU}), its meaning is
 the asymmetry in the contribution from  natural and unnatural-parity
 exchange parts
\begin{eqnarray}
  \rho^{1}_{1-1}=\frac12\,
   \frac{|I^{\rm N}_0|^2 - |I^{\rm U}_0|^2}
  {|I^{\rm N}_0|^2 + |I^{\rm U}_0|^2}.
   \label{rho11-1-1}
\end{eqnarray}
Therefore, it is considered as a good tool to extract the relative
contributions of the unnatural-parity-exchange processes from the
angular distribution $W^L(\Phi - \Psi)$. However, the existence of
spin-flip processes violates  this identity and instead of
Eq.~(\ref{rho11-1-1}) one has to use
\begin{eqnarray}
  \rho^{1}_{1-1}=\frac{1}{2}
   \frac{|I^{\rm N}_0|^2 - |I^{\rm U}_0|^2  + |I^{1-1}_1|^2}
  {|I^{\rm N}_0|^2 + |I^{\rm U}_0|^2 +|I^{10}|^2 + |I^{1-1}_2|^2 },
   \label{rho11-1-2}
\end{eqnarray}
 where $|I^{10}|^2={\rm Tr }[I_{\alpha;10}I^\dag_{\alpha;10}]$ is the
 spin-contribution for transitions
 $\lambda_\gamma\to\lambda_\phi=0$;
 $|I^{\alpha;1-1}_1|^2={\rm Tr}
 [I_{\alpha;1-1}I^\dag_{\alpha;-11}]$ and
 $|I^{\alpha;1-1}_2|^2={\rm Tr} [I_{\alpha;1-1}I^\dag_{\alpha;1-1}]$
 are the spin-flip contributions for transitions
 $\lambda_\gamma\to\lambda_\phi=-\lambda_\gamma$.
 Only at $\theta=0$ ($|t|=|t|_{\rm min}$) and $\rho^0_{00}\simeq0$
 equations (\ref{rho11-1-1})
 and (\ref{rho11-1-2}) are equivalent to each other.

 For the pure Pomeron-exchange amplitude the contribution of the
 double-spin-flip can be estimated as
\begin{eqnarray}
 && \frac{1}{N}{|I^{1-1}_1|^2}\simeq
  \frac{1}{N}{|I^{1-1}_2|^2}\simeq
   \frac{1}{2(1-\rho^0_{00})}(\rho^0_{1-1})^2\ll\rho^{0}_{00},
\label{R1-1}
\end{eqnarray}
where  $N$ is the normalization factor, defined by
Eq.~(\ref{Norm}). Therefore, for the pure Pomeron-exchange
amplitude we get the relation
\begin{eqnarray}
  \rho^{1}_{1-1}\simeq\frac12\,(1-\rho^0_{00})
   \label{rho11-1-a}
\end{eqnarray}
which is confirmed  by the explicit calculation of $\rho^0_{00}$
and $\rho^1_{1-1}$, see in Figs.~\ref{fig:6}(a) and
\ref{fig:8}(a): $\rho^{1}_{1-1}$ decreases from $\frac12$ at
$|t|=|t|_{\rm min}$ to 0.23 at $|t|=|t|_{\rm max}$.

For the forward-angle photoproduction, where the contribution of
the resonant channel to $|I^{1-1}|^2$ remains negligible, we get
the following relation between $\rho^{1}_{1-1}$ and the relative
contribution of the  spin-conserving unnatural-parity-exchange
amplitude
  \begin{eqnarray}
  |\alpha^U|^2\simeq
   \frac{|I^{\rm U}_0|^2}
  {|I^{\rm N}_0|^2 + |I^{\rm U}_0|^2 + |I^{10}|^2},
  \label{alpha-U}
 \end{eqnarray}
which can be rewritten as
\begin{eqnarray}
  |\alpha^U|^2\simeq\frac12\left(1-2\rho^1_{1-1}-\rho^0_{00}\right).
   \label{rho11-1-f}
\end{eqnarray}
This means that for evaluation of the relative contribution of the
unnatural-parity exchange  part from the data with a linearly
polarized beam at small momentum transfers one has to account for
$\rho^0_{00}$, which in turn is extracted from the analysis of
$W^0(\cos\Theta)$. At large $|t|$, $\rho^1_{1-1}$ is determined by
the interplay between  the resonant and all other channels and has
no simple meaning.

Using the definition of spin-density matrices in
Eqs.~\ref{rhomatrix} one can get the following relation
\begin{eqnarray}
 -{\rm Im}\rho^2_{1-1}\simeq \rho^1_{1-1}
 - \frac{\left(\rho^0_{1-1}\right)^2}{1-\rho^0_{00}}.
 \label{rho12}
\end{eqnarray}
 Therefore, in Eq.~(\ref{WL-0}),
$\bar{\rho}^1_{1-1}\simeq\rho^1_{1-1}$
 and  $\Delta_{1-1}\simeq 0$. So,
 the term proportional to $\cos[2(\Phi+\Psi)]$ in Eq.~(\ref{WL-0}) is negligible.
\begin{figure}[ht] \centering
 \includegraphics[width=50mm]{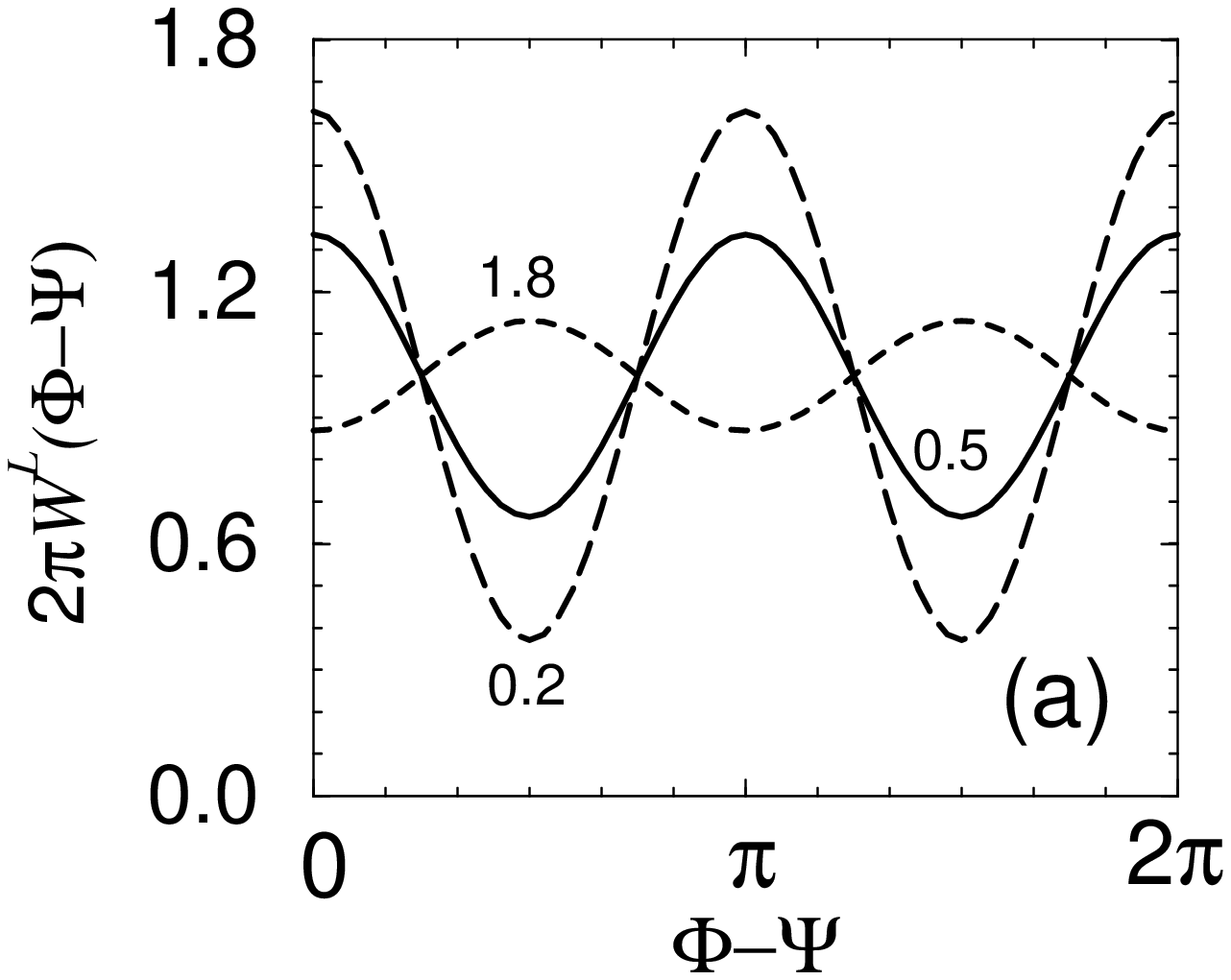}\qquad\qquad
 \includegraphics[width=50mm]{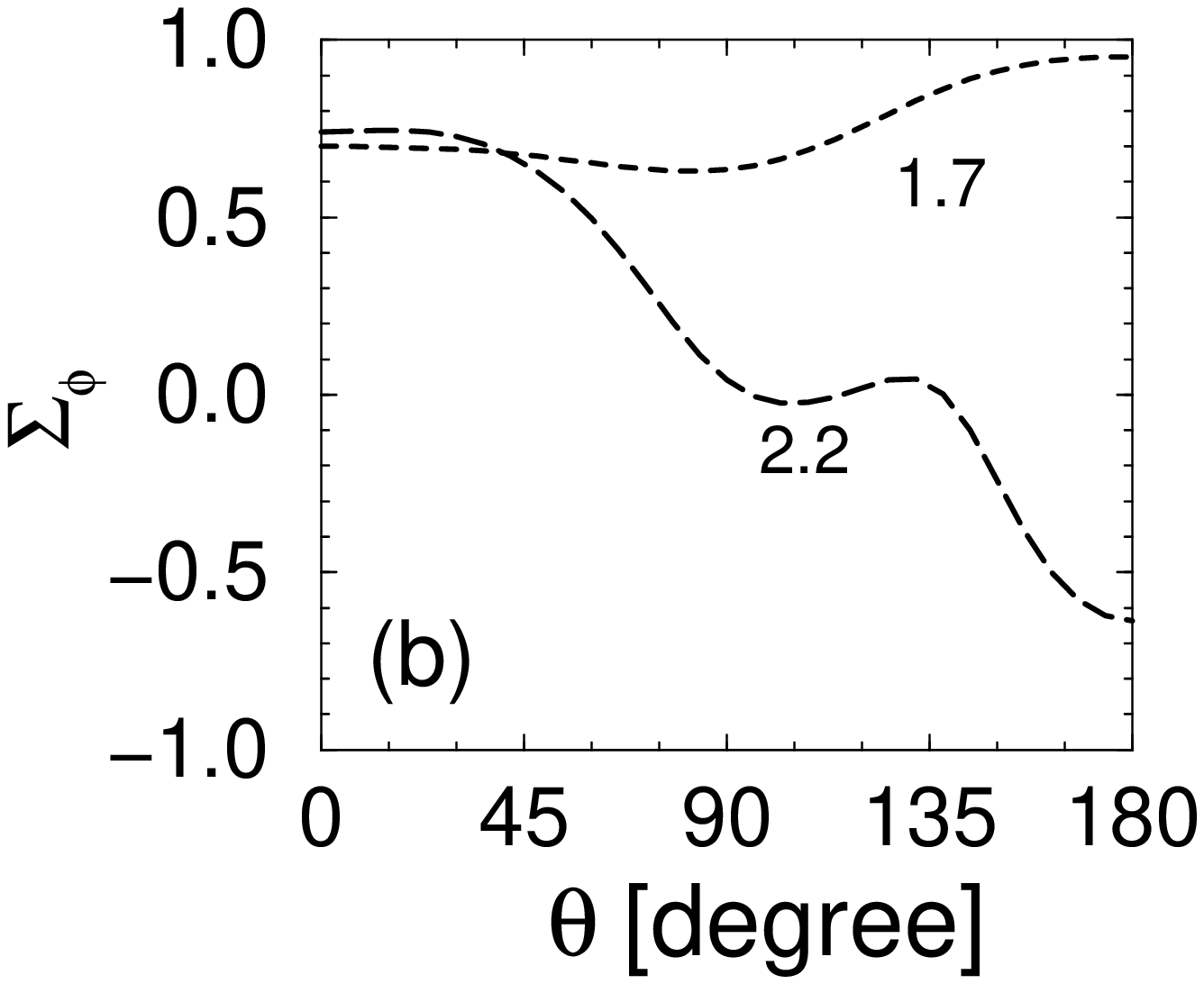}
 \caption{(a): The angular distribution of the $\phi$-meson
 decay as a function of $\Phi-\Psi$
 in reaction $\gamma p\to\phi p$ with linear polarized photon beam
 at $E_\gamma=2.2$ GeV
 and  $-t=$ 0.2, 0.5 and 1.8 GeV$^2$.
 b:  The $\phi$-meson decay  asymmetry for $\gamma p\to \phi p$ reaction at
  $E_\gamma=1.7$ and  2.2 GeV.}
 \label{fig:10}
\end{figure}

 Fig.~\ref{fig:10}(a) shows the angular distribution $W^L(\Phi-\Psi)$
 at $|t|=0.2,\,0.5$ and
 $1.8$ GeV$^2$ ($E_\gamma=2.2$ GeV). The amplitude of modulation
 of this distribution is equal
 to $2P_\gamma\, {\rho}^1_{1-1}$. In calculation we use
 $P_\gamma=0.95$ which is reasonable  for the highly-polarized
 photon beam at the LEPS of SPring-8~\cite{LEPS}.
 The amplitude of modulation has a maximum value at forward
 angles and decreases with increasing $|t|$.

 Fig.~\ref{fig:10}(b) displays the $\phi$-meson decay asymmetry
 $\Sigma_\phi$ (\ref{sigma_v}), as a function of production
 angle $\theta$.
 It is determined by
 the matrix elements $\rho^{1,0}_{1-1}$ and $\rho^{1,0}_{11}$.
 For the pure helicity-conserving
 amplitude there exist an identity: $\Sigma_\phi=2\rho^1_{1-1}$.
 But in spin-flip processes this relation is violated. The scale
 of the violation increases with $|t|$. At large $|t|$, the shape of
 $\Sigma_\phi$  is sensitive to the resonant amplitude.

\begin{figure}[ht] \centering
 \includegraphics[width=50mm]{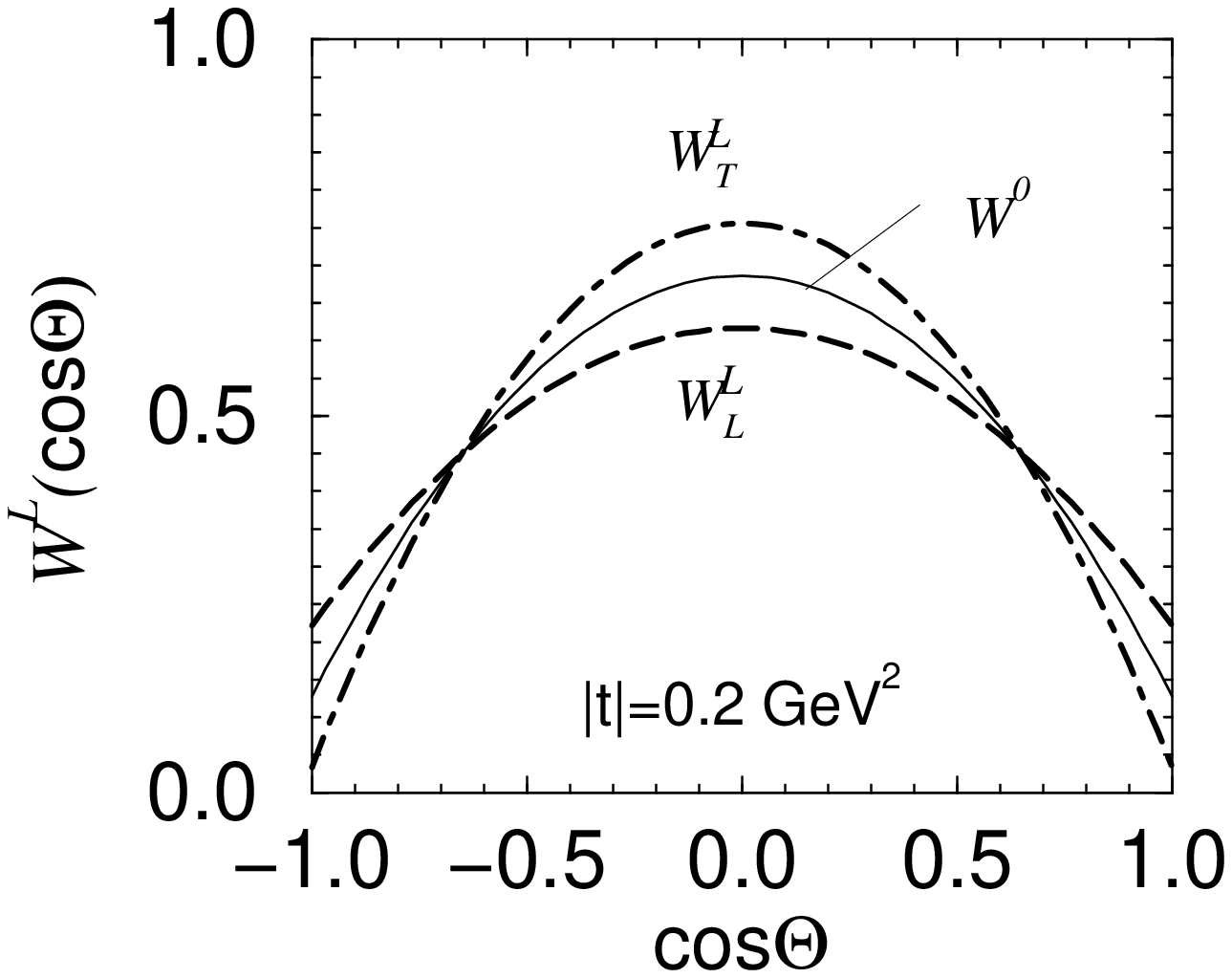}\qquad\qquad
 \includegraphics[width=50mm]{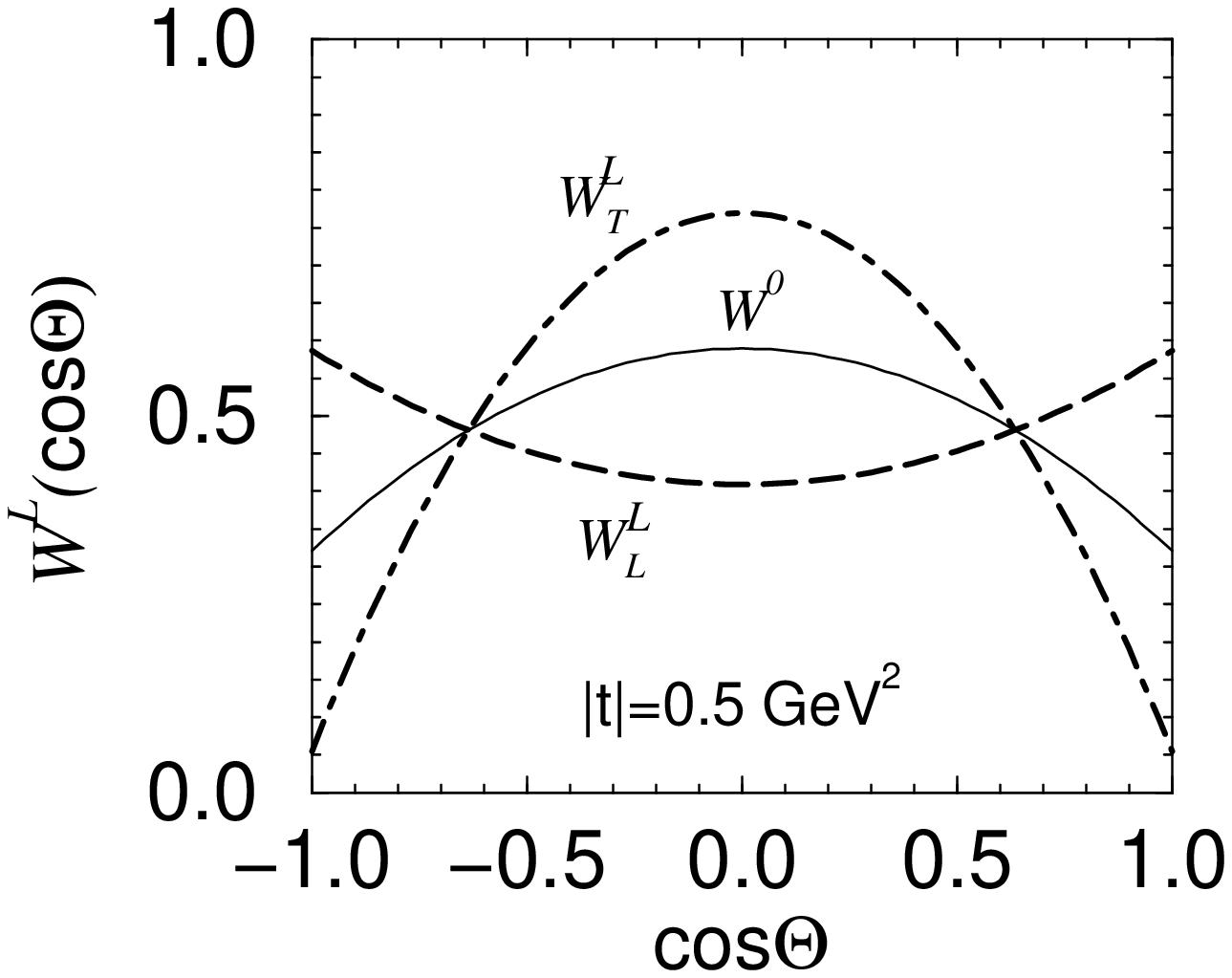}
 \caption{The  $\phi$-meson decay
  distribution in reaction $\gamma p\to\phi p$
  with linear polarized photon beam
 at $E_\gamma=2.2$ GeV
 for vertical beam polarization with $\Psi=\frac\pi2$ ($W^L_T$)
 and horizontal polarization with $\Psi=0$ ($W^L_L$)
 at $|t|=$ 0.2 (a) and 0.5 GeV$^2$ (b).}
 \label{fig:add}
\end{figure}
 The decay distribution as a function of $\cos\Theta$ for
 the linearly polarized beam depends on the beam-polarization
 angle $\Psi$ and  matrix elements $\rho^1_{11}$ and
 $\rho^1_{00}$ (cf. Eq.\ref{WL-0}).
  In Fig.~\ref{fig:add} we show result for the calculation of
  $W^L_L(\cos\Theta)\equiv W^L(\cos\Theta,\Psi=0)$ and
  $W^L_T(\cos\Theta)\equiv W^L(\cos\Theta,\Psi=\frac\pi2)$ at
  $|t|=0.2$ GeV$^2$ and $|t|=0.5$ GeV$^2$:  left (a) and
  right (b) panels, respectively. Also,  for
  comparison, results for the angular distributions with an unpolarized
  beam are shown (thin solid lines). The difference between
  $W^L_L$ and $W^L_T$ disappears at $|t|\simeq |t|_{\rm min}$, but
  becomes very large at finite $|t|$. Using these distributions,
  one can extract $\rho^1_{00}$ and  $\rho^1_{11}$ from the data:
  \begin{eqnarray}
  \rho^1_{00}=\frac{1}{3P_\gamma}(W^L_T(\Theta=0)-W^L_L(\Theta=0))\nonumber\\
  \rho^1_{11}=\frac{1}{3P_\gamma}
  (W^L_T(\Theta=\frac\pi2)-W^L_L(\Theta=\frac\pi2)).
  \end{eqnarray}

  The decay distribution as a function of the beam-polarization
 angle $\Psi$ is defined by the matrix elements $\rho^1_{11}$ and
 $\rho^1_{00}$. These matrix elements, taken separately, are
 finite at  $|t|\neq |t|_{\rm min,}\,|t|_{\rm max}$.
 But the absolute value of the sum
 $2\rho^1_{11}+\rho^1_{00}$ in Eqs.~(\ref{WLPSI})
 is very small at forward-angle photoproduction
 \begin{eqnarray}
 2\rho^1_{11}+\rho^1_{00}\simeq0, \label{rho11-00}
 \end{eqnarray}
 and becomes
 sizeable only at large momentum transfers. Eq.~(\ref{rho11-00})
 allows to express the angular distribution for the linearly polarized
 beam $W^L(\cos\Theta,\Psi)$ similarly to the distribution for
 unpolarized beam $W^0(\cos\Theta)$
\begin{eqnarray}
 {W}^L (\cos\Theta,\Psi)=\frac32\left(
 \frac{1}{2}(1-\rho^{\rm eff}_{00})\sin^2\Theta
  + \rho^{\rm eff}_{00}\cos^2\Theta\right),
 \label{WL-eff}
 \end{eqnarray}
where
\begin{eqnarray}
 \rho^{\rm eff}_{00}=
 \rho^{0}_{00}-P_\gamma\rho^1_{00}\cos(2\Psi).
 \label{rho-eff}
 \end{eqnarray}
\begin{figure}[ht] \centering
 \includegraphics[width=50mm]{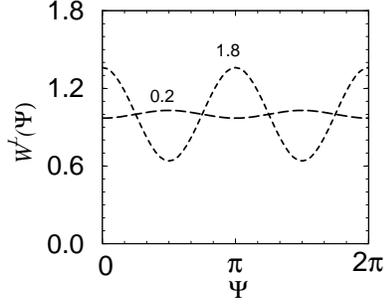}\qquad\qquad
 \caption{The total $\phi$-meson decay
  distribution in reaction $\gamma p\to\phi p$
  with linear polarized photon beam
 at $E_\gamma=2.2$ GeV
 and  $-t=$ 0.2 and 1.8 GeV$^2$
as a function of  the angle between the beam polarization and
production planes.}
 \label{fig:11}
\end{figure}

 Fig.~\ref{fig:11} shows the calculation of the total decay
 distribution as a function of the beam polarization angle $\Psi$.
 It is defined by sum
 $2\rho^1_{11}+\rho^1_{00}$ (cf. Eq.~\ref{WLPSI}),
 which is finite only at large momentum transfers. Here the sign and
 the amplitude of $W^L(\Psi)$ is determined by the
 resonant channels and prediction in this region is sensitive to the
 underlying theoretical model for the resonance part.
 \begin{figure}[ht] \centering
 \includegraphics[width=50mm]{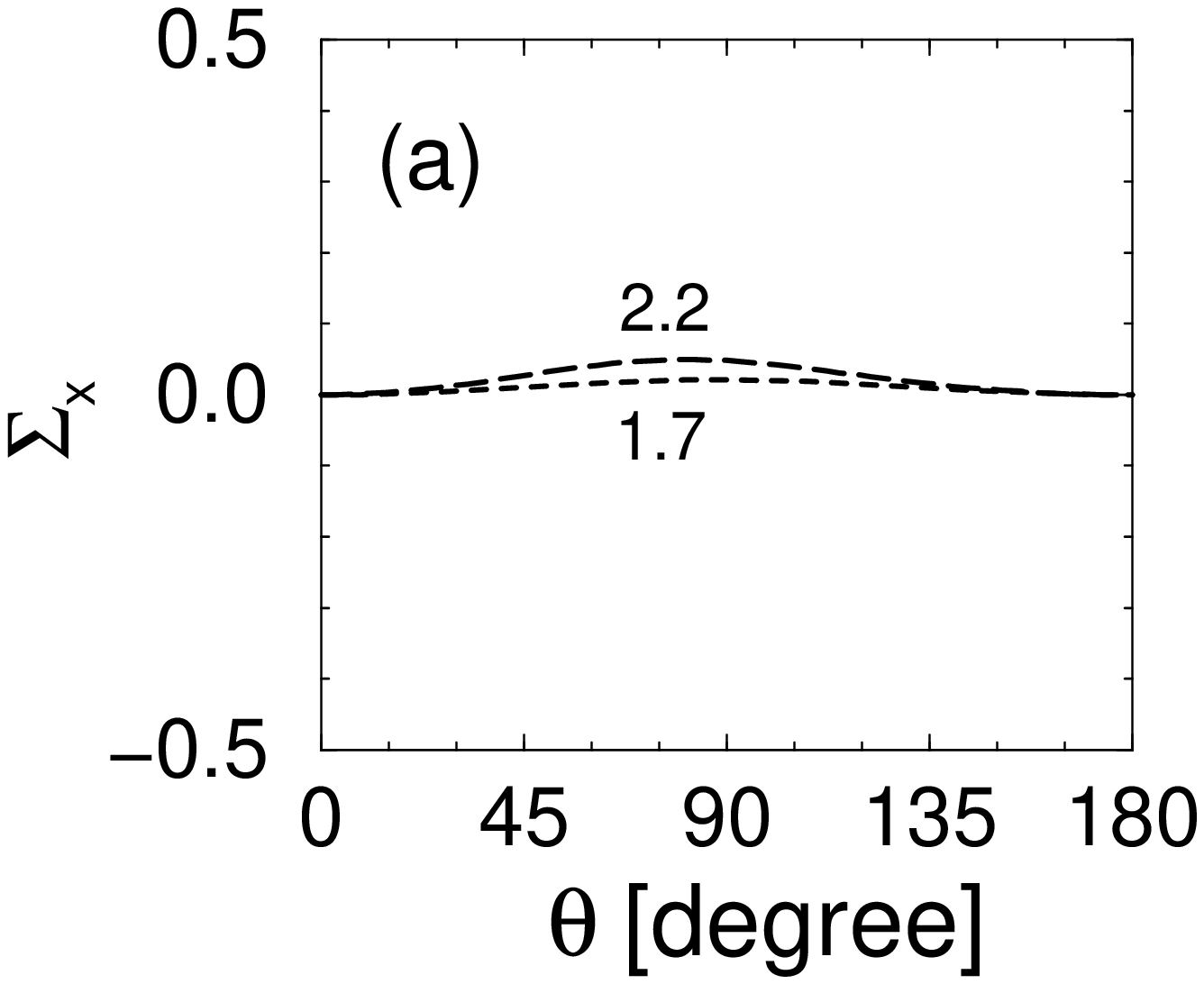}\qquad
 \includegraphics[width=50mm]{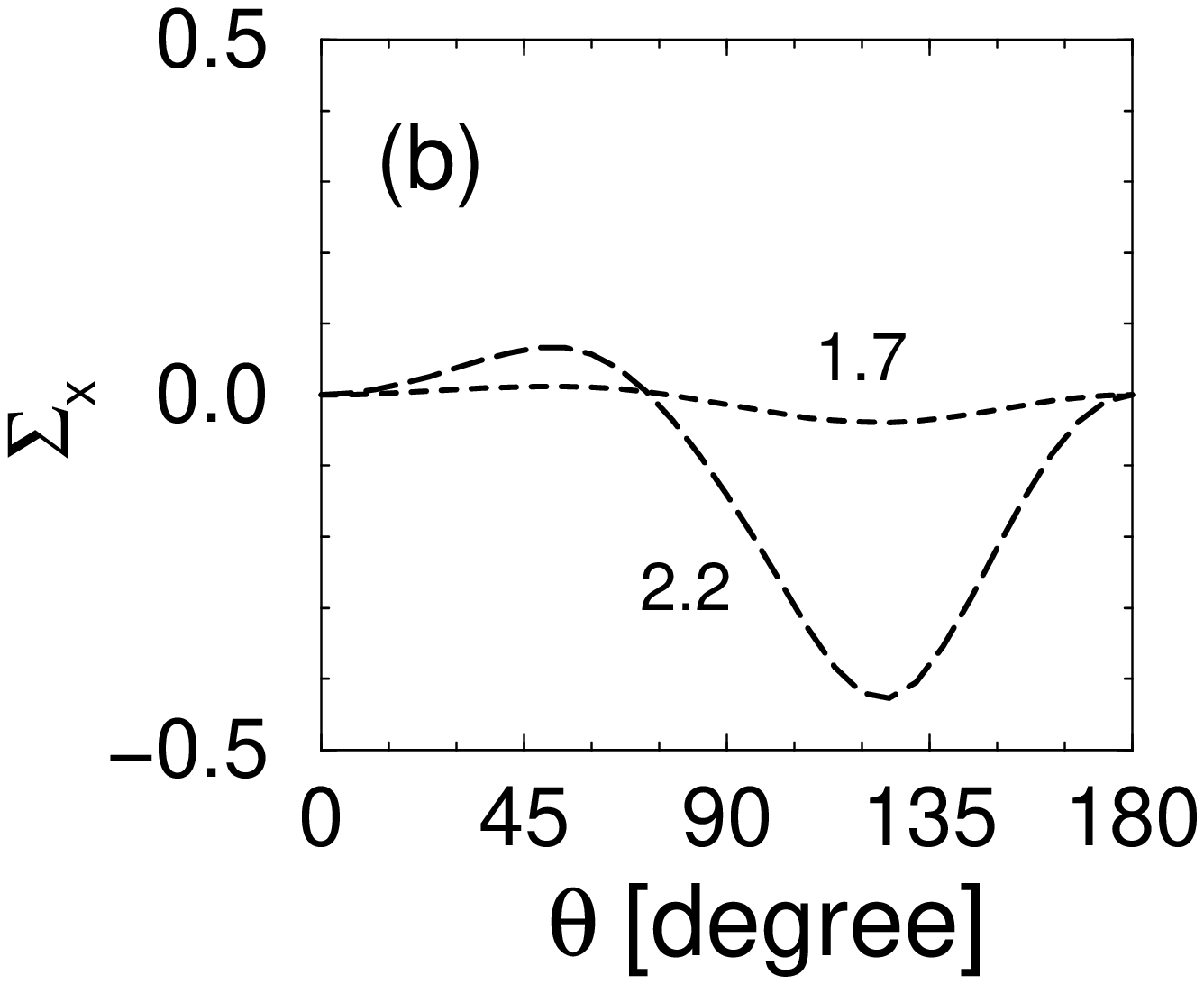}
 \vspace{0.5cm}
 \caption{
  The beam asymmetry for $\gamma p\to \phi p$ reaction at
  $E_\gamma=1.7$ and 2.2 GeV. (a): Result  for the  Pomeron exchange
  channel;
  (b): Result for the full model.}
 \label{fig:12}
\end{figure}
 The beam asymmetry $\Sigma_x$ of Eq.~(\ref{sigma_x}) is
 related to $W^L(\Psi)$ as
\begin{eqnarray}
\Sigma_x =\frac{W^L(\frac\pi2)- W^L(0)}{W^L(\frac\pi2)+ W^L(0)}
\label{sigma-xW}
\end{eqnarray}
and for the pure Pomeron exchange channel is defined by the
interference of the first and third terms in (\ref{G2g-fi}) and
can be evaluated in c.m.s. as
\begin{eqnarray}
\Sigma_x \simeq \frac{q^2\sin^2\theta}{2{\bar s}}.
 \label{sigma-xP}
\end{eqnarray}
 It is positive and increases near threshold with energy as $q^2$
(proportional to increase of the phase space),
 but remains small: $\Sigma_x\ll 1$.
 For the full model, the dominant contribution to
 $\Sigma_x$ at large $|t|$ comes from the resonant channel. Thus,
 at  $|t|\sim 1.8$ GeV ($E_\gamma=2.2$ GeV),  $\Sigma_x$ becomes to
 be negative with large absolute value. This is illustrated in
 Fig.~\ref{fig:12} where we show $\Sigma_x$, calculated for the
 pure Pomeron exchange amplitude and for the full model at the
 left (a) and the right (b) panels, respectively.

\begin{figure}[ht] \centering
 \includegraphics[width=55mm]{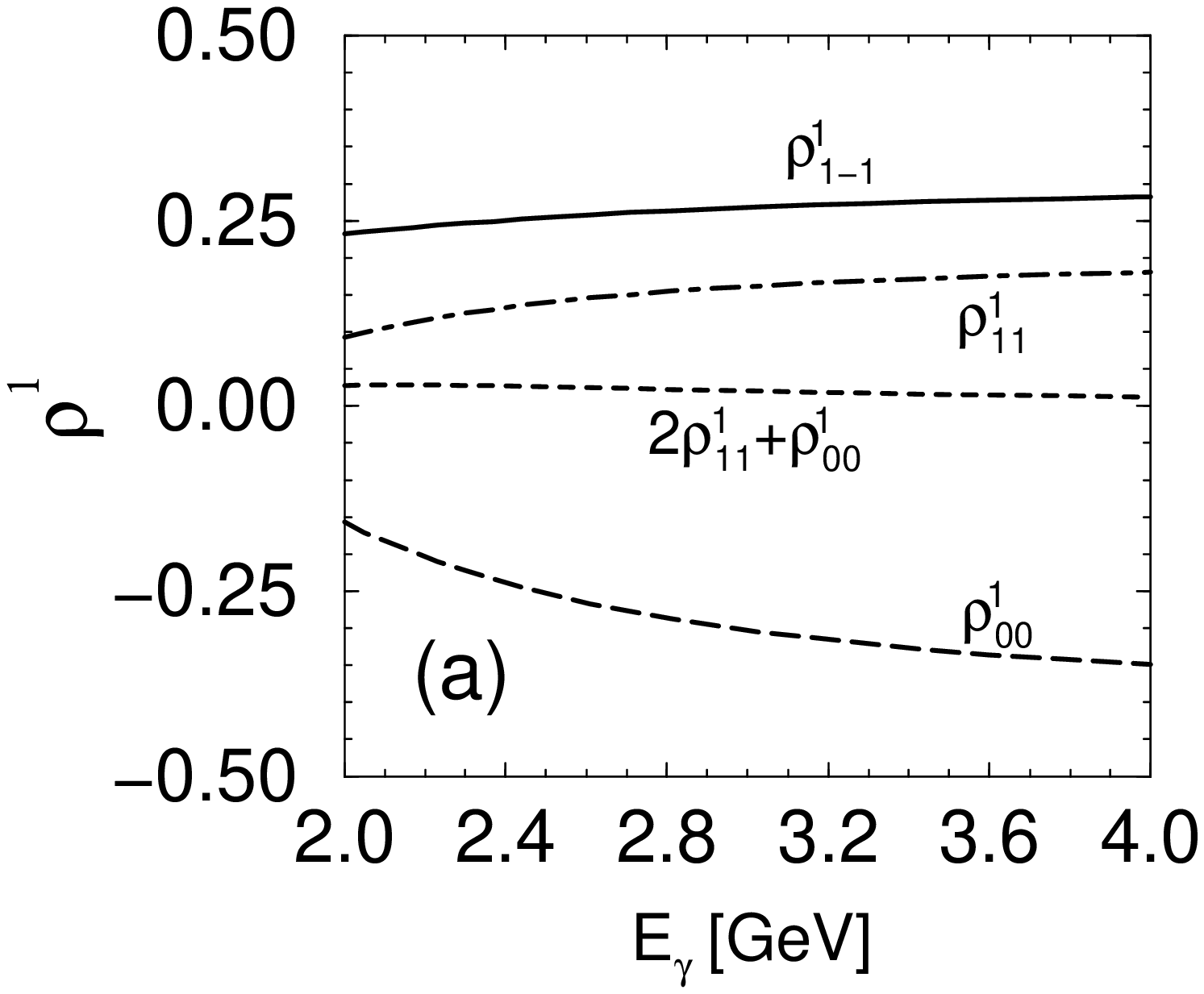}\qquad\qquad
 \includegraphics[width=55mm]{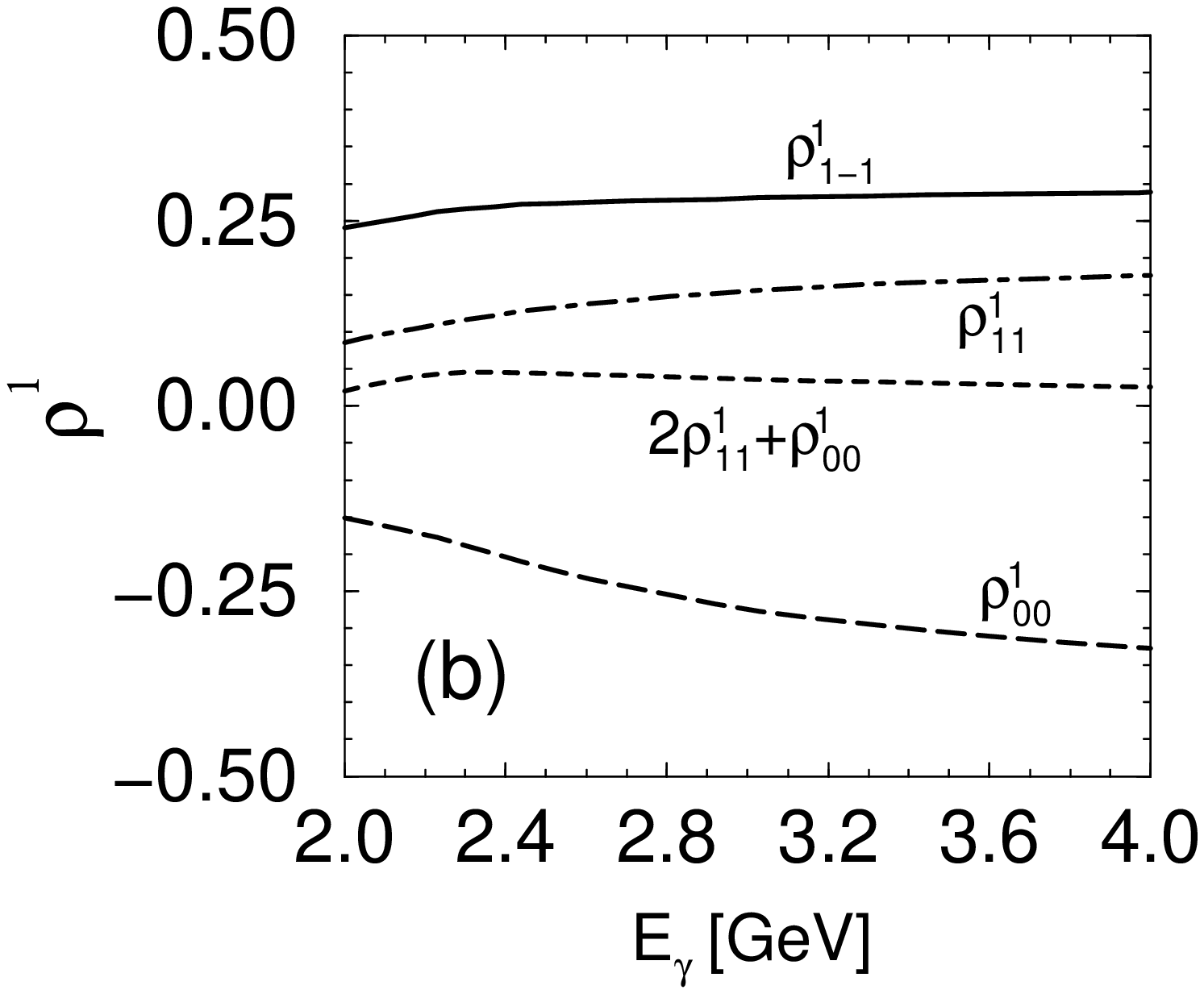}
 \caption{Spin-density matrix elements
 $\rho^1_{00}$, $\rho^1_{11}$, $\rho^1_{1-1}$,  and
 sum of $2\rho^1_{11}+\rho^1_{00}$
 for reaction $\gamma p\to\phi p$ as a function of $E_\gamma$
 at $|t|=0.4$ GeV$^2$.
 (a): Result for the Pomeron exchange amplitude;
 (b): Result for the full model. }
 \label{fig:13-1}
\end{figure}
Fig.~\ref{fig:13-1} displays the energy dependence of the most
important matrix elements, which define the angular distributions
of $\phi\to K^+K^-$-decay  with the linearly polarized photon beam
and the beam asymmetry. The left (a)
 and right (b) panels correspond to the calculation for the Pomeron-exchange
 and for the full amplitudes, respectively.
 The energy dependence of $\rho^1_{00}$ and $\rho^1_{11}$
 in both cases is similar to each other. The difference in
 $\rho^1_{1-1}$ is explained by the contribution of unnatural
 parity $\pi-\eta$-meson exchange in the total amplitude.
 The calculation results in some decrease of
 $\rho^1_{1-1}$ and  $\rho^0_{11}$; increase of  $\rho^1_{11}$, and
 almost constant value for $2\rho^1_{11}+\rho^1_{00}\simeq0$.

 Unfortunately, the available experimental data
 on the spin-density matrix elements
 $\rho^{1,2}$ in $\phi$-meson photoproduction
 at $E_\gamma\sim2-5$ GeV~\cite{Ballam73} are
 of a pure accuracy to make some definite conclusion about the photoproduction
 mechanism. To increase statistics they are combined
 at two energies (2.8 and 4.7 GeV) with momentum transfers $0.02 \leq
 |t|\leq 0.8$ GeV$^2$. The data are given in helicity frame were the
 spin-density matrix elements have additional kinematical
 dependence on the momentum transfers compared to the GJ-system~\cite{TLTS99}.
 In Table~1 we show the comparison of this
 data with our calculation for the "central" point $E_\gamma=3.75$ GeV and
 $|t|=0.4$ GeV$^2$ in the helicity frame.
 One can see that the theoretical values are within experimental
 accuracy. Exception is $\rho^1_{1-1}$, were the calculated
 value (0.44) exceeds the  experimental one ($0.18\pm0.13$) by two
 standard deviations. But close inspection shows some inconsistent with the data.
 Really, following the identity of Eq.~(\ref{rho12})
 and using experimental values of
 $\rho^0_{00},\,\rho^0_{1-1})$, one could expect $\rho^1_{1-1}\sim {-\rm
 Im}\rho^2_{1-1}$, i.e it must be close to $0.5$ which is also supported by
 Fig.~29 of~\cite{Ballam73}. Nevertheless, it is clear, that for better
 understanding of details of the photoproduction processes the
 more precise experimental data in a wide kinematical region are desired.

\begin{figure}[ht] \centering
 \includegraphics[width=55mm]{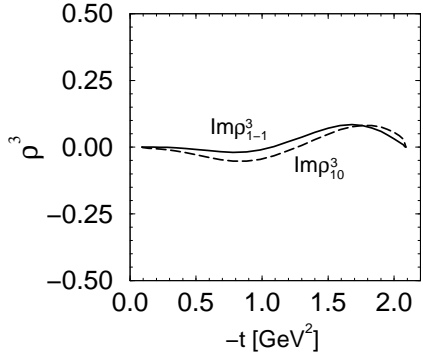}
 \caption{Spin-density matrix elements
 Im$\rho^3_{10}$ and  Im$\rho^3_{1-1}$
 for the reaction $\gamma p\to\phi p$ as a function of $-t$
 at $E_\gamma=2.2$ GeV for the full model.}
 \label{fig:14}
\end{figure}

The spin-density matrix elements $\rho^3$, responsible for the
angular distribution with  the circularly  polarized beam are
shown in Fig.~\ref{fig:14}.
 They reach their maximum values ${\rm Im}\rho^3_{1-1(10)}\simeq
 0.1$ at $|t|\sim 1.8$ GeV$^2$. The finite value of ${\rm
 Im}\rho^3_{1-1}$ generates an additional term in the
 angular distribution  according to Eq.~(\ref{WC}): $\Delta
 W^\pm(\Phi)= \pm (P_\gamma/\pi)\,{\rm Im}\rho^3_{1-1}\sin2\Phi$. At forward
 photoproduction angles this term is rather weak and
 $W^\pm(\Phi)\simeq W^0(\Phi)$.

 \begin{figure}[ht] \centering
 \includegraphics[width=55mm]{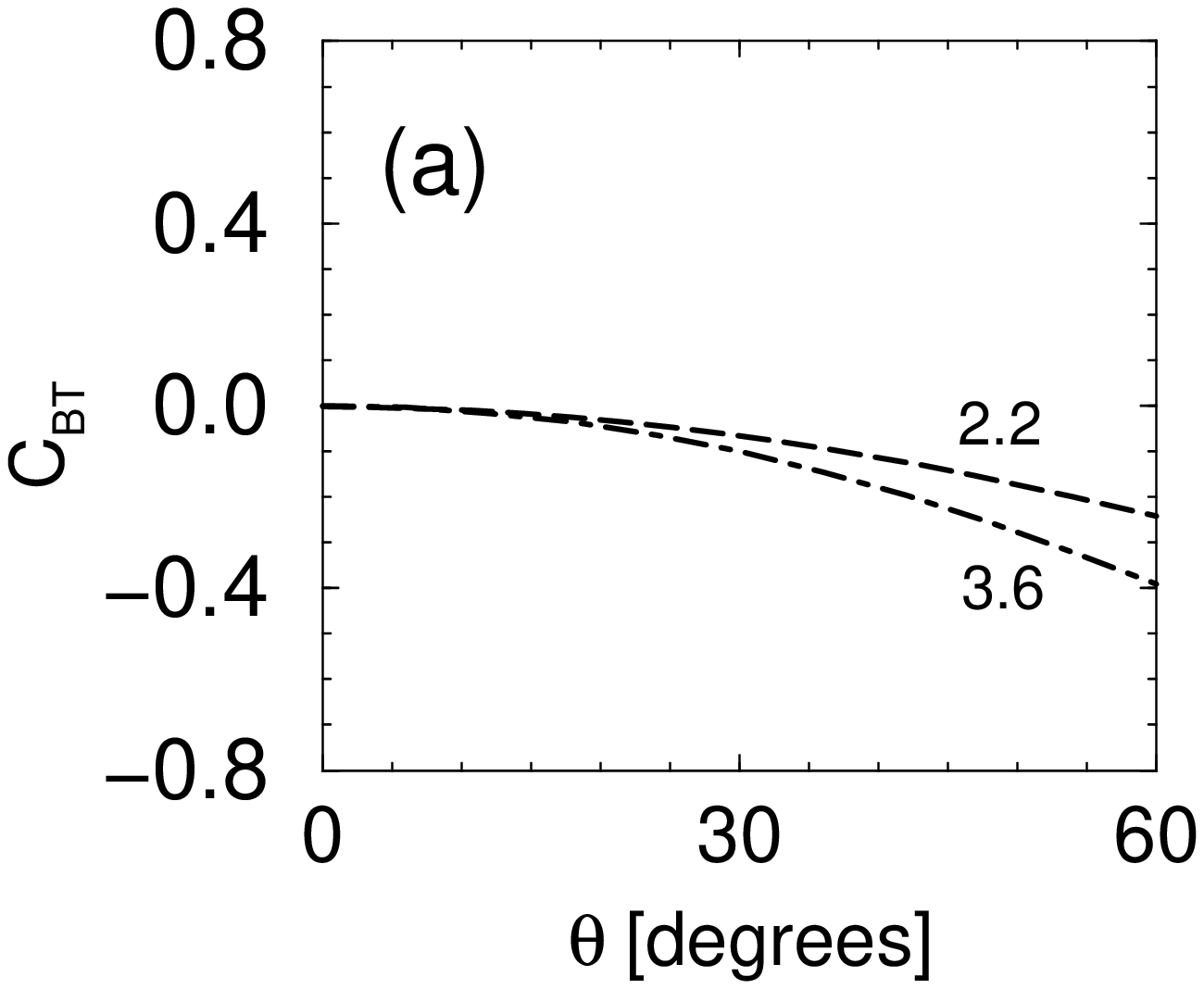}\qquad
 \includegraphics[width=55mm]{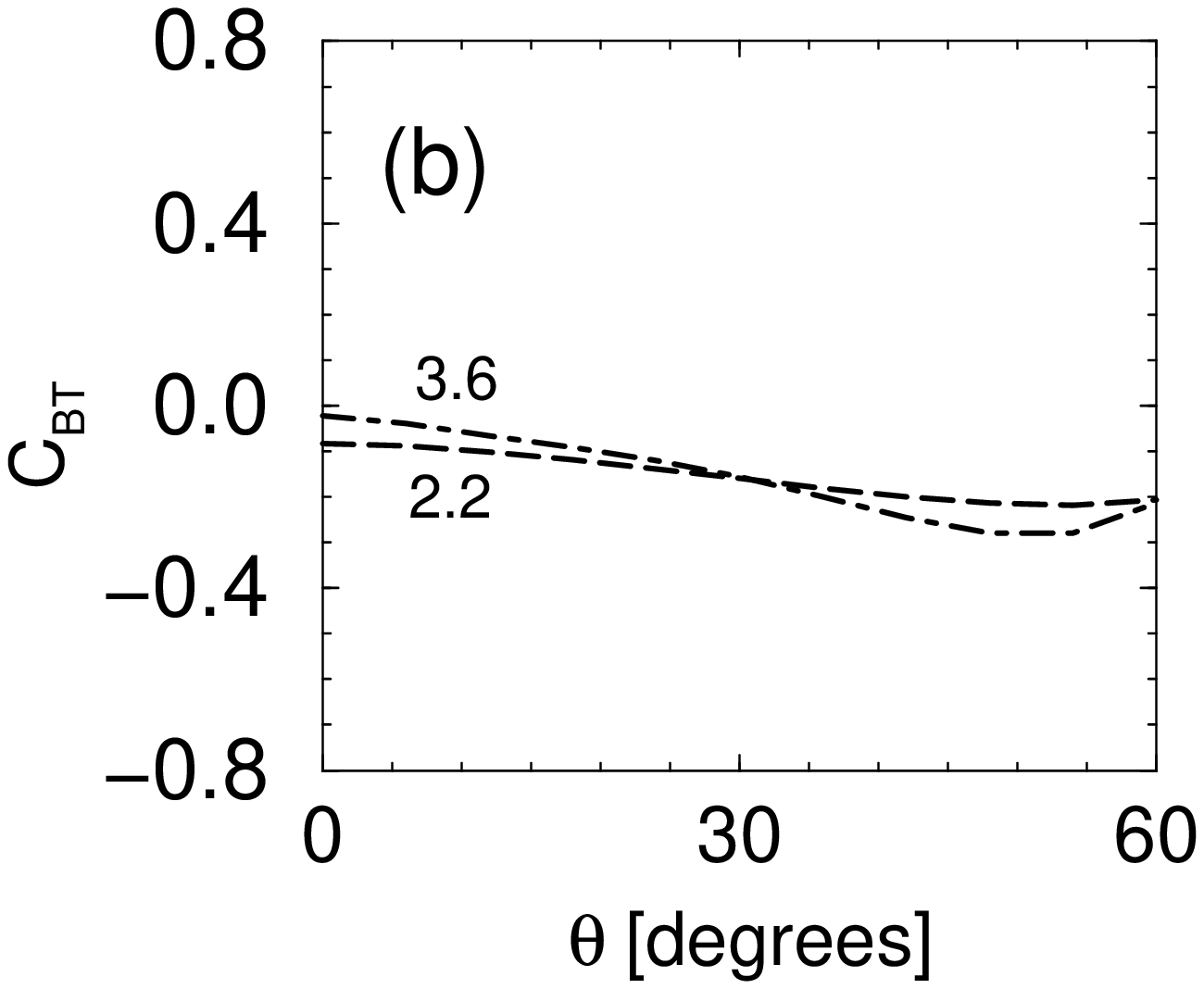}
 \vspace{0.5cm}
 \caption{
 The beam-target
 asymmetry for $E_\gamma=2.2$ and $3.6$ GeV at forward photoproduction angles.
 (a): Result for the Pomeron exchange amplitude;
 (b): Result for the full model. }
 \label{fig:15}
\end{figure}

 Next, we investigate   the beam-target asymmetry which may be studied in
 reactions with a circularly-polarized beam and polarized target.
 Using the notation of
 Eq.~(\ref{C-BT}), the expression for helicity-conserving amplitudes
 (\ref{I_NU}) and neglecting spin-flip processes,
 we can estimate $C_{BT}$ at forward-angle
 photoproduction as
 \begin{eqnarray}
 C_{BT}(t_{\rm max})\simeq
 2 |\alpha^U|\cos(\delta_{\rm N}-\delta_{\rm U}),
  \label{CBT1}
 \end{eqnarray}
  where $\delta_{\rm N},\,\delta_{\rm U} $ are the phases of the natural
  and unnatural-parity-exchange amplitudes, respectively.
  From this expression it looks very attractive to use $C_{BT}$ as a tool
  for studying contribution from exotic processes with
  unnatural-parity-exchange, because the asymmetry depends on $\alpha^U$, but not
  on $|\alpha^U|^2$~\cite{TOYM98}.

  However, analysis  of the pure Pomeron-exchange amplitude
  shows that the second term of Eq.~(\ref{G2g-fi})
  gives some contribution to $C_{BT}$ even without admixture
  of an unnatural-parity exchange component. This term contains a
  part which describes the interaction of the photon and proton spins which
  is governed  by condition
  \begin{eqnarray}
  {\bf s}_p+{\bm\lambda}_\gamma ={\bf s}_{p'},
  \end{eqnarray}
  and select the initial state with the total spin $|{\bf s}_i| = \frac12$.
  As a result, the beam-target asymmetry has an additional contribution
  \begin{eqnarray}
   \Delta C_{BT}&\simeq&
   - \frac{(E_p+M_N)(E_{p'}+M_N)(n-n'\cos\theta)^2(q(M_\phi^2+|t|^2 +kM_\phi))^2}
   {4{\bar s}^2(M_\phi(E_\phi+M_\phi))^2},\nonumber\\
   n&=&\sqrt{\frac{E_p-M_N}{E_p+M_N}},
   \qquad\qquad n'=\sqrt{\frac{E_{p'}-M_N}{E_{p'}+M_N}}.
  \label{C-BTP}
  \end{eqnarray}
 Fortunately, at $E_\gamma=2-3$ GeV and for $|t|\simeq |t|_{\rm min}$
 this term is small  and disappears with increasing $E_\gamma$.
 However, it is quite large at large momentum transfers. This
 is illustrated in Fig.~\ref{fig:15} where we show calculations for
 the Pomeron-exchange amplitude and  for the full model
 for $E_\gamma=2.2$ and 3.6 GeV.
 At small $|t|$ and $E_\gamma\sim2-3$ GeV
 the conventional processes
 considered above do not contribute to $C_{BT}$ and it may be
 used as a tool to study the non-diffractive component with unnatural-parity
 exchange, like $s\bar s$-knockout~\cite{TOYM98}, etc.
 At large $|t|$ the beam-target asymmetry is defined by the interplay
 of all channels and is very sensitive to the production mechanism.

\section{summary}

 In this paper we have discussed several  topics of
 current interest for the $\phi$ meson photoproduction at low energies.
 In particular, we found that the  spin-dependent interaction in
 the diffractive (Pomeron exchange) amplitude is responsible
 for the spin-flip transitions which are suppressed completely
 in the  helicity-conserving processes. These transitions
 give sizeable contributions to the spin-density matrix elements
 and may be measured via the angular distributions of $K^+K^-$ decays
 in reactions with unpolarized and polarized photons.
 Of  special interest is the finite and large
 value of the $\rho^0_{1-1}$ matrix element generated by the double spin-flip
 transition. It is caused
 by the spin-orbital interaction  inherent to the two-gluon exchange
 processes in the diffractive channel.
 This matrix element generates
 the $\Phi$-dependence of the decay-angular distribution
 in reactions with unpolarized photons at forward angles.
 That is, the spin observables at small $|t|$ may be used as a tool
 for studying diffractive mechanism in detail.

  Combined study of $\phi$ and $\omega$ photoproduction at large
  angles allows the analysis of the status of OZI-rule for $\phi NN^*$
  -interactions relative to the standard estimation based
  on the violation of $\phi-\omega$ mixing from its ideal value.
  We found a large (factor of 4) scale of this violation which agrees
  with other independent indications for this effect.

  We also have shown that spin observables at large momentum
  transfers are due to
  the interplay between the resonant and all other channels, and
  therefore, may be used to test the resonance excitation
  mechanism, which is a issue of current interest.

  It would be interesting to extend our analysis of spin-density
  matrix elements to the $\omega$-photoproduction too. This needs
  to include into consideration additional channels like initial
  and final state interactions~\cite{OhLee}, direct
  quark exchange~\cite{JLabomega},
  contribution of the conventional meson
  trajectories~\cite{Laget2000} and others. This will be subject
  of future study.

  Another interesting problem we have not investigated in this work
  is to use the spin observables to
  extract information about the exotic isoscalar processes, like $s\bar
  s$-knockout~\cite{Henley,TOYM98}, $G$-poles~\cite{Kochelev},
  etc. with unnatural-parity-exchange properties. The most
  challenging question is to exclude the contribution of the
  pseudoscalar $\pi$-meson exchange. This problem may be solved
  in a combined study of the spin-density  matrix elements measured using
  linearly-polarized photons and the beam-target and/or beam-recoil double
  polarizations measured using circularly-polarized photons on the proton
  and deuteron targets~\cite{TFL02}. It is then possible to determine
  the absolute value and the phase of the "exotic channel".
  The corresponding theoretical estimations will be presented in our
  forthcoming paper.

\acknowledgments

 We thanks S. Date, H.~Ejiri, M. Fujiwara, K.~Hicks, T.~Hotta,
 K. Mawatari, T.~Mibe, T.~Nakano, D.J. Tedeschi and
 R.G.T.~Zegers for fruitful discussion.
 This work was supported in part
 by U.S. Department of
 Energy, Nuclear Division, Contract N$^o$ W-31-109-ENG-38.

\appendix
\section{Comparison studies of diffractive mechanisms}

 Up to now we have discussed  diffractive photoproduction, where the
dominance contribution comes from the Pomeron exchange and the
additional trajectories are added to improve  the total
unpolarized cross section at low energy. Let us denote this
diffractive model as a model "P". As we discussed in Sect.~III the
finite value of threshold parameter $a_P\sim s_P$ in
(\ref{s_thresh}) eliminates the Pomeron contribution at
$E_\gamma\sim2$ GeV,
 which must be compensated by an increase of strength
 of the additional trajectories.
 The validity of different assumptions must be checked
 in study of the polarization
 observables in diffractive region, because the vertex functions
 for the Pomeron and  other trajectories in Eqs.~(\ref{h-munu}) - (\ref{h-f2})
 lead to different predictions.
 By way of illustration we shell compare
 the model~"P", with the model~"$2^+$" which represents
 the $f_2'$ Regge trajectory and PS-meson
 exchange
 and  the model "$0^+$", which  is the PS-meson exchange
 and the scalar (glueball-exchange)  trajectory. The  two last
 cases are realized if one choose the threshold parameter
 in (\ref{s_thresh}) ${a_P}\simeq s_R\sim(M_N+M_\phi)^2$ and $a_{R\neq P}=0$.
 \begin{figure}[h] \centering
 \includegraphics[width=0.3\columnwidth]{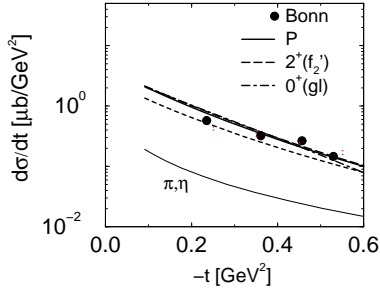}
 \caption{
 (a): The differential cross section of  $\gamma p\to \phi p$
  reaction at $E_\gamma=2.2$ GeV for the models
  "P", $"2^+"$ and "$0^+$" indicated by solid dashed,
  and dot-dashed curves, respectively. The contribution of the
  pseudoscalar $\pi,\eta$-exchange is shown by the solid thin
  curve. The short dashed curve corresponds to the case when
  diffractive channel represents Pomeron and PS-exchange.
  Data are taken from Ref.~\protect\cite{Bonn}}.
 \label{fig:18}
\end{figure}
 \begin{figure}[h] \centering
 \includegraphics[width=0.3\columnwidth]{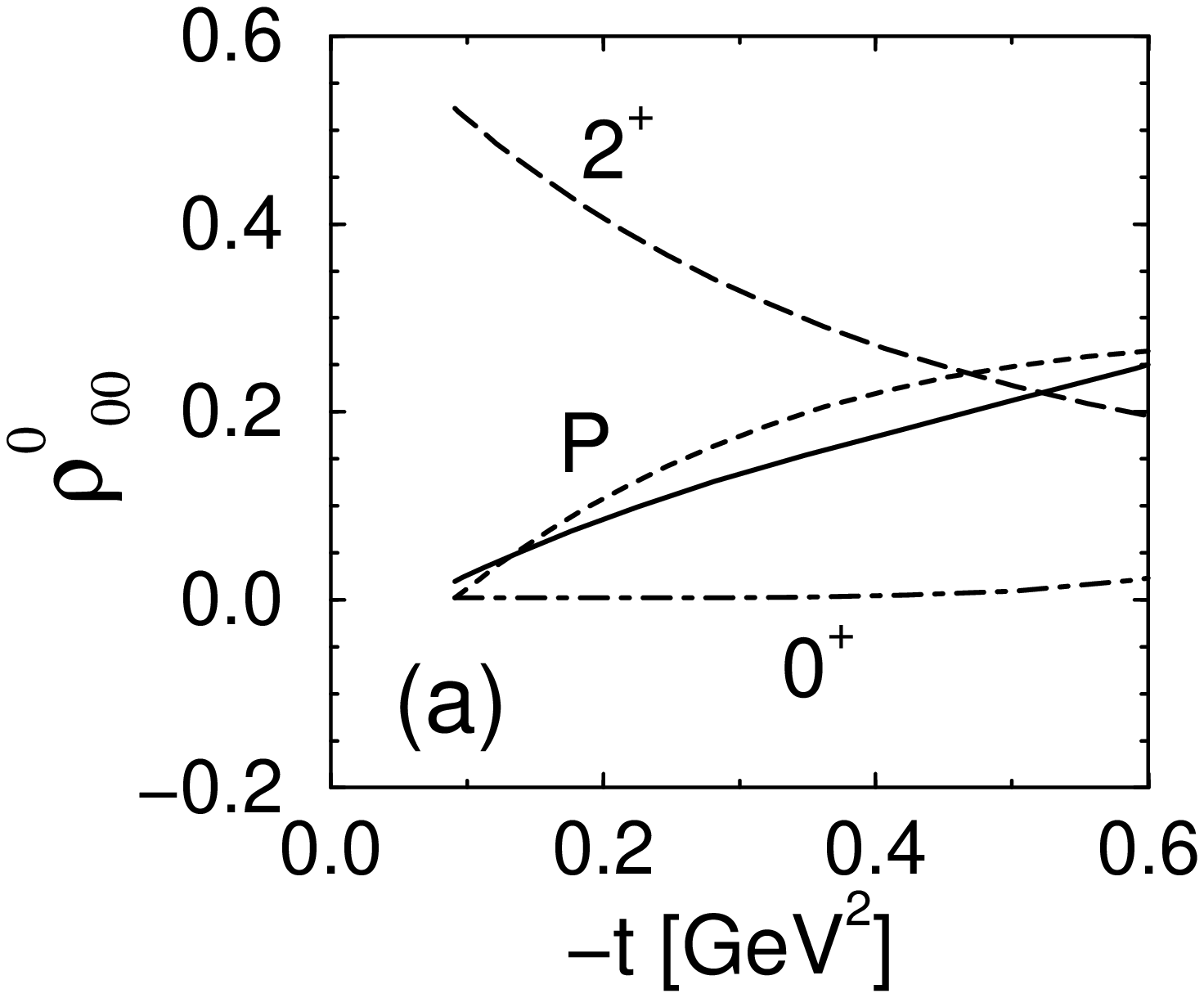}\qquad
 \includegraphics[width=0.31\columnwidth]{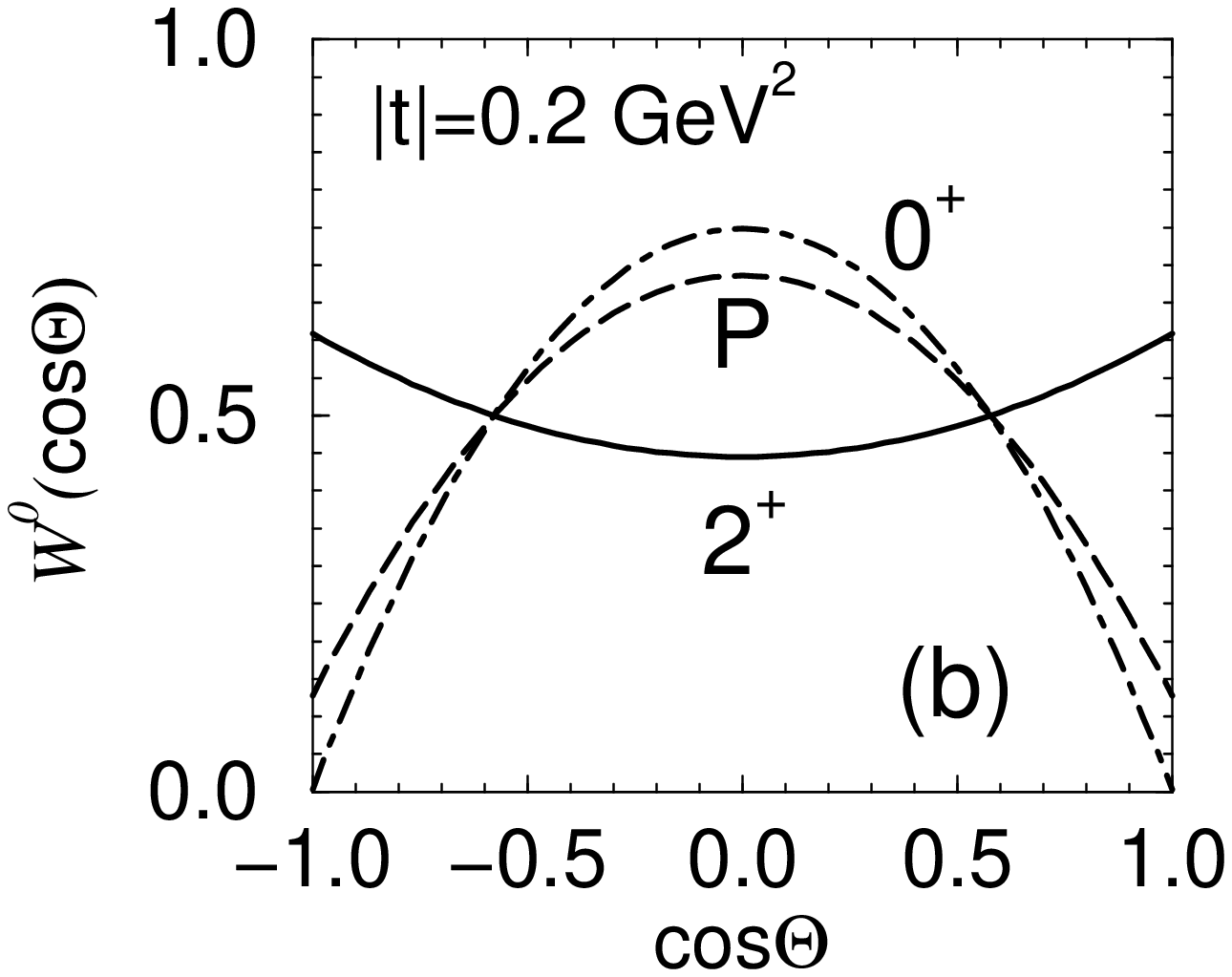}
 \caption { The spin-density matrix element
 $\rho^0_{00}$ (a) and  the angular distribution $W^0(\cos\Theta)$
 (b) for the three models of diffractive $\phi$-meson
 photoproduction.
 Notation is the same as in Fig.~\protect\ref{fig:18}.}
 \label{fig:19}
\end{figure}
 \begin{figure}[h] \centering
 \includegraphics[width=0.3\columnwidth]{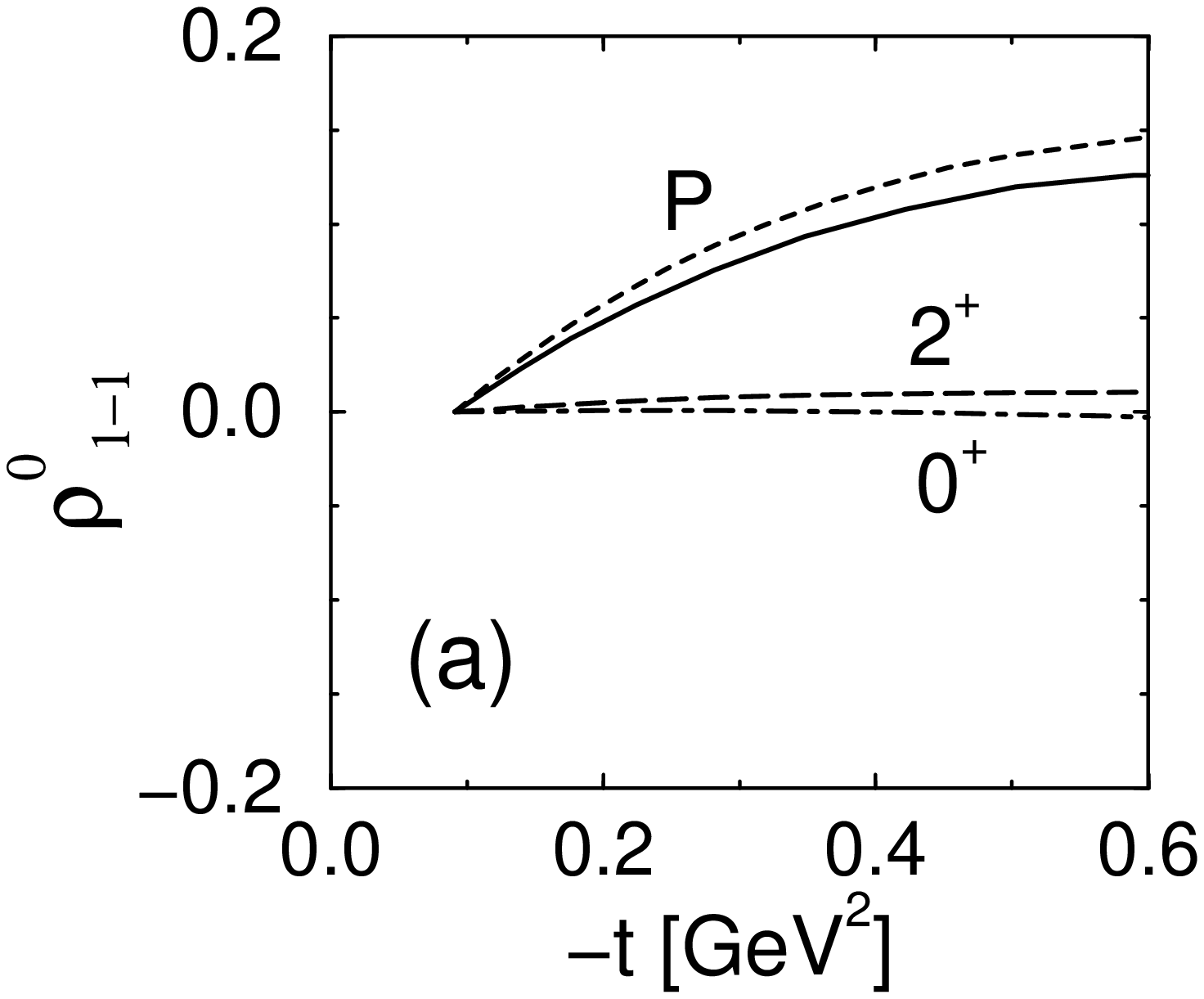}\qquad
 \includegraphics[width=0.31\columnwidth]{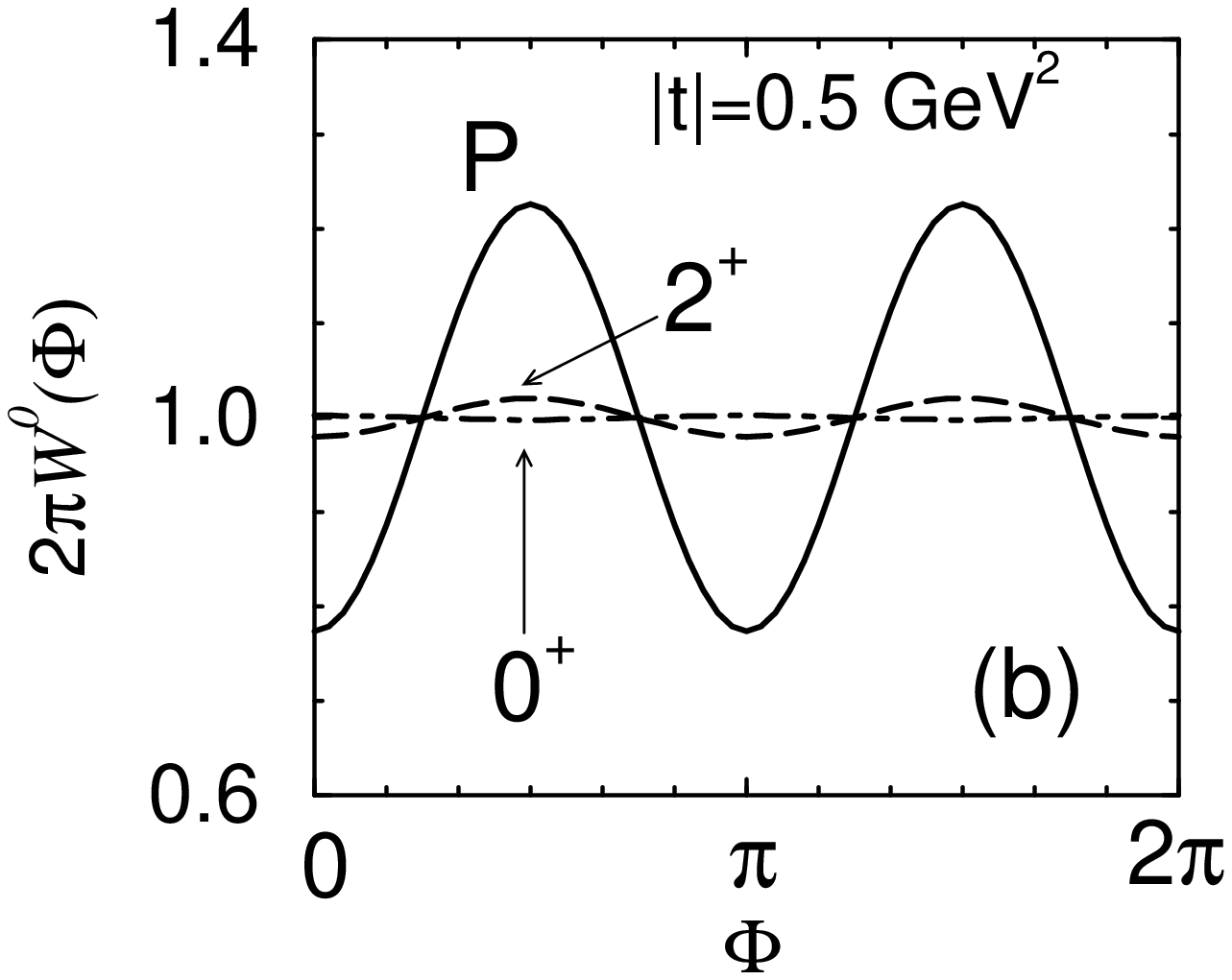}
 \caption { The spin-density matrix element
 $\rho^0_{1-1}$ (a) and  the angular distribution $W^0(\Phi)$ (b)
 for the three models of diffractive $\phi$-meson
 photoproduction.
 Notation is the same as in Fig.~\protect\ref{fig:18}.}
 \label{fig:20}
\end{figure}
 \begin{figure}[h] \centering
 \includegraphics[width=0.3\columnwidth]{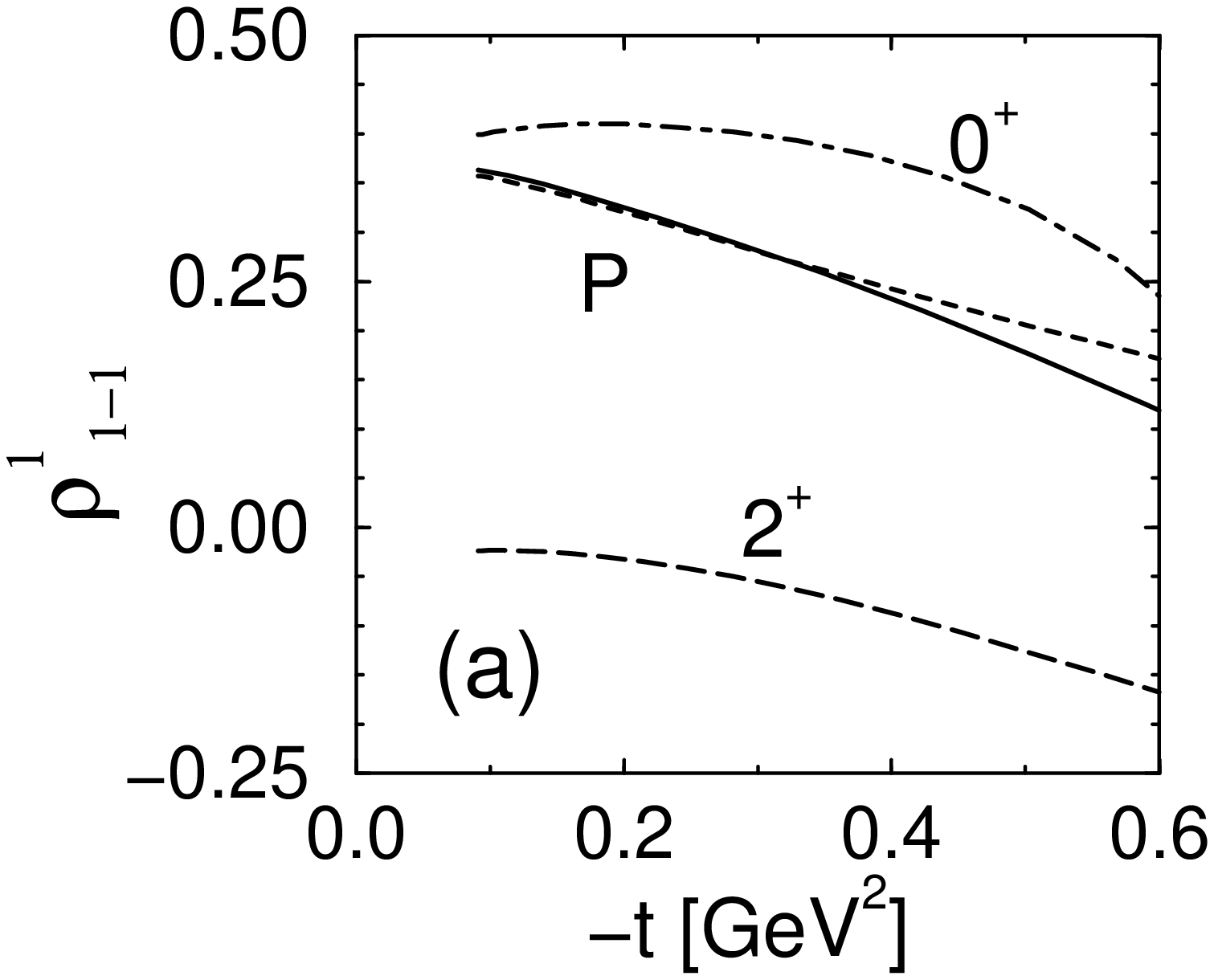}\qquad
 \includegraphics[width=0.31\columnwidth]{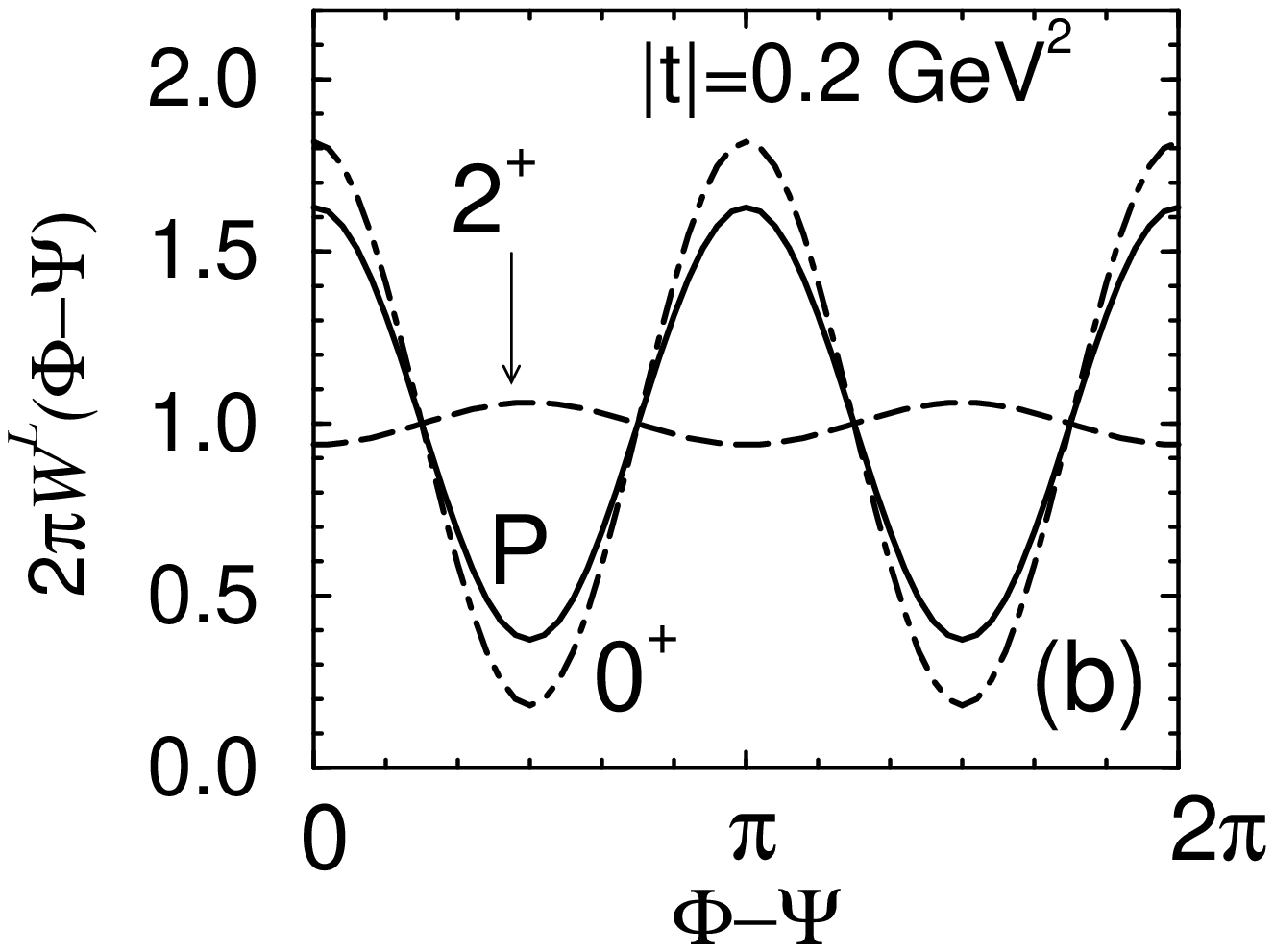}
 \caption { The spin-density matrix element
 $\rho^1_{1-1}$ (a) and  the angular distribution
 $W^L(\Phi-\Psi)$ (b)
 for the three models of diffractive $\phi$-meson
 photoproduction with linearly polarized photons.
 Notation is the same as in Fig.~\protect\ref{fig:18}.}
 \label{fig:21}
\end{figure}

 In Fig.~\ref{fig:18} we show the differential cross section of
 $\gamma p\to \phi p$ reaction as a function of $-t$ at $E_\gamma=2.2$ GeV for
 the three models together with the available experimental
 data~\cite{Bonn}. Parameters of the Pomeron exchange amplitude
 are fixed from the high energy region, for other trajectories
 we use $g_{2^+}=0.82\,g_{P}$, and
 $g_{0^+}=3.9\,g_{P}$.
 The contribution of the
 pseudoscalar $\pi,\eta$-exchange is also shown for completeness.
 We also show case when the diffractive channel is built from the
 Pomeron and PS-exchanges. In this case the calculation is slightly below
 data. But all other three models are close to each other and it is difficult
 to distinguish between them
 using only unpolarized differential cross section.
 Situation reverses if we look at spin observables.

 Fig.~\ref{fig:19}(a) shows
 $\rho^0_{00}$ for different models as a function of $-t$
 at $E_\gamma=2.2$ GeV. Thus, for the
 $0^+$-exchange $\rho^0_{00}\simeq 0$, for the  Pomeron exchange it
 increases monotonically with $|t|$.  For $2^+$-exchange
 $\rho^0_{00}$ decreases with $t$, starting from a large value at
 $|t|=|t|_{\min}$, because of the spin-flip terms.
 They are the first and  the third terms in the square brackets
 of (\ref{h-f2}). These terms generate a finite value of $\rho^0_{00}$
 even at forward angle photoproduction.
 The  difference in $\rho^0_{00}$ leads to the difference in
 decay distribution $W^0(\cos\Theta)$ shown in
 Fig.~\ref{fig:19}(b) for $|t|=0.2$ GeV$^2$.
 For $0^+$-exchange,  $\phi$-mesons produced to be transversely
 polarized, the Pomeron
 exchange process results in partial $\phi$-meson "depolarization".
 In case of $2^+$-exchange we get rather strong $\phi$-meson depolarization
 with an enhancement of the longitudinal polarization at $|t|=|t|_{\min}$.
 One can also see, that the difference between "hybrid" model "P"
 (solid curve) in Fig.~\ref{fig:19}(a) and pure Pomeron-exchange model (short
 dashed curve) is negligible.

 In Fig.~\ref{fig:20}(a) we show $\rho^0_{1-1}$ generated by the
 the double spin-flip transition $\lambda_\gamma\to
 \lambda_\phi=-\lambda_\gamma$. This matrix element is proportional
 to interference of helicity conserving and double spin-flip
 transition amplitudes and almost equal to zero for $0^+$-model.
 For "P" and $2^+$ models it  increases with $|t|$
 monotonically, but in $2^+$-exchange it is much smaller.

 As we have discussed above the matrix element $\rho^1_{1-1}$ depends
 on the contribution of unnatural parity exchange components and
 strength of the single-spin-flip component or $\rho^0_{00}$ (cf.
 Eq.(\ref{rho11-1-f})). In  $2^+$-model the spin-conserving component
 or $\rho^0_{11}$ is dominated by PS-exchange and therefore the relative
 contribution of unnatural parity exchange is smaller then in other cases.
 Also $\rho^0_{00}$ is greater. This leads to strong decreasing of $\rho^1_{1-1}$
 at forward angles up to the negative values,
 which is illustrated in Fig.~\ref{fig:21}(a). At small and finite
 $|t|$ this matrix element is the biggest for $0^+$ - model, since
 there are no  spin-flip processes in this case.
The corresponding angular distributions $W^L(\Phi - \Psi)$ at
$|t|=0.2$ GeV$^2$ are shown in Fig.~\ref{fig:21}(b). One can see
strong difference between $2^+$-exchange and other models.
Existing data~\cite{Ballam73}  support for the small value for
$\rho^0_{00}$ and finite value for $\rho^1_{1-1}$. This eliminates
large component of $2^+$-exchange. In order to distinguish between
Pomeron and $0^+$-exchange one needs at least data on
$\rho^0_{1-1}$-matrix element (cf. Fig.~20) which is crucial for
these models on the qualitative level.



 \begin{table}
  \caption{ Comparison of calculated spin density matrix elements
   in  helicity system
   with experimental data of~\protect\cite{Ballam73} at
   $E_\gamma=2.8$ and 4.7 GeV  for $0.02\leq|t|\leq 0.8$ GeV$^2$.
   Theoretical prediction is done
   at $E_\gamma=3.75$ and for $|t|=0.4$ GeV$^2$.}
  \label{tab:I}
 \begin{ruledtabular}
 \begin{tabular}{lrrc}
 $ \rho_{\lambda\lambda'} $&\qquad Exper. &\qquad Calc.\\ 
 $ \rho^0_{00} $  &  $-0.04\pm 0.06$ & $ 0.061$ &         \\ 
 $ {\rm Re}\rho^0_{10} $  & $-0.00\pm 0.06$ & $- 0.067$ & \\ 
 $ \rho^0_{1-1} $  & $-0.04\pm 0.10$ & $ 0.042$ &         \\ 
 $ \rho^1_{00} $  & $-0.13\pm 0.09$ & $ 0.010$ &          \\ 
 $ \rho^1_{11} $  & $-0.06\pm 0.11$ & $ 0.018$ &          \\ 
 $ {\rm Re}\rho^1_{10} $  & $0.00\pm 0.09$ & $ 0.063$ &   \\ 
 $ \rho^1_{1-1} $  & $0.18\pm 0.13$ & $ 0.44$ &           \\ 
 $ {\rm Im}\rho^2_{10} $  & $-0.02\pm 0.10$ & $ -0.052$   \\ 
 $ {\rm Im}\rho^2_{1-1} $  & $-0.51\pm 0.16$ & $ -0.44$   \\ 
 \hline
\end{tabular}
\end{ruledtabular}
 \vspace*{0.5cm}
\end{table}

\end{document}